\documentclass[aps,
               pra,
               floatfix,
               twocolumn,
               showpacs,
               amsmath,
               amssymb,
               reprint,
               10pt,
              ]{revtex4-2}

\usepackage{graphicx}
\usepackage{dcolumn}
\usepackage{bm,amssymb,amsmath,dsfont}   
\usepackage{lipsum}
\usepackage{physics}
\usepackage[T1]{fontenc}
\usepackage{txfonts}
\usepackage{bbold}
\usepackage{float}
\usepackage{hyperref}
\hypersetup{
    colorlinks=true,
    linkcolor=blue,
    citecolor=blue,
    urlcolor=blue}

\usepackage{pstricks}
\usepackage{tikz}
\usepackage{color}
\usepackage{xcolor}
\DeclareFontFamily{OT1}{cmm}{}
\DeclareMathAlphabet{\mathcm}{OML}{cmm}{m}{it}
\makeatletter
\renewcommand{\vec}[1]{\mathbf{#1}}
\newcommand{\eq}[1]{(\ref{eq:#1})}
\newcommand{\Eq}[1]{Eq.\,\eqref{eq:#1}}

\newcommand{\Fig}[1]{Fig.~\ref{fig:#1}}
\newcommand{\fig}[1]{\ref{fig:#1}}

\newcommand{\App}[1]{App.~\ref{app:#1}}

\renewcommand{\i}{\text{i}}

\definecolor{applegreen}{rgb}{0.55, 0.71, 0.0}
\definecolor{byzantine}{rgb}{0.74, 0.2, 0.64}

\newcommand{\phis}{\varphi_\mathrm{s}}
\newcommand{\mf}{m_\mathrm{F}}
\newcommand{\ncone}{\tilde{\rho}|c_1|}
\newcommand{\totd}{\tilde\rho}
\newcommand{\sided}{\rho}
\newcommand{\dsided}{\delta\sided}
\newcommand{\sidedm}{n}
\newcommand{\e}{\mathrm{e}}
\renewcommand{\ncone}{\totd|c_1|}

\makeatletter
\let\cat@comma@active\@empty
\makeatother

\begin{document}

\preprint{APS/123-QED}

\title{Double sine-Gordon class of universal coarsening dynamics in a spin-1 Bose gas}

\author{Ido Siovitz}
\author{Anna-Maria E. Glück}
\author{Yannick Deller}
\author{Alexander Schmutz}
\author{\\ Felix Klein}
\author{Helmut Strobel}
\author{Markus K. Oberthaler}
\author{Thomas Gasenzer}
\email{t.gasenzer@uni-heidelberg.de}
\affiliation{Kirchhoff-Institut f\"{u}r Physik, Ruprecht-Karls Universit\"{a}t Heidelberg, Im Neuenheimer Feld 227, 69120 Heidelberg, Germany}

\date{\today}

\begin{abstract}
Far from equilibrium, universal dynamics prevails in many different situations, from pattern coarsening to turbulence.
A central longstanding problem concerns the development of a theory of coarsening that rests on the microscopic properties of the system and allows identifying the interaction mechanisms underlying a possible overarching universality class of the associated scaling dynamics.
In quantum systems, this is complicated by the existence of nonlinear and topological excitations due to the compact nature of phase degrees of freedom.
We show that the double sine-Gordon model as a noncompact low-energy effective model of the spin-1 Bose gas accounts for subdiffusive coarsening dynamics, identifying field configurations spread over multiple wells of the sinusoidal potential as a precondition for the slow scaling.
This is in contrast to diffusion-type scaling which the model is known to exhibit as well, where field configurations are seen to not extend over more than two wells. 
Experimental observations of a spinor BEC support these characteristics, thus constituting a platform for the investigation of sine-Gordon dynamics.
Our results point to a path towards a classification of pattern coarsening in many-body systems on the basis of  microscopic models.
\end{abstract}

\maketitle


\section{\label{sec:Intro} Introduction}
%
Universal dynamics of quantum systems far from equilibrium has garnered significant attention in modern research, ranging from 
superfluid turbulence \cite{Vinen2006a,Tsubota2008a}, 
and wave turbulence \cite{Zakharov1992a,Nazarenko2011a}
to nonthermal fixed points \cite{Berges:2008wm,Schole:2012kt,PineiroOrioli:2015dxa,Chantesana:2018qsb.PhysRevA.99.043620,Mikheev:2018adp}.
The aim is to develop a unified framework for characterizing and classifying far-from-equilibrium scaling evolution, inspired by, but going beyond universality in and near equilibrium \cite{Hohenberg1977a,Janssen1979a,Diehl1986a,Janssen1992a}.
Long studied scaling phenomena include pattern coarsening and phase-ordering kinetics \cite{Bray1994a.AdvPhys.43.357,Puri2019a.KineticsOfPT,Cugliandolo2014arXiv1412.0855C},
for which a unifying scaling theory is lacking.
Such a theory should ideally result from the underlying microscopic dynamics of the considered system, e.g.~as a reduction to effective, relevant degrees of freedom. 
It would potentially lead beyond generalized diffusion models and provide a scaling theory defining the universality class the coarsening process belongs to. 
A growing demand as well as the potential for advancing such a framework is underlined by the extensive recent experimental 
\cite{%
Henn2009a.PhysRevLett.103.045301,
Gring2011a,
AduSmith2013a,
Langen2015b.Science348.207,
Navon2015a.Science.347.167N,
Navon2016a.Nature.539.72,
Rauer2017a.arXiv170508231R.Science360.307,
Gauthier2019a.Science.364.1264,
Johnstone2019a.Science.364.1267,
Eigen2018a.arXiv180509802E,
Prufer:2018hto,
Erne:2018gmz,
Navon2018a.Science.366.382,
Glidden:2020qmu,
GarciaOrozco2021a.PhysRevA.106.023314,
Lannig2023a.2306.16497,
Martirosyan:2023mml,
Gazo:2023exc,
MorenoArmijos2024a.PhysRevLett.134.023401,
Martirosyan:2024rxm}
and theoretical efforts 
\cite{%
Berges:2004ce,
Kodama:2004dk,
Barnett2011a,
Marcuzzi2013a.PhysRevLett.111.197203,
Langen2013a.EPJST.217,
Nessi2014a.PhysRevLett.113.210402,
Gagel2014a.PhysRevLett.113.220401,
Bertini2015a.PhysRevLett.115.180601,
Babadi2015a.PhysRevX.5.041005,
Buchhold2015a.PhysRevA.94.013601,
Berges:2008sr,
Nowak:2012gd,
Hofmann2014a,
Maraga2015a.PhysRevE.92.042151,
Williamson2016a.PhysRevLett.116.025301,
Williamson2016a.PhysRevA.94.023608,
Bourges2016a.arXiv161108922B.PhysRevA.95.023616,
Chiocchetta:2016waa.PhysRevB.94.174301,
Karl2017b.NJP19.093014,
Schachner:2016frd,
Walz:2017ffj.PhysRevD.97.116011,
Schmied:2018upn.PhysRevLett.122.170404,
Mazeliauskas:2018yef,
Schmied:2018osf.PhysRevA.99.033611,
Williamson2019a.ScPP7.29,
Schmied:2019abm,
Spitz2021a.SciPostPhys11.3.060,
Gao2020a.PhysRevLett.124.040403,
Wheeler2021a.EPL135.30004,
Gresista:2021qqa,
RodriguezNieva2021a.arXiv210600023R,
Preis2023a.PhysRevLett.130.031602,
Liu:2022rss,
Heinen:2022rew,
Heinen2023a.PhysRevA.107.043303,
Siovitz:2023ius.PhysRevLett.131.183402,
Gliott2024a.PhysRevLett.133.233403,
Noel2024:PhysRevD.109.056011,
Noel:2025mtb},
see also
\cite{%
Berges:2015kfa,
Langen:2016vdb,
Schmied:2018mte,
Mikheev:2023juq}, 
exploring the nature of universal space-time scaling, to a large part in the field of ultra-cold atoms.

Coarsening and phase-ordering kinetics generically mean that order increases in a self-similar manner, characterized by the spatio-temporal scaling of order-parameter correlations.
For example, in spinor quantum gases, which we focus on here, subdiffusive 
\cite{Schmied:2018osf.PhysRevA.99.033611,Siovitz:2023ius.PhysRevLett.131.183402,Lannig2023a.2306.16497}
as well as diffusion-type coarsening 
\cite{Prufer:2018hto,Schmied:2019abm,Lannig2023a.2306.16497}
has been found in the structure of magnetic order.
The task is to isolate the relevant degrees of freedom and their interactions that account for the specific scaling dynamics.
For multi-component Bose gases with interaction suppressed density fluctuations, a low-energy effective theory (LEEFT), takes the form of a nonlinear Luttinger-liquid type model of the phase excitations \cite{Mikheev:2018adp}.
Assuming the absence of topological excitations, this effective theory makes it particularly easy to account for the scaling exponents \cite{Mikheev:2018adp}, which are confirmed numerically \cite{PineiroOrioli:2015dxa} and experimentally \cite{Prufer:2018hto,Lannig2023a.2306.16497}, while their direct derivation from the full nonlinear Schrödinger model is analytically cumbersome \cite{PineiroOrioli:2015dxa,Chantesana:2018qsb.PhysRevA.99.043620}.
If the pattern coarsening, however, involves topological excitations, one must take into account the compact phase of the quantum field in the statistical description of scaling. 
This is in particular the case for multi-component systems allowing inter-species exchange, such as spinor gases.
Nonlinear excitations prevail, thus preventing an analytical scaling analysis so far \cite{Schmied:2018osf.PhysRevA.99.033611,Schmied:2019abm,Siovitz:2023ius.PhysRevLett.131.183402}.
Hence, the reduction to an effective model explaining pattern coarsening is desirable, which accounts for the topological excitations in the underlying system.

\begin{figure*}[t]
    \centering
    \includegraphics[width=1\textwidth]{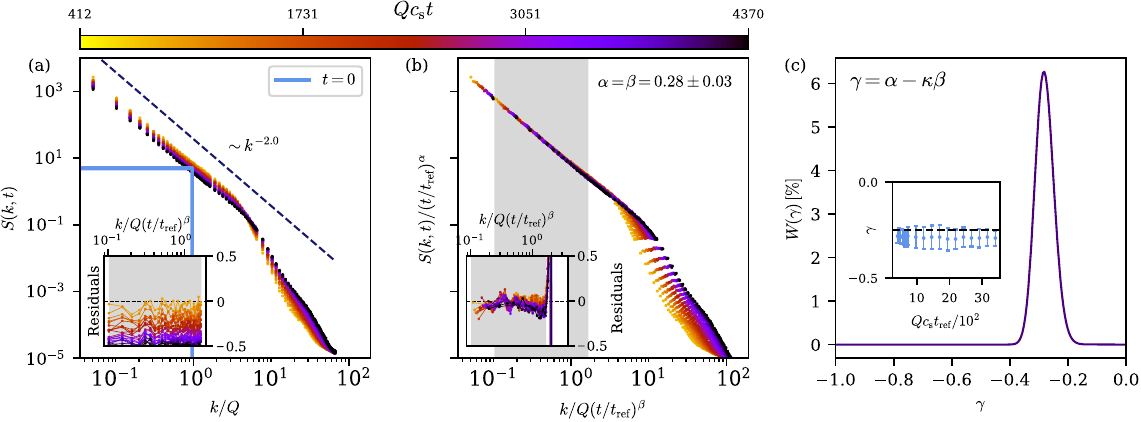}
    \caption{
    Subdiffusive self-similar scaling evolution near a nonthermal fixed point of the DSG model in (1+1)D.
    (a) Time evolution of the structure factor 
    $S(k,t) = \langle|\varphi(k,t)|^2\rangle$
     of the real scalar field.
     The initial $S(k,0)$ (blue line) is a box with cutoff $Q$. 
     At long times, the redistribution of excitations towards the IR leads to a power-law shape $S(k,t)\sim k^{-\kappa}$ at large wave lengths.
    (b) The collapse of the curves to the universal scaling function according to $S(k,t)=(t/t_\mathrm{ref})^{\alpha}S([t/t_\mathrm{ref}]^{\beta}k,t_\mathrm{ref})= (t/t_\mathrm{ref})^{\alpha-\kappa\beta} k^{-\kappa}$, to the reference time $t_{\mathrm{ref}} = 412/(Qc_\mathrm{s})$, with $c_\mathrm{s}$ denoting the free speed of sound, exhibits the spatio-temporal scaling of the correlator in the regime of low wave numbers, $k\ll k_{\xi_\mathrm{s}} = (2M\ncone)^{1/2} \approx 4\,Q$, with a resulting subdiffusive exponent $\alpha=\beta=0.28(3)$, and with $\kappa\simeq2.0$.
    The inset shows the residuals of the spectra w.r.t.~the reference spectrum, calculated as the relative difference of the rescaled spectra and the spectrum at $t_\mathrm{ref}$, with the equal distribution of errors confirming the self-similarity of the scaling.
    (c) Likelihood function of the scaling exponent $\gamma = \alpha-\kappa\beta = (1-\kappa)\beta$, from which the error of the exponents was extracted. The inset shows the center and width of the likelihood function for different reference times. }
    \label{fig:1DScaling}
\end{figure*}

Here, we show that the double sine-Gordon (DSG) model exhibits subdiffusive self-similar scaling far from equilibrium.
Focusing on the spatial field patterns and dynamics in the DSG model, we find that the spread of the field over many minima of the sinusoidal potential is a crucial characteristic of the subdiffusive coarsening, other than domain-size growth alone. 
This provides valuable insight into the features of the nonequilibrium universality class.

We show that the experimentally accessible spin-1 Bose gas in the easy-plane ferromagnetic phase can be mapped, in the infrared (IR), onto the DSG model as a low-energy effective theory.
By integrating out the weak density fluctuations in the fundamental Bose fields 
$\psi_{\mf} = \sqrt{\rho_{\mf}} \exp{\i\phi_{\mf}}$,
for the magnetic sublevels $\mf=0,\pm 1$, the resulting theory consists of dynamical equations for the relative phase angles, viz.~of the Larmor phase, 
$\varphi_\mathrm{L}=(\phi_1 - \phi_{-1})/2$,
and spinor phase,
$\phis  = \phi_1+\phi_{-1}-2\phi_0$.
Focusing on the easy-plane phase, we identify the spinor phase as the relevant degree of freedom. 
We conclude that the effective theory leads to the same far-from-equilibrium spatio-temporal, subdiffusive scaling as the spin-1 gas in one spatial dimension.

Remarkably, the unwrapping of the cyclic phase degree of freedom of the spinor gas translates topological information into the \emph{noncompact} DSG field.  
This provides a basis for the analytical classification and handling of scaling characteristics of the system.
We present numerical and experimental evidence of the validity of this sine-Gordon-type theory for a quasi-one-dimensional condensate of $~^{87}\mathrm{Rb}$ atoms in the $F=1$ ma\-ni\-fold.
Finally, we show for different choices of parameters and initial conditions, that the DSG also accounts for diffusion-type scaling evolution.

\section{Universal scaling dynamics of the\newline Double Sine-Gordon model}
\subsection{Self-similar scaling dynamics in (1+1)D}
We show the subdiffusive scaling dynamics of the double sine-Gordon (DSG) model, with $\varphi\in\mathbb{R}$,
\begin{align}
    \ddot{\varphi} = c_\mathrm{s}^2 \Delta \varphi -\lambda \sin\varphi + \lambda_\mathrm{s}\sin(2\varphi),
\end{align}
where $c_\mathrm{s}$ denotes the free speed of sound and $\lambda, \, \lambda_\mathrm{s}$ are the DSG couplings, cf.~\App{TWA} for details.
We choose the initial condition of $S(k,t) = \langle|\varphi(k,t)|^2\rangle$ to reflect a box distribution in momentum space with cutoff $Q$ (\Fig{1DScaling}a, blue line), and center the distribution around $\expval{\varphi} = \pi$, i.e., at a maximum of the cosine potential. 
This allows the system to randomly decay to the adjacent and further minima, accumulating in either of them at later times.
At $t \gtrsim 412/(Qc_\mathrm{s})$, the system enters a self-similar scaling regime, with the structure factor exhibiting a pure power-law, $S(k,t)\sim k^{-\kappa} \sim k^{-2}$, i.e., fractal form in the IR, as expected for the correlator of a phase angle \cite{Mikheev:2018adp,Mikheev2023a}. 
Hence, we may rescale $S(k,t)=(t/t_\mathrm{ref})^{\alpha}S([t/t_\mathrm{ref}]^{\beta}k,t_\mathrm{ref}) = (t/t_\mathrm{ref})^{\alpha-\kappa\beta} k^{-\kappa}$
by means of fitting $\gamma = (d-\kappa)\beta$, where we take $\alpha=d\beta$, corresponding to conservation of the momentum integral over $S$.
We find 
$\alpha=\beta=0.28(3)$ (see \Fig{1DScaling}b),
confirming distinctly subdiffusive ($\beta < 1/2$) scaling.
The inset of \Fig{1DScaling}c shows the independence of the exponent on the reference time. 

\subsection{Domain growth versus self-similar scaling}
\begin{figure*}[t]
    \centering
    \includegraphics[width=\linewidth]{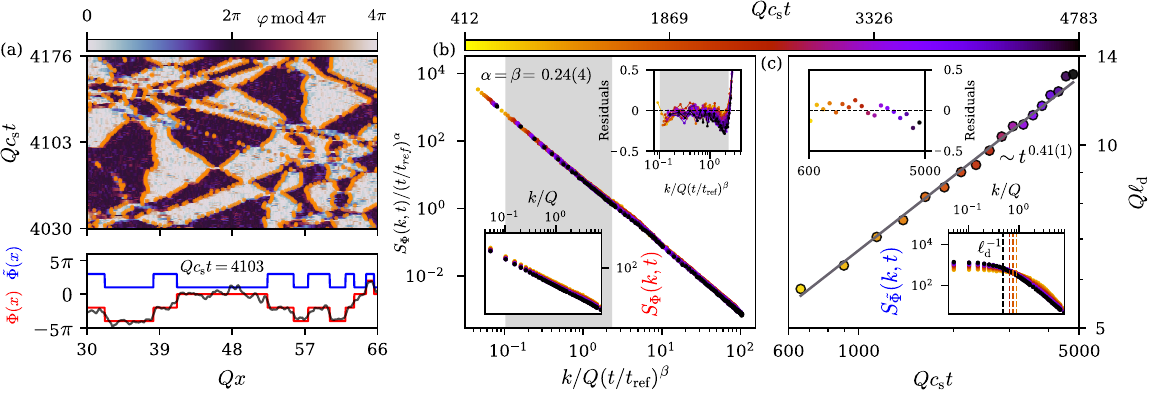}
    \caption{
    Defect coarsening in the dynamics of the DSG.
    (a) (upper panel) Excerpt of the time evolution of the DSG field $\varphi$ in a single TW run with full system length $L=122\,Q^{-1}$. 
    For better visibility, we plot it modulo $4\pi$.
    Domain walls are detected and denoted by orange markers.
    (lower panel) A function $\Phi(x)$ (red) is constructed by jumping by $2\pi$ ($-2\pi$) for each detected (anti-)kink.
    A function $\tilde{\Phi}(x)=\cos(\Phi(x)/2)$ (blue) alternates between $\pm 1$ for each detected defect, regardless of their signs, encoding the length-scale of domains alone. The blue curve is scaled and shifted for better visibility. 
    The fundamental field $\varphi$ is shown in grey.
    (b) Spatio-temporal evolution of defect correlations.
    The spatial correlation in Fourier space is calculated as 
    $S_\Phi(k,t) = \expval{\Phi(k)\Phi(-k)}$
    and averaged over $10^3$ realizations.
    In the IR, the correlation function shows, within the error bounds, the same scaling evolution as the spectra in \Fig{1DScaling}.
    Upper inset: Residuals, i.e., the relative difference of the rescaled spectra and the reference spectrum at $t_\mathrm{ref}=412/(Qc_\mathrm{s})$.
    Lower inset: Snapshots of the unscaled data.
    (c) Spatio-temporal evolution of domain sizes. 
    We extract the IR length-scale associated with domain walls $\ell_\mathrm{d}$ (cf.~yellow to black dashed lines in lower inset) from the correlation function $S_{\tilde{\Phi}}(k,t) = \langle\tilde{\Phi}(k)\tilde{\Phi}(-k)\rangle$  and observe a power law coarsening with $t^{0.41(1)}$, differing from the scaling of the spectra in (b).}
    \label{fig:Fig3}
\end{figure*}

In the DSG dynamics, the system decays into the various minima of the periodic potential, with domains of the respective field values forming dynamically. Relatively sharply defined cross-over regions are found separating them, as seen in \Fig{Fig3}a.

We investigate the spatial configuration of these domains by constructing a function $\Phi(x)$, which jumps by $2\pi$ ($-2\pi$) at each kink (anti-kink), thus isolating the effects of the domains themselves from other excitations, see \Fig{Fig3}b.
The structure factor of $\Phi$, 
$S_\Phi(k,t)=\expval{\Phi(k)\Phi(-k)}$,
is found to exhibit a power-law spectrum with $\kappa\approx2$ in the IR.
Using the same rescaling algorithm as in \Fig{1DScaling}, we obtain the scaling exponents 
$\alpha = \beta=0.24(4)$ (cf.~\Fig{Fig3}b).
The residuals in the upper inset of \Fig{Fig3}c indicate that this scaling is concentrated in the IR and that thus fluctuations across separate wells of the potential contribute to the overall universal dynamics.

We emphasize, though, that $\Phi(x)$ encodes more than the size of the domains seen in \Fig{Fig3}a. It captures the sequence of orientations of the kinks and thus the rescaling of the fractal pattern of steps as illustrated in the lower panel of \Fig{Fig3}a. 
Hence, it embodies the overall long-wave structure of the DSG field, which is possible due to the periodic symmetry of the noncompact DSG potential.
This becomes clear when reducing the field to $\tilde{\Phi}(x)=\cos(\Phi(x)/2)$, which alternates between $\pm1$ and thus encodes the domain length only. 
The time evolution of this length exhibits a power law exponent distinctly different from $\beta$, see \Fig{Fig3}c.
This shows that the subdiffusive coarsening of the DSG field is a more intricate phenomenon than domain coarsening and underlines the relevance of the long-range structure of $\varphi$, which spreads over several to many minima of the sinusoidal potential.
\subsection{Diffusion-type scaling vs. nonlinear equations}
\begin{figure*}[t]
    \centering  \includegraphics[width=\textwidth]{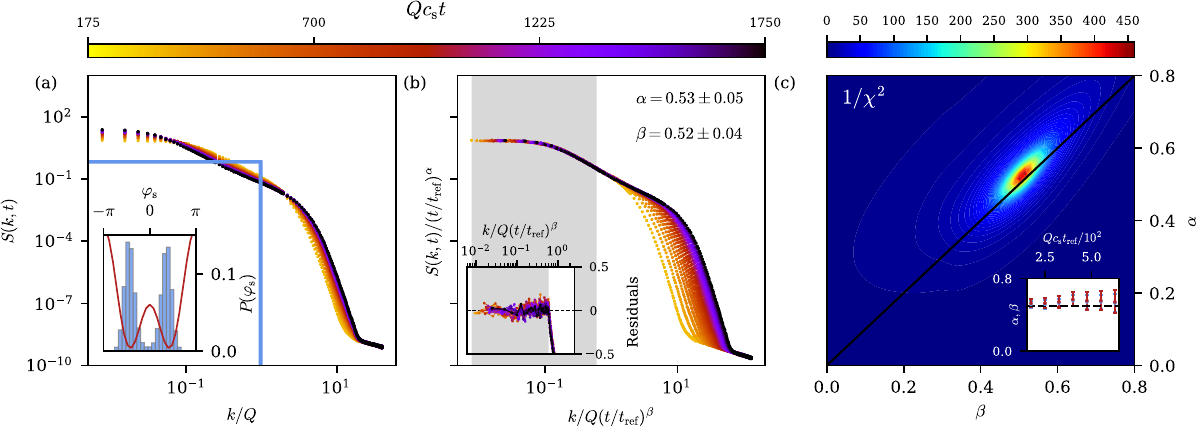}
    \caption{Self-similar scaling of the DSG model in (1+1)D with only two minima occupied.
    (a) Unscaled structure factor $S(k,t) = \langle \varphi(k,t)\varphi(-k,t) \rangle$ of the DSG dynamics, starting from the momentum box indicated by the blue line.  
    The form differs from that of \Fig{1DScaling} and shows a plateau, hinting at a dominant coarsening length scale in the system. The inset shows the PDF (blue bars), which is centered only around two minima of the shown bare potential (red).
    (b) Rescaled structure factor. Using the same algorithm as in \Fig{1DScaling}, we obtain diffusion-type scaling exponents $\alpha=0.53(5)$ and $\beta=0.52(4)$, obeying $\alpha=d\beta$ within errors, here for $d=1$. 
    The inset shows the residuals with an even distribution implying strict self-similarity.
    (c) Inverse $\chi^2$ distribution showing the most likely distribution. 
    The inset shows the stability of scaling w.r.t.~the reference time, with the dashed line indicating the value $0.5$.
    }
    \label{fig:1Dbeta05}
\end{figure*}

As a contrast to the subdiffusive scaling discussed above, the DSG model in $d=1$ spatial dimensions is found to also exhibit diffusion-type self-similar dynamics of the field correlations, see \Fig{1Dbeta05}. 
To achieve diffusion-type scaling, we chose the couplings such that the potential landscape changes significantly from the subdiffusive case, see \App{TWA} for details. The potential landscape now shows a \textit{local} maximum at $\expval{\varphi}=2\pi\mathbb{Z}$, degenerate \textit{global} maxima at $(2\mathbb{Z}+1)\pi$ and degenerate minima between them.
We initiate a momentum box and center the DSG field around $\expval{\varphi}=0$.
Thus, the system decays from the local maximum into the minima, but does not have enough energy to overcome the potential barrier at $\varphi = \pm \pi$, where the global maxima are. The system hence only occupies two minima of the sinusoidal potential.
Consequently, we obtain diffusion-type exponents of $\alpha=0.53(5)$ and $\beta=0.52(4)$, indicating that the number of occupied minima is of great importance to the scaling behavior of the system, cf.~\Fig{1Dbeta05} for our results.
The DSG model hence comprises both subdiffusive ($\beta < 1/2$) and diffusion-type ($\beta = 1/2$) scaling behaviors.\\

We note that subdiffusive and diffusive scaling do not imply that the evolution is governed by a simple diffusion-type equation as is it is often chosen for the phenomenological description of pattern coarsening \cite{Bray1994a.AdvPhys.43.357,Puri2019a.KineticsOfPT,Cugliandolo2014arXiv1412.0855C}.
There, a diffusion equation is used to describe self-similar scaling with $\beta=1/2$, reflecting the combination of a first-order time derivative with a second-order spatial derivative \cite{Bray1994a.AdvPhys.43.357}, and, e.g., the Cahn-Hilliard equation governs scaling with $\beta=1/4$, as it contains a fourth-order spatial derivative as a result of an additional conservation law \cite{Cahn1958a.JChemPhys.28.258}.
We emphasize, though, that the diffusion-type as well as the subdiffusive scaling observed in our numerical simulations and considered in our work is not to be identified automatically with pattern coarsening phenomenologically or microscopically described by either of these diffusion-type equations.
We rather point out that the description we aim at, in line with the microscopic description of the scaling close to a nonthermal fixed point, 
results in a description of the corresponding scaling on the grounds of the full nonlinear evolution of the system.
This typically requires an effective-theory description as introduced in the present work as well as a (non)perturbative approach to the scaling analysis of such a model.
For sine-Gordon-type models, such an analysis has been given in Refs.~\cite{Heinen2023a.PhysRevA.107.043303,Koeper2023a} and resulted in two different possible exponents, depending on the type of field configuration spreading in the periodic field potential of the model.
While diffusion-type scaling requires the field to remain within two minima of the model, corresponding to a simple $Z_2$ symmetry breaking with relevant coupling term $\lambda\varphi^4$, subdiffusive scaling requires the field to spread across many minima of the potential.
%

\section{Low-energy effective theory of the spin-1 Bose gas}
\subsection{Mapping the spin-1 Bose gas to DSG model}
In the following, we show that the far-from equilibrium dynamics of the spin-1 Bose gas in the easy-plane phase belongs to a sine-Gordon-type nonequilibrium universality class.
The spin-1 Lagrangian is given by
\begin{align}
    \mathcal{L} 
    &= \frac{\i}{2}\left( \psi_a^* \partial_t \psi_a - \psi_a \partial_t \psi_a^*\right) 
    - \frac{1}{2M}\nabla\psi_a^*\nabla \psi_a - q (f^{z})^{2}_{ab}\psi_a^*\psi_b 
    \nonumber \\
    &\ \ \ - \frac{c_0}{2}(\psi_a^*\psi_a)^2 - \frac{c_1}{2}\sum_{i\in\{x,y,z\}}(\psi_a^* f^i_{ab}\psi_b)^2,
    \label{eq:spin-1_lagrange_main}
\end{align}
where summation over the same indices is implied, 
$\psi_a$, $\mf=a\in \{-1,0,1\}$
represent the bosonic fields corresponding to the respective Zeeman magnetic sub-level $\mf$, $M$ is the particle mass and $q$ is the quadratic Zeeman shift, which induces an effective shift in the energies of the $\mf$ components relative to the $\mf=0$ component, $(f^{z})^{2}_{ab}=\delta_{ab}(1-\delta_{a0})$.
The linear Zeeman shift is absorbed into the fields by considering a rotating frame of reference. The term
$c_0 (\psi_a^*\psi_a)^{2}$
describes density-density interactions, whereas the term
$c_1 (\psi_a^* f_{ab}^i \psi_a)^{2}$
accounts for spin-dependent interactions, with $f^i$, $i \in \{x,y,z\}$ being the generators of the $\mathfrak{so}(3)$ Lie algebra in the three-dimensional, $F=1$  fundamental representation, cf.~\App{Hamiltonian} and Ref.~\cite{Kawaguchi2012a.PhyRep.520.253}. 

We reparametrize the Lagrangian \eq{spin-1_lagrange_main} in terms of the total local density of particles, the sum and the difference of the $\mf=\pm 1$ densities
\begin{align}
 \totd=\sum_{a}\psi_a^*\psi_a = \sum_{a}\rho_{a}, \quad  \sided = \frac{\rho_1+\ \rho_{-1}}{2}, \quad \epsilon=\frac{\rho_1-\rho_{-1}}{2},
\end{align}
as well as the overall, Larmor and spinor phases 
\begin{align}
    \quad \theta = \phi_1+ \phi_{-1}\,, \quad 
    \varphi_\text{L} = \frac{\phi_1 - \phi_{-1}}{2}\,, \quad 
    \phis = \theta - 2\phi_0\, . 
    \label{eq:relphases_app}
\end{align}
In  experimentally realistic settings for $^{87}\mathrm{Rb}$, the density interactions dominate over spin-changing collisions as 
$c_0\gg |c_1|$.
As a result, the total density $\totd$ of the condensate can be considered to be constant, yielding
\begin{align}
    \psi_{\pm 1} = \sqrt{\sided\pm\epsilon}\ \e^{{\i}(\theta/2\,\pm\,\varphi_\text{L})}\,, \quad 
    \psi_{0} = \sqrt{\totd-2\sided}\ \e^{{\i}( \theta\,-\,\phis )/{2}}\, . 
    \label{eq:coordinatetrf_main}  
\end{align}
Inserting the expressions \eq{coordinatetrf_main} into the Lagrangian \eq{spin-1_lagrange_main} gives
\begin{widetext}
\begin{align}
    \mathcal{L} 
    =\  & - \frac{\totd}{2}\left(\dot\theta-\dot{\varphi}_\mathrm{s}\right)
    -2 \epsilon\,\dot{\varphi}_\text{L} 
    - \sided\,\dot{\varphi}_\mathrm{s} 
     - \frac{1}{8M}\Big\{ ({\totd-2\sided}) (\nabla\theta-\nabla\phis)^2  
    +2{\sided} (\nabla\theta)^{2}
    +8{\sided} (\nabla\varphi_\text{L})^{2} 
    + 8{\epsilon} \nabla\varphi_\text{L}\,\nabla\theta\Big\}
    \nonumber \\
    & - \frac{1}{8M}\left\{(\sided-\epsilon)\left[\nabla \ln(\sided-\epsilon)\right]^2 
    + (\sided+\epsilon)\left[\nabla \ln(\sided+\epsilon)\right]^2 
    + (\totd-2\sided)\left[\nabla \ln(\totd-2\sided)\right]^2\right\} 
    \nonumber \\
    & -2q\sided - \frac{c_0}{2}\totd^2 - 2c_1\Big[-2\sided^2 +\epsilon^2 + \sided\totd + \sqrt{\sided^2 - \epsilon^2} (\totd-2\sided)\cos\phis\Big]\, .
\label{eq:lagrange_coord_new_main}
\end{align}

The dynamics in the spin-1 gas is then expected to be dominated by  large phase excitations, while density fluctuations around the mean-field values $n$ and $\bar{\epsilon}$, 
    $\dsided(x,t) = \sided(x,t) - \sidedm$
and
    $\delta\epsilon(x,t) = \epsilon(x,t) - \bar\epsilon$,
will be small due to the energy and symmetry constraints from the interaction terms in  \eq{spin-1_lagrange_main}.
Here the mean background magnetization vanishes in the easy plane, $\langle F_{z}\rangle=2\bar\epsilon=0$.
We expand the Lagrangian in the density fluctuations up to second order and obtain 
\begin{align}
    \mathcal{L}^0 
    &= - \frac{\sidedm}{M} \left(\nabla \varphi_\text{L}\right)^2 
        - \frac{n}{4M}\left(1-\frac{2\sidedm}{\totd}\right) 
            \left(\nabla \phis\right)^2 
        - 2qn 
        - \frac{c_0}{2}\totd^2
        - 2c_1\sidedm(\totd-2\sidedm)(1+\cos\phis)
    \,,\\
    \mathcal{L}^1 
    &=  \begin{pmatrix}
            -\dot{\varphi}_\mathrm{s} -2q  
            -2c_1(\totd-4\sidedm)(1+\cos\phis) \,,
            &
            -2\dot{\varphi}_\text{L} 
        \end{pmatrix}  
        \begin{pmatrix}
            \dsided \\
            \delta\epsilon
        \end{pmatrix}
    \,,\\
    \mathcal{L}^2 
    &=  \begin{pmatrix}
           \dsided\,, & \delta\epsilon
        \end{pmatrix}
        \begin{pmatrix}
            \frac{\nabla^2}{4Mn}\frac{\totd}{\totd-2n} + 4c_1(1+\cos\phis) & 0 \\
            0 & \frac{\nabla^2}{4Mn}
                -c_1\left[2+\left(2-{\totd}/{\sidedm}\right)\cos\phis\right]
        \end{pmatrix}
        \begin{pmatrix}
            \dsided \\
            \delta\epsilon
        \end{pmatrix}
    \,. \label{eq:L2}
\end{align}
\end{widetext}

\begin{figure}[H]
    \centering \includegraphics[width=0.48\textwidth]{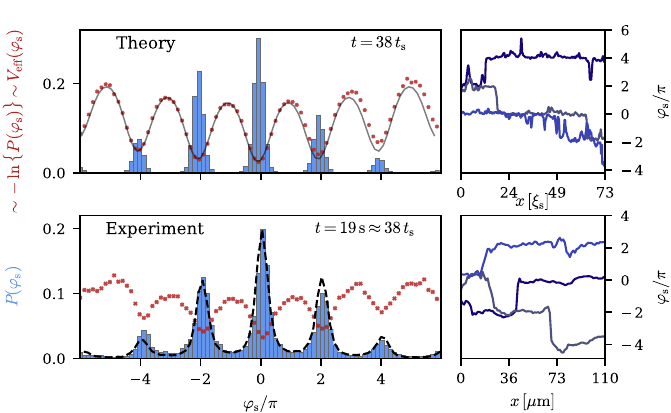}
    \caption{Probability distribution histogram (blue bars) of the spinor phase $\phis$ after a quench from the polar phase to the easy plane. (Upper left panel) Numerical result. 
    After preparing the system in the polar phase, $q_\mathrm{i}>2\,\ncone$, we quench the quadratic Zeeman shift to $q_\mathrm{f}=0.9\,\ncone$,
    after which $\phis$ settles quickly into the minima of its effective potential (red crosses), thus underlining the reduction to the DSG model. 
    This potential is extracted in a Boltzmann approximation as $V_{\mathrm{eff}}(\phis) \sim -\ln(P(\phis))$.
    The solid grey line is the analytical expression \Eq{DSGLagrangian}. Notice a small mean-field shift which forms dynamically and raises the potential for higher $\phis$, see \App{EarlyTime} for more details. We observe the occupation of many minima of the effective DSG potential.
    (Lower left panel) Experimentally extracted distribution of $\phis$, having prepared $\sim 10^5$ atoms in a quasi-one-dimensional cigar-shaped trap with hard walls in the longitudinal direction,
    in the $\mf=0$ state with quadratic Zeeman shift 
    $q_\mathrm{i} \gg {2}{\totd |c_1| } $
    and quenching to 
    $q_\mathrm{f} \approx {\totd |c_1| }$.
    The corresponding oscillating effective potential (red crosses) is evaluated after an evolution time $t=19\mathrm{s} \approx 38\,t_\mathrm{s} = 38 \cdot 2\pi/(\ncone)\approx 115\,(Qc_\mathrm{s})^{-1}$. 
    The pedestal of the histogram can be attributed to the employed measurement scheme. 
    The dashed line shows the theoretical PDF using the same extraction method as in the experiment, and taking into account a systematic calibration offset, see \App{ExptMethods} for details.
    The upper and lower right panels show the spatial configuration of $\phis$ for three different realizations denoted each by a different shade of blue. One observes that the field configuration interpolates between the DSG minima via localized phase kinks that cause the field to spread over several minima.
}   
    \label{fig:numhist}
    
\end{figure} 

%
We perform the Gaussian integral over the density fluctuations and focusing on IR dynamics, we neglect the kinetic terms in \Eq{L2}. We expand around 
$\varphi_\mathrm{s} = 2\pi \mathbb{Z}$
to obtain an effective Lagrangian in the phase angles, cf.~\App{DSGCalc} for details,
\begin{align}
    &\mathcal{L}_{\mathrm{eff}}(\varphi_\mathrm{s}) 
    = -\frac{1}{32c_1}\dot{\varphi}_\mathrm{s}^2 
    - \frac{n(\totd-2\sidedm)}{4M\totd}(\nabla\varphi_\mathrm{s})^2 
    \nonumber \\
    &\quad -\ \left[2c_1\sidedm(\totd-2\sidedm)
    -\frac{q^2}{16c_1}\right] \cos\varphi_\mathrm{s} 
    + \frac{q^2}{32c_1}\sin^2\varphi_\mathrm{s}
    \,.
\label{eq:DSGLagrangian}
\end{align}
Hence, the LEEFT takes the form of the DSG model for the spinor phase.
The periodic potential derives from the local spin-spin interactions, in contrast to standard cases, where it is caused by a linear coupling due to an external field transverse to magnetization \cite{Gritsev2007a.PhysRevB.75.174511,Coleman1975a.PRD11.2088} or results in a description dual to a 2D Coulomb gas \cite{Kosterlitz1973a,Amit1980a.JPA13.585,Giamarchi2003a,Cazalilla2011a}.
In \cite{Schmied:2018osf.PhysRevA.99.033611,Siovitz:2023ius.PhysRevLett.131.183402}, the same subdiffusive coarsening as in the DSG model has been numerically observed in the structure factor
$S_{F_\perp}(k,t) = \langle|F_\perp(k,t)|^2\rangle$
of the transverse spin degree of freedom 
$F_\perp = F_x + \i F_y$
(cf. \App{Hamiltonian})
during the post-quench dynamics of the spin-1 gas.
For that, the system is prepared in the polar phase, where all the atoms macroscopically occupy the $\mf=0$ component, and is then quenched into the easy-plane phase, via a sudden change of the quadratic Zeeman shift to a value $q_\mathrm{f}=0.9\,\ncone$. 
\begin{figure*}[t]
    \centering \includegraphics[width=\textwidth]{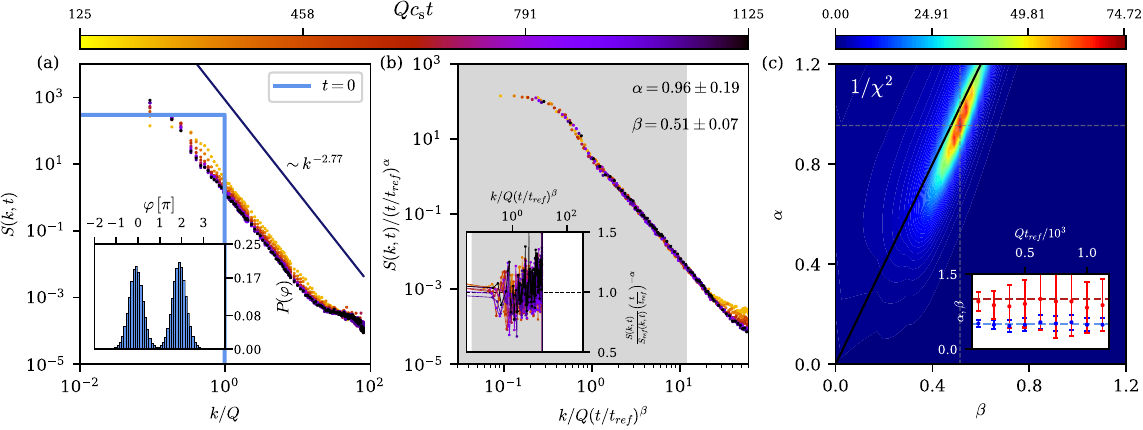}
    \caption{Self-similar scaling of the DSG model in (2+1)D.
    (a) Time evolution of the structure factor $S(k,t) = \langle|\varphi(k,t)|^2\rangle$.
    The initial condition (blue line) is a box with cutoff $Q$. 
    The redistribution of excitations in the system leads to a power law in the IR, for momenta greater than a characteristic scale $k_{\Lambda}(t)\sim t^{-\beta}$.
    The inset shows that the field configurations occupy only two minima of the sinusoidal potential.
    (b) The collapse of the curves to the universal scaling function, with reference time $t_{\mathrm{ref}} = 100\,(Q c_\mathrm{s})^{-1}$, shows the spatio-temporal scaling of the correlator with exponents
    $\alpha = 0.98(20)$ and $\beta = 0.51(7)$. The inset shows the residuals of the spectra w.r.t.~the reference spectrum.
    The equal distribution of errors confirms self-similarity of the evolution.
    (c) Inverse $\chi^2$ distribution showing the most likely scaling exponents.
    Notice the proximity of the scaling exponents to the $\alpha=d\beta=2\beta$ line.
    The inset shows the stability of the scaling of $\alpha$ (red) and $\beta$ (blue) w.r.t.~the reference time. The blue and red dashed lines show the value $0.5$ and $1$, respectively.}
    \label{fig:2DScaling}
\end{figure*}

Our results demonstrate that the DSG dynamics of the spinor phase alone accounts for the subdiffusive scaling exhibited by the full spin-1 system, while there is no need to take into account the Larmor and total phases as, e.g., in \cite{Mikheev:2018adp}.

\subsection{Comparison with experimental results}
\Fig{numhist} (left panels) shows the post-quench long-time probability distribution function (PDF) of the spatially resolved spinor phase profiles inferred from $10^3$ truncated Wigner (upper) and $140$ experimental runs (lower panel), the latter for a quasi-one-dimensional condensate of $\sim 10^5$ $^{87}\mathrm{Rb}$ atoms, see \App{ExptMethods}.
The PDF is localized at multiples of $2\pi$, corresponding to the minima of the effective DSG potential. 
We conclude that the approximation of small 
$\phis - 2\pi\mathbb{Z}$,
chosen in the derivation of the LEEFT \eq{DSGLagrangian}, is experimentally confirmed, thus underlining the reduction of the dynamics to a DSG model.
The logarithm of the PDF, which in equilibrium is proportional to the free energy of the system, provides an estimate of the effective potential, coinciding qualitatively for the simulated and measured distributions. 
Most importantly, we find field configurations to spread over many minima of the periodic potential, as exemplarily shown in the right panels for 3 realizations each. 

From the derivation of our LEEFT we can infer that field configurations interpolate between the minima in a spatially localized manner, on the order of the spin healing length $\xi_\mathrm{s}=(2M\ncone)^{-1/2}$ only. 
The reason for this is that if only density fluctuations at \(k=0\) are taken into account as in the derivation of Eq.\,\eqref{eq:DSGLagrangian}, the Green's function of the integrated-out density fluctuations diverges at $\phis\simeq(2\nu+1)\pi$, $\nu\in\mathbb{Z}$, i.e. at the maxima of the potential.
However, for excitations of nonzero momenta $k>0$, the energy gap regularizes these divergences. 
It turns out that for momenta on the order of the spin healing length,
$k\sim k_{\xi_{\mathrm{s}}}=1/\xi_\mathrm{s}$,
the gap modifies the potential in a way that the DSG model is restored around $\phis=(2\nu+1)\pi$, albeit with different coupling parameters, cf.~\App{TWA} for details.
In that case, field configurations prevail, which interpolate between adjacent DSG potential minima within the short length scale on the order of $k_{\xi_{\mathrm{s}}}^{-1}$, cf.~the right panels of \Fig{numhist}.

Such interpolations are indeed observed in the universal scaling dynamics of the full spin-1 model as space-time vortex defects in $\varphi_\mathrm{L}$ and $\phis$, cf.~\cite{Siovitz:2023ius.PhysRevLett.131.183402}.
Hence, we observe that the existence of topological charges, such as winding numbers seen in the spinor gas, is translated to a noncompact effective theory, which does not enforce the $2\pi$ periodicity of the phase.
Remarkably, the DSG model reproduces the subdiffusive scaling of the spinor gas despite the absence of topological information, thus ultimately allowing for a further analytical study of scaling characteristics, as such approaches generically require the absence of topological excitations.

\subsection{Scaling evolution according to the two-dimensional\newline DSG model}
In contrast to the one-dimensional case, scaling dynamics resulting in a diffusion-type exponent in a two-dimensional spin-1 system \cite{Schmied:2019abm} has been attributed to the dynamics of spin vortex patterns. 
To compare with this setting, we simulate the DSG model in two dimensions, preparing again a momentum box of DSG field about a mean value $\langle\varphi\rangle=\pi$ chosen at a maximum of the cosine potential. 
An analysis of the ensuing evolution of the $\varphi$ distribution in this case reveals that the DSG field $\varphi$ is concentrated mainly across two minima of the periodic effective potential, see inset of \Fig{2DScaling}a.
This corresponds to the formation of spin-type magnetic domains as seen in \Fig{2DHistogram}b in \App{TWA}.
At long evolution times, these domains coarsen, i.e., grow in size, corresponding to universal dynamical scaling evolution with $\beta\approx 0.5$, cf.~\Fig{2DScaling}. 
The time evolution and scaling collapse of the spectra $S(k,t)$ are shown in Figs.~\fig{2DScaling}a,b. 
The presence of a weak plateau in the spectra allows us to rescale the spectra while optimising $\alpha$ and $\beta$ independently, with larger errors on $\alpha$ than $\beta$ due to the smallness of the plateau, see panel c. 
We obtain $\beta = 0.51(8)$, $\alpha = 0.98(20)\simeq d\beta$ and $\kappa=2.76(1)$, corroborating the spin-1 results from \cite{Schmied:2019abm} within the error bounds. Once more we find indications that the spread of the DSG field across the potential is crucial for the type of scaling found in the system.
Clarifying the closer relation of this scaling with the phenomenology of spin vortices in the full spin-1 gas would be desirable but is beyond the scope of the present work. 

\section{Conclusions}
Universal dynamics of the spin-1 Bose gas after a parameter quench can be recaptured by a real scalar field theory, which takes the form of a double sine-Gordon model for the spinor phase.
This effective description is consistent with numerical and experimental observations regarding the probability distribution function of $\phis$.
The far-from-equilibrium dynamics of the effective model shows pattern coarsening in the IR regime of wave numbers $k\ll  k_{\xi_\mathrm{s}}$, of the subdiffusive ($\beta < 1/2$) as well as the diffusion type ($\beta = 1/2$), 
consistent with previous findings of \cite{Schmied:2018osf.PhysRevA.99.033611,Siovitz:2023ius.PhysRevLett.131.183402, Schmied:2019abm} for the full spin-1 gas.
The subdiffusive and diffusion-type scaling are associated with field configurations spreading over  many, or few minima of the sinusoidal potential, respectively.
These results corroborate analytical findings of Refs.~\cite{Heinen2023a.PhysRevA.107.043303,Koeper2023a}.
Our results are crucial to the understanding of the dominant mechanisms leading to self-similar scaling far from equilibrium, by the reduction to a noncompact field theory of a single real scalar field, towards the identification of far-from-equilibrium universality classes. 
They open a perspective for classifying also in other systems subdiffusive  \cite{Schole:2012kt,Karl2017b.NJP19.093014,Erne:2018gmz,Johnstone2019a.Science.364.1267,Spitz2021a.SciPostPhys11.3.060,Noel:2025mtb} vs.~diffusion-type scaling \cite{PineiroOrioli:2015dxa,Walz:2017ffj.PhysRevD.97.116011,Johnstone2019a.Science.364.1267,Prufer:2018hto,Schmied:2019abm,Gresista:2021qqa,Gazo:2023exc,Martirosyan:2024rxm}. 
They furthermore open the possibility to use spinor Bose gases for experimentally investigating fundamental sine-Gordon dynamics, such as soliton collisions \cite{blakie_2021;PhysRevResearch.3.023043}, breathers and $n$-bounce solutions \cite{campbell_1986,
dmitriev_2008;PhysRevE.78.046604}.

\begin{acknowledgments}
The authors thank S.~Lannig for inspiring discussions and N.~Antolini for experimental assistance and discussions. 
They thank P.~Heinen, W.~Kirkby, H.~K{\"o}per, L.~M.~A.~H.~Le, A.~N.~Mikheev, V.~Noel, A.-M.~Oros, J.~M.~Pawlowski, N.~Rasch, J.~Siefker and Y.~Werner for discussions and collaboration on related topics. 
They in particular thank I.~Moss for useful, further clarifying discussions.
They acknowledge support 
by the Deutsche Forschungsgemeinschaft (DFG, German Research Foundation), through 
SFB 1225 ISOQUANT (Project-ID 273811115), 
grant GA677/10-1, 
and under Germany's Excellence Strategy -- EXC 2181/1 -- 390900948 (the Heidelberg STRUCTURES Excellence Cluster), 
and by the state of Baden-W{\"u}rttemberg through bwHPC and the DFG through 
grants 
INST 35/1503-1 FUGG, INST 35/1597-1 FUGG, and INST 40/575-1 FUGG
(SDS, Helix, and JUSTUS2 clusters).
A.E.G. acknowledges support from Cusanuswerk Bischöfliche Studienförderung.
\end{acknowledgments}

\begin{appendix}


\renewcommand{\theequation}{A\arabic{equation}}
\setcounter{equation}{0}
\makeatletter
\newcommand{\sectionterm}{appendix}

\newcommand{\nc}{\totd|c_1|}
\newcommand{\qbar}{\bar{q}}
\newcommand{\kxis}{k_{\xi_\mathrm{s}}}

\section{Field theory of the spin-1 Bose gas}
In this \sectionterm, we introduce the quantum field model and give a short overview of the relevant ground-state mean-field characteristics of the spin-1 Bose-Einstein condensate. 

\subsection{Model Hamiltonian and Lagrangian and their parametrisation}
\label{app:Hamiltonian}
The classical spin-1 Hamiltonian reads, in $d$ dimensions,
\begin{align}
    H=\int \dd{\vec{x}} &\left[
    \vb*{\Psi}^\dagger(\vec{x},t) \left(-\frac{1}{2M}\nabla^{2}
    +qf_z^2\right)\vb*{\Psi}(\vec{x},t)\right. \nonumber \\
    &\hspace{1.5cm}\left.+\frac{c_0}{2}\totd(\vec{x},t)^2 
    + \frac{c_1}{2}\abs{\vb*{F}(\vec{x},t)}^2\right]
    \,,
    \label{eq:Spin1Hamiltonian}
\end{align}
where $M$ is the atomic mass, $q$ represents the quadratic Zeeman shift and the term $\sim c_0\totd^{2}$ describes $\mathrm{U}(3)$-symmetric density-density interactions, with total density $\totd=\sum_a\psi_a^*\psi_a$.
Spin-dependent interactions are governed by the term $\sim{c_1} |\vb*{F}|^{2}$, with
$\vb*{F}={\psi}^*_{a} \vb*{f}_{ab} {\psi}_{b}$, $a,b\in\{+1,0,-1\}$ denoting the magnetic-sub-level indices in the spin-1 manifold.
The $3\times3$ generator matrices $\vb*{f}=(f_x,f_y,f_z)$ of the $\mathfrak{so}(3)$ Lie algebra in the fundamental representation are defined as
\begin{align}
  f_{x}=\mqty(0 & 1 & 0 \\ 1 & 0 & 1 \\ 0 & 1 & 0)\,,\quad
  f_{y}=\mqty(0 & -\i & 0 \\ \i & 0 & -\i \\ 0 & \i & 0)\,,\quad
  f_{z}=\mqty(1 & 0 & 0 \\ 0 & 0 & 0 \\ 0 & 0 & -1)\,.
  \label{eq:su3generators}
\end{align}
The field spinors are defined as
\begin{align}
    \vb*{\Psi}(\vec{x},t) 
    = \mqty(\psi_1(\vec{x},t) \\ 
            \psi_0(\vec{x},t) \\ 
            \psi_{-1}(\vec{x},t))
    \,,
    \label{eq:Spinor}
\end{align}
in terms of the magnetic field components $\psi_{m_F}$, $m_F\in\{0,\pm1\}$.
For the derivation of the low-energy effective theory it will be more convenient to start from the corresponding spin-1 Lagrangian \eq{spin-1_lagrange_main},
\begin{align}
    \mathcal{L} = \frac{\i}{2}&\left( \psi_a^* \partial_t \psi_a - \psi_a \partial_t \psi_a^*\right) 
    - \frac{1}{2M}\nabla\psi_a^*\nabla \psi_a - q (f^{z})^{2}_{ab}\psi_a^*\psi_b \nonumber \\
    & \hspace{0.5cm}- \frac{c_0}{2}(\psi_a^*\psi_a)^2 - \frac{c_1}{2}\sum_{i\in\{x,y,z\}}(\psi_a^* f^i_{ab}\psi_b)^2\,.
    \label{eq:sm:spin-1_lagrange}
\end{align}
Here and in the following we suppress the space-time arguments of all fields.
The magnetic field components can be expressed in terms of the respective densities $\rho_{m_F}$ and phase angles $\phi_{m_F}$ as $\psi_{m_F}=\sqrt{\rho_{m_F}}\exp{\i\phi_{m_F}}$.
Rewriting these by means of the total density $\totd$, as well as the mean side-mode density $\rho$ and $z$-magnetization $F_z=2\epsilon$,
\begin{align}
    \totd = \rho_{-1}+\rho_0+\rho_{1}\,,
    \quad  \sided = \frac{\rho_1 + \rho_{-1}}{2}\,,
    \quad &\epsilon = \frac{\rho_1 - \rho_{-1}}{2}\,, 
    \label{eq:reldensities}
\end{align}
and of the overall phase $\theta$, the Larmor phase $\phi_\text{L}$, and the spinor phase $\phis$, 
\begin{align}
    \quad \theta = \phi_1+ \phi_{-1}\,, \quad 
    \varphi_\text{L} = \frac{\phi_1 - \phi_{-1}}{2}\,, \quad 
    \phis = \theta - 2\phi_0\,, 
    \label{eq:relphases}
\end{align}
yields the phase-density representations of the three magnetic field components,
\begin{align}
    \psi_{\pm 1} = \sqrt{\sided\pm\epsilon}\ \e^{{\i}(\theta/2\,\pm\,\varphi_\text{L})}\,, \quad 
    \psi_{0} = \sqrt{\totd-2\sided}\ \e^{{\i}( \theta\,-\,\phis )/{2}}\,.
    \label{eq:coordinatetrf}
\end{align}
\begin{widetext}
We proceed by inserting the expressions \eq{coordinatetrf} into the Lagrangian \eq{sm:spin-1_lagrange}, which gives
\begin{align}
    \mathcal{L} 
    =\  & - \frac{\totd}{2}\left(\dot\theta-\dot{\varphi}_\mathrm{s}\right)
    -2 \epsilon\,\dot{\varphi}_\text{L} 
    - \sided\,\dot{\varphi}_\mathrm{s} 
    \nonumber \\
    & - \frac{\totd-2\sided}{8M} (\nabla\theta-\nabla\phis)^2  
    -\frac{\sided}{4M} (\nabla\theta)^{2}
    -\frac{\sided}{M} (\nabla\varphi_\text{L})^{2} 
    - \frac{\epsilon}{M} \nabla\varphi_\text{L}\,\nabla\theta
    \nonumber \\
    & - \frac{1}{8M}\left\{(\sided-\epsilon)\left[\nabla \ln(\sided-\epsilon)\right]^2 
    + (\sided+\epsilon)\left[\nabla \ln(\sided+\epsilon)\right]^2 
    + (\totd-2\sided)\left[\nabla \ln(\totd-2\sided)\right]^2\right\} 
    \nonumber \\
    & -2q\sided - \frac{c_0}{2}\totd^2 - 2c_1\left[-2\sided^2 +\epsilon^2 + \sided\totd + \sqrt{\sided^2 - \epsilon^2} (\totd-2\sided)\cos\phis\right].
\label{eq:lagrange_coord_new}
\end{align}
%

\subsection{Ground states of the polar and easy-plane phase}
\label{sec:GSpolareasyplane}
The energy term describing the quadratic Zeeman shift $q$ competes with the spin-spin interactions  proportional to $c_1$. This competition gives rise to different ground states in the system depending on the chosen point in the $q$-$c_1$-plane.
Our focus is set on simulating the dynamics of a ferromagnetic system, i.e., for $c_1<0$, where two phases are separated by a second-order quantum phase transition controlled by $q$. 
For $q>2\nc$, the system resides in the so-called polar phase, which is characterized by a vanishing magnetization and described by the following mean-field spinor,
\begin{align}
    \vb*{\Psi}_\mathrm{P} = \sqrt{\totd}\,\e^{\i \phi_0}\,\mqty(0 \\ 1 \\ 0)
    \,.
    \label{eq:PolarMFGS}
\end{align}
In the ground state, \eq{PolarMFGS} describes the mean field, with the freedom of a global U$(1)$ phase of the condensate.
The easy-plane phase, on the other hand, is reached when tuning $q$ below this critical line, i.e., for $0<q<q_{\text{c}}\equiv2\nc$ for any given $c_1<0$. 
Spontaneous symmetry breaking here gives rise to a magnetization in the $F_x$--$F_y$-plane, with mean-field spinor
\begin{align}
    \vb*{\Psi}_{\mathrm{EP}} 
    = \frac12\sqrt{\totd}\,\e^{\i \theta/2}\,
    \mqty(
    \e^{\i\varphi_\mathrm{L}}\sqrt{1-\qbar} \\ 
    \e^{-\i\phis/2}\sqrt{2+2\qbar} \\
    \e^{-\i \varphi_\mathrm{L}}\sqrt{1-\qbar})
    \,,
    \label{eq:EPMFGS}
\end{align}
which depends on the quadratic Zeeman shift $q$ relative to its critical value, $q_{\text{c}}=2\totd|c_{1}|$, 
\begin{align}
\qbar = \frac{q}{q_{\text{c}}} =\frac{q}{2\totd|c_{1}|}
=1-\frac{4n}{\totd}\,.
\label{eq:qbarDef}
\end{align}
The last equation results from the mean-field relation \(q = 2|c_1|(\totd-4\sidedm)\), which follows from assuming the fields to take their homogeneous ground state values $\totd=$ const., $\sided=\sidedm=$ const., $\epsilon=0$, $\phis=0$, $\varphi_\mathrm{L}=$ const., and solving the Euler-Lagrange equation $\partial\mathcal{L}/\partial n=0$ for the Lagrangian 
\eq{lagrange_coord_new}.

The emergent transverse magnetization gives rise to a complex order parameter $F_\perp = F_x + \mathrm{i}F_y = \sqrt{2}(\psi_1^* \psi_0 + \psi_0^* \psi_{-1})$, exhibiting a total magnetization $\abs{F_\perp}=[1-\qbar^2]^{1/2}$. One can also write $\abs{F_\perp}$ in terms of the mean-field background-solution relative phases and densities as $\abs{F_\perp} = 2\sqrt{n(\totd -n) \,(1+\cos{\phis})}$.

\section{Double sine-Gordon low-energy effective theory}
\label{app:DSGCalc}
\renewcommand{\theequation}{B\arabic{equation}}
\setcounter{equation}{0}
\noindent 
In this \sectionterm, we sketch the approximate mapping between the spin-1 Lagrangian \eq{sm:spin-1_lagrange} and the low-energy effective Double Sine-Gordon (DSG) Lagrangian \eq{DSGLagrangian}. 
\subsection{Expansion of the Lagrangian about constant mean-field densities}
We start from the Lagrangian density in the form \eq{lagrange_coord_new}.
In the regime of low-energy excitations, density fluctuations are strongly suppressed. 
Hence, we assume the density fields to be given by their mean-field background values with small fluctuations added,
\begin{align}
\totd(\vec{x},t) = \totd = \text{const.} \,, \quad 
\sided(\vec{x},t) = \sidedm + \delta \sided(\vec{x},t)\,, \quad 
\epsilon(\vec{x},t) = \bar{\epsilon} + \delta\epsilon(\vec{x},t)\,.
\label{eq:fluctansatz}
\end{align}
Since $|c_1|\ll c_0$, we neglect flucutations of the total density $\totd$.
As we expand about a mean-field ground state of a homogeneous system in the easy-plane phase, we take \(\sidedm=const.\) and the mean-field background solution of the density difference to vanish, $\bar{\epsilon}=0$. 
Before inserting this ansatz into the Lagrangian, we rewrite the terms, in \eq{lagrange_coord_new}, which contain time and spatial derivatives in such a way that the coupling of the overall phase $\theta$ to the Larmor and spinor phases takes the following form,
\begin{align}
    \mathcal{L} 
    =\  & - \frac{\totd}{2}\left[\dot\theta
    -\left(1-\frac{2\sided}{\totd}\right)\dot{\varphi}_\mathrm{s}
    +4 \frac{\epsilon}{\totd}\,\dot{\varphi}_\text{L}\right] 
    \nonumber  \\
    & - \frac{\totd}{8M}\left[\nabla\theta
    -\left(1-\frac{2\sided}{\totd}\right)\nabla\phis
    +4 \frac{\epsilon}{\totd}\,\nabla\varphi_\text{L}\right]^2  
    -\frac{\sided}{4M}\left(1-\frac{2\sided}{\totd}\right)(\nabla\phis)^2
    -\frac{\sided}{M} (\nabla\varphi_\text{L})^{2} 
    +\frac{2\epsilon^2}{M\totd} (\nabla\varphi_\text{L})^{2} 
    -\frac{\epsilon}{M}\left(1-\frac{2\sided}{\totd}\right) \nabla\varphi_\text{L}\,\nabla\phis
    \nonumber \\
    & - \frac{1}{8M}\left\{(\sided-\epsilon)^{-1}\left[\nabla(\sided-\epsilon)\right]^2 
    + (\sided+\epsilon)^{-1}\left[\nabla(\sided+\epsilon)\right]^2 
    + (\totd-2\sided)^{-1}\left[\nabla(\totd-2\sided)\right]^2\right\} 
    \nonumber \\
    & -2q\sided - \frac{c_0}{2}\totd^2 - 2c_1\left[-2\sided^2 +\epsilon^2 + \sided\totd + \sqrt{\sided^2 - \epsilon^2} (\totd-2\sided)\cos\phis\right].
\label{eq:lagrange_coord_new_theta-rewritten}
\end{align}
We then redefine the total phase by shifting it by the spinor and Larmor phases, each multiplied with a constant, as
\begin{align}
  \theta \to \tilde\theta 
  = \theta 
  - \left(1-\frac{2n}{\totd}\right)\phis
  + 4\frac{\bar{\epsilon}}{\totd}\varphi_\mathrm{L}
  \,.
\label{eq:redeftotalphase}
\end{align}
Note that the gradient of $\tilde\theta$, for constant densities $\totd$, $n$, and $\bar\epsilon$, is proportional to the total current in mean-field approximation, i.e., neglecting the fluctuations in \eq{fluctansatz},
\begin{align}
    \tilde{\vec{j}}
    = -\frac{\i}{2M}\left(\psi^*_a\nabla\psi_a-\text{c.c.}\right)
    = \frac{\totd}{2M}\,\nabla\tilde\theta
    +\mathcal{O}(\delta\rho,\delta\epsilon)
    \,.
\end{align}
Inserting \eq{fluctansatz} and \eq{redeftotalphase} into \eq{lagrange_coord_new_theta-rewritten}, one finds that, in leading order, the spinor and Larmor phases decouple from the total phase, leaving only couplings between the gradients of the  phases, which are linear in $\delta\rho$ and $\delta\epsilon$,
\begin{align}
    \mathcal{L} 
    =\  
    & - \frac{\totd}{2}\dot{\tilde\theta}
    -\delta\sided\,\dot{\varphi}_\mathrm{s}
    -2\delta\epsilon\,\dot{\varphi}_\text{L} 
    \nonumber \\
    & - \frac{\totd}{8M}\left(\nabla{\tilde\theta}\right)^2
    - \frac{n}{4M}\left(1-\frac{2n}{\totd}\right)\left(\nabla\phis\right)^2
    -\frac{n}{M} (\nabla\varphi_\text{L})^{2} 
    \nonumber \\
    &
    - \frac{\delta\sided}{4M}\left[
        \left(1-\frac{4n}{\totd}\right)(\nabla\phis)^2
        + 4(\nabla\varphi_\text{L})^{2} 
        + 2\nabla{\tilde\theta}\,\nabla\phis
    \right]
    - \frac{\delta\epsilon}{M}\left[ 
        \nabla{\tilde\theta}\,\nabla\varphi_\text{L}
        +\left(1-\frac{2n}{\totd}\right)
        \nabla\varphi_\text{L}\,\nabla\phis
    \right]
    \nonumber \\
    &     
    - \frac{1}{8M}\left\{
        \frac{(\nabla\delta\rho)^2}{\totd-2n-2\delta\rho}
        +\frac{2(n+\delta\rho)}{(n+\delta\rho)^2-(\delta\epsilon)^2}
            \left[(\nabla\delta\rho)^2+(\nabla\delta\epsilon)^2\right]
        -\frac{4\delta\epsilon}{(n+\delta\rho)^2-(\delta\epsilon)^2}
            \nabla\delta\rho\,\nabla\delta\epsilon
    \right\} 
    \nonumber \\
    & -2q\delta\rho 
    -2c_1\left[(\totd -4n)\delta\rho
        -2\delta\rho^2 +\delta\epsilon^2 
        + \sqrt{\sided^2 - \delta\epsilon^2} 
            (\totd-2n-2\delta\rho)\cos\phis
    \right]
    \nonumber \\
    & -2qn 
    - \frac{c_0}{2}\totd^2
    -2c_1n\left(\totd-2n\right)    
    \,.
\label{eq:lagrange_coord_new_final}
\end{align}
Approximating the above Lagrangian to leading order in the density fluctuations, neglecting terms of $\mathcal{O}(\delta\rho_\alpha\nabla\delta\rho_\beta\nabla\delta\rho_\gamma)$, with $\delta\rho_{\alpha,\beta,\gamma}\in\{\delta\rho,\delta\epsilon\}$, as well as terms of order $\mathcal{O}(\delta\rho_\alpha\nabla\varphi_\beta\nabla\varphi_\gamma)$, with $\delta\rho_{\alpha}\in\{\delta\rho,\delta\epsilon\}$, $\varphi_{\beta,\gamma}\in\{{\tilde\theta},\phis,\varphi_\text{L}\}$, we can write it in the following form,
\begin{align}
    \mathcal{L} 
    = \mathcal{L}_{\tilde\theta} + \mathcal{L}^0 + \mathcal{L}^1 
    + \mathcal{L}^2  + \mathcal{O}(\delta\rho_\alpha\nabla\delta\rho_\beta\nabla\delta\rho_\gamma,
    \delta\rho_\alpha\nabla\varphi_\beta\nabla\varphi_\gamma)
    \,,
\end{align}
with 
\begin{align}
    \mathcal{L}_{\tilde\theta}
    &=  - \frac{\totd}{2}\dot{\tilde\theta}
        - \frac{\totd}{8M}\left(\nabla{\tilde\theta}\right)^2
    \,,\\
    \mathcal{L}^0 
    &= - \frac{\sidedm}{M} \left(\nabla \varphi_\text{L}\right)^2 
        - \frac{n}{4M}\left(1-\frac{2\sidedm}{\totd}\right) 
            \left(\nabla \phis\right)^2 
        - 2qn 
        - \frac{c_0}{2}\totd^2
        - 2c_1\sidedm(\totd-2\sidedm)(1+\cos\phis)
    \,,\\
    \mathcal{L}^1 
    &=  \begin{pmatrix}
            -\dot{\varphi}_\mathrm{s} -2q  
            -2c_1(\totd-4\sidedm)(1+\cos\phis) \,,
            &
            -2\dot{\varphi}_\text{L} 
        \end{pmatrix}  
        \begin{pmatrix}
            \dsided \\
            \delta\epsilon
        \end{pmatrix}
    \,,\\
    \mathcal{L}^2 
    &=  \begin{pmatrix}
           \dsided\,, & \delta\epsilon
        \end{pmatrix}
        \begin{pmatrix}
            \frac{\nabla^2}{4Mn}\frac{\totd}{\totd-2n} + 4c_1(1+\cos\phis) & 0 \\
            0 & \frac{\nabla^2}{4Mn}
                -c_1\left[2+\left(2-{\totd}/{\sidedm}\right)\cos\phis\right]
        \end{pmatrix}
        \begin{pmatrix}
            \dsided \\
            \delta\epsilon
        \end{pmatrix}
    \,.
\end{align}
This constitutes the approximate Lagrangian, which is our starting point for integrating out the density fluctuations in the following, reducing it to the low-energy effective DSG model.

\subsection{Reduction to a low-energy effective theory for the phases}
\noindent
As we disregard fluctuations of the total density, we may also neglect the contribution $\mathcal{L}_{\tilde\theta}$, which in the chosen approximation decouples from the remaining Lagrangian.
Thus, the approximate Lagrangian takes the form of 
\begin{align}
    \mathcal{L}=\mathcal{L}^0 + \bm{J}\cdot \bm{\dsided} + \frac{1}{2} \bm{\dsided}^T \cdot \bm{G}^{-1} \cdot \bm{\dsided} + \mathcal{O}(\dsided^3, \delta\epsilon^3)\,,
\end{align}
where $\bm{\dsided}=(\dsided,\delta\epsilon)^T$ and
\begin{align}
    \bm{J} 
    &= \left(-\dot{\varphi}_\mathrm{s} - 2q 
        - 2c_1(\totd-4n)(1+\cos\phis), 
        -2\dot{\varphi}_\mathrm{L}\right)
    \,, 
    \\
    \bm{G}^{-1} 
    & = \mqty(\frac{\nabla^2}{2Mn}\frac{\totd}{\totd-2n} 
        + 8c_1(1+\cos\phis) & 0 \\
            0 & \frac{\nabla^2}{2Mn} 
        - 2c_1\left(2+ \left(2-\frac{\totd}{n}\right)
            \cos\phis\right))
    \,.
    \label{eq:InvGreensFunc}
\end{align}
The quadratic form allows us integrating out the density fluctuations by carrying out the Gaussian integrals for $\dsided$ and $\delta\epsilon$ according to 
\begin{align*} 
    Z 
    &= \int \mathcal{D}\dsided\,\mathcal{D}\delta\epsilon\,\mathcal{D}\phis\,\mathcal{D}\varphi_\text{L} 
    \exp{\i \int_{t, x} \left(
    \mathcal{L}^0+\bm{\dsided}^{T}\bm{J}
    +\frac{1}{2}\bm{\dsided}^{T}\bm{G}^{-1} \cdot \bm{\dsided}\right)}
    \\
    &= C \int \mathcal{D}\phis\,\mathcal{D}\varphi_\text{L} 
    \underbrace{\exp{\i \int_{t, x} \left[\mathcal{L}^0-\frac{1}{2}\bm{J^T}\bm{G}\bm{J}\right]-\frac{1}{2}\ln\det \bm{G}^{-1}}}_{=\,\exp{\i S^\mathrm{eff}}} 
    \,,
\end{align*}
and collecting the result in the effective action
\begin{align}
    S^\mathrm{eff} 
    = \int_{t, x} \left[\mathcal{L}^0-\frac{1}{2}\bm{J^T}\bm{G}\bm{J}\right]
    - \frac{\i}{2} \ln\det \bm{G}
    \,.
\end{align}
Neglecting, furthermore, irrelevant constant terms in $\mathcal{L}^0$, this procedure yields the following real part of the effective Lagrangian, where the denominators containing derivatives are implied to denote the respective Green's functions:
\begin{align}
    \Re\mathcal{L}^\mathrm{eff} 
    =\ 
    &- \frac{n}{M} \left(\nabla \varphi_\mathrm{L}\right)^2 
    - \frac{n}{4M}\left(1-\frac{2\sidedm}{\totd}\right) 
        \left(\nabla\phis\right)^2 
    - 2c_1n(\totd-2n)\cos\phis
    \\
    &- \frac{1}{2} \left\{\,
        \dot{\varphi}_\mathrm{L}\,
        \frac{4}{\frac{\nabla^2}{2Mn}-2c_1[2+(2-\totd/n)\cos\phis]}\,
    \dot{\varphi}_\mathrm{L} 
    \right. 
    \nonumber\\
    & \left. 
    \qquad 
    + \left[\dot{\varphi}_\mathrm{s}
    + 2q + 2c_1(\totd-4n)(1+\cos\phis)\right]
        \frac{1}{\frac{\nabla^2}{2Mn} 
        \frac{\totd}{(\totd-2n)} +8c_1(1+\cos\phis)}
    \left[\dot{\varphi}_\mathrm{s}
    + 2q + 2c_1(\totd-4n)(1+\cos\phis)\right] 
    \,\right\}. \nonumber
    \label{eq:LeffkInoee}
\end{align}
The imaginary part contains the functional determinant and results as 
\begin{align}
    \Im\mathcal{L}^\mathrm{eff}
    &=\frac{1}{2\Delta t(\Delta x)^d}
    \ln\left(\frac{(1+\cos\phis) 
    \left[2+\left(2-{\tilde{\rho}}/{n}\right)\cos\phis\right]}{2\left(4-{\tilde{\rho}}/{n}\right)}\right)
    \,,
\end{align}
where \(\Delta t\) and \(\Delta x\) are the time and length scales relevant for regularization, defined by \(\sum_{t, x} = (\Delta t)^{-1}(\Delta x)^{-d} \int_{t, x}\). 
As such, they are related to the system's volume in Fourier space. 
Moreover, we have normalized the imaginary part to vanish at \(\phis=0\), using that overall constants do not change the generating functional. 
As the imaginary part of $\mathcal{L}$ only leads to an overall damping of \(Z\), we will focus on discussing the real part in the following. 

It is, furthermore, useful to express the Lagrangian in terms of dimensionless space, time, and energy density, 
\begin{align}
   \mathbf{x} = \bar{\mathbf{x}}/\kxis\,,\qquad
   t = \bar{t}\,\frac{2M}{\kxis^{2}}\,,\qquad
   \mathcal{L}^{\mathrm{eff}} = \bar{\mathcal{L}}^{\mathrm{eff}}\totd\frac{\kxis^{2}}{2M}\,,
\end{align}
where the spin healing wave number is defined as
\begin{align}
\kxis  =(2M\ncone)^{1/2}\,.
\end{align}
In terms of $\bar{x}$, $\bar{t}$ and $\qbar$, cf.~\Eq{qbarDef}, the real part of the effective Lagrangian in the easy-plane phase, i.e., for $c_{1}<0$, $0<\qbar\leq1$, takes the form
\begin{align}
    \Re\bar{\mathcal{L}}^\mathrm{eff} 
    =\ 
    &- \frac{1}{8}\left[
        4({1-\qbar}) \left(\nabla_{\bar{x}} \varphi_\mathrm{L}\right)^2 
        + \frac12({1-\qbar^2}) (\nabla_{\bar{x}}\phis)^2 
        - {2}(1-\qbar^{2})\cos\phis
    \right]
    \nonumber\\
    &- \frac{1}{2} \left\{\,
    \partial_{\bar{t}}{\varphi}_\mathrm{L}\,
    \frac{1-\qbar}
        {\nabla_{\bar{x}}^2+1-\qbar-({1+\qbar})\cos\phis}\,
    \partial_{\bar{t}}{\varphi}_\mathrm{L} 
    \right. \nonumber\\
    & \qquad\left. 
    + \left[\partial_{\bar{t}}{\varphi}_\mathrm{s}
    +4\bar{q}-2\bar{q}(1+\cos\phis)\right]
    \frac{(1-\qbar^{2})/8}{\nabla_{\bar{x}}^2-(1-\qbar^{2})(1+\cos\phis)}
    \,\left[\partial_{\bar{t}}{\varphi}_\mathrm{s}
    +4\bar{q}-2\bar{q}(1+\cos\phis)\right]
    \right\}. 
    \label{eq:Leffqbar}
\end{align}
\end{widetext}
\subsection{Reduction to a double sine-Gordon model}
\renewcommand{\theequation}{B\arabic{equation}}
\noindent
For our low-energy effective theory, we consider only momenta which are much lower than the healing momentum of the system.
Hence, we will eventually omit the momentum dependence of $\mathcal{L}^2$, such that the matrix elements of the Green's function $\mathbf{G}$ are given by the respective inverses of the matrix elements of $\mathbf{G}^{-1}$, \Eq{InvGreensFunc}.
Yet, the resulting effective theory would be divergent for $\phis = \pi$ and, depending on the ratio ${\totd}/{\sidedm}$ and thus $\bar{q}$, in general also at different values of $0<|\phis|<\pi$.
This can be seen as a manifestation of a constraint for the system: the spinor phase $\phis$ cannot simultaneously `hop' between degenerate ground states across the entire system.
In the following, we will argue that, despite this constraint, there can be nevertheless such hopping between adjacent minima as long as this occurs locally, i.e., in higher momentum modes of the field.

We may now consider two limiting cases: A lowest-energy theory of very low momenta $k\approx 0$, where the field configuration is concentrated around $\phis\approx 2\pi N$, with $N\in \mathbb{Z}$, and a theory around the spin healing momentum $k=\kxis$, where we can also perform the expansion around $\phis\approx \pi N $. 
We first turn to the former. 
In this case, we assume 
\begin{align}
k^{2}\ll
4k^{2}_{\xi_\mathrm{s}} \,,
\qquad\mathrm{i.e.}\quad
0\approx\bar{k}^{2}\ll4\,,
\end{align}
$\bar{k}=k/\kxis$, such that we can effectively neglect the Laplacian term in the denominators in the second and third lines of \Eq{Leffqbar}. 
The dynamics of the spin-1 gas in the easy-plane phase are then characterized by a weakly fluctuating spin length, which corresponds to \(\phis\) fluctuating around one of its mean values \(2\pi N\), with \(N\in\mathbb{Z}\), which correspond to a fully elongated spin vector in the $F_{x}$-$F_{y}$-plane. 
Therefore, we can use
\begin{align}
    1+\cos\phis= 2\left[1-\sin^2({\phis}/{2})\right]
\end{align}
and expand the denominators in the effective Lagrangian in powers of \(\sin^2(\phis/2)\) up to order \(\sin^4(\phis/2)\). 
Moreover, together with this assumption and motivated by numerical results, we may also neglect any terms of order \(\dot{\varphi}_j\sin^2(\phis/2)\) and \((\nabla\varphi_j)^2\sin^2(\phis/2)\), $j\in\{$L,s$\}$.
With these approximations, we find that the effective actions for \(\varphi_\text{L}\) and \(\phis\) decouple and take the form
\begin{widetext}
\begin{align}
    \mathcal{L}^{\mathrm{eff}}_{\phis}  
    &= -\frac{1}{32c_1} \dot{\varphi}_\mathrm{s}^2 - \frac{n(\totd-2n)}{4M\totd}(\nabla\phis)^2 
    - \left(2c_1n(\totd-2n) - \frac{q^2}{16c_1}\right)\cos\phis+ \frac{q^2}{32c_1} \sin^2\phis
    \,,     
    \nonumber\\
    \mathrm{i.e.}\quad
    \bar{\mathcal{L}}^{\mathrm{eff}}_{\phis}  
    &= \frac{1}{4}\,\left[
    \frac{1}{8}\left(\partial_{\bar{t}}{\varphi}_\mathrm{s}\right)^2
    -\frac{1-\qbar^2}{4}\left({\nabla_{\bar{x}}\phis}\right)^2
    +(1-2\qbar^{2})\cos\phis
    -\frac{\qbar^{2}}{2}\sin^2\phis
    \right]\,,    
    \label{eq:LagrangianphiS}
\end{align}
and 
\begin{align}
    \mathcal{L}^\mathrm{eff}_{\varphi_\text{L}} 
    &= \frac{2n}{q} \dot{\varphi}_\mathrm{L}^2 - \frac{n}{M}(\nabla\varphi_\mathrm{L})^2
    \,,
    \qquad
    \mathrm{i.e.}\quad
    \Re\bar{\mathcal{L}}^\mathrm{eff}_{\varphi_\text{L}} 
    = 
   \frac{1-\qbar}{4\qbar}\left(\partial_{\bar{t}}{\varphi}_\mathrm{L}\right)^2
   -\frac{1-\qbar}{2}\left({\nabla_{\bar{x}}\varphi_\mathrm{L}}\right)^2
    \,.
    \label{eq:LeffphiL}
\end{align}
\end{widetext}
Thus, the effective theory for the Larmor phase is a quadratic, free model, while the spinor phase $\phis$ is described by a Double Sine-Gordon (DSG) Lagrangian, which exhibits, compared with a pure SG model, a distorted periodic potential for the phase field. 
At the same time, the Larmor phase $\varphi_{\mathrm{L}}$ decouples and follows the pure massless Klein-Gordon model of the free wave equation.
We emphasize that the presence of the $\sin^2\phis$ term was found to be crucial for achieving scaling behavior far from equilibrium, even if its relative amplitude is much smaller than that of the $\cos\phis$ term. Truncating the expansion at the leading order would lead to a pure sine-Gordon model, yet all performed numerical simulations have shown that, in one spatial dimension, the power spectra remain static in that case. 

It becomes, however, clear from  \Eq{Leffqbar} that this DSG model cannot be valid for \(\phis\approx (2n+1)\pi\), \(n\in\mathbb{Z}\), because, in the limit $k\to0$, the denominator in the terms involving a shifted $\dot\phis$ vanishes in that case.
Moreover, for $\cos\phis=(1-\qbar)/(1+\qbar)$, the denominator of the $\dot\varphi_\mathrm{L}$-dependent term vanishes, which is possible in the easy-plane phase ($0\leq\qbar\leq1$).
As a result, long-wave-length fluctuations of the spinor phase, with $k\to0$, will not interpolate between adjacent minima of the cosine potential, forcing these fluctuations to stay near its minima.

Hence, in order for the DSG model to be applicable for all values of $\phis$, one needs to consider fluctuations with sufficiently large momenta, such that no divergences can appear in the above model. 
Superficially, one can estimate, from the denominator in the spinor-phase dependent terms of \eq{LeffkInoee}, \eq{Leffqbar} that, in the easy-plane phase, one needs $k^{2}\gtrsim2\kxis^{2}$ in order for the denominators to be regular throughout.
For this estimate, we consider the most basic approximation, where one replaces the Laplacian in \Eq{LeffkInoee} by $k^{2}\sim \kxis^{2}$, (in \eq{Leffqbar} by $\bar{k}=k/\kxis=1$) neglecting therewith that the Green's function also depends non-linearly on the spinor phase.
In this approximation, one thus assumes that only the derivative terms show a momentum dependence, while $\cos\phis$ is taken to be set by its constant mean-field value.
After replacing the Laplacian in the denominators of \Eq{Leffqbar}, $\nabla_{\bar{x}}^{2}\to-\bar{k}^{2}=-1$, we may again expand these denominators, however now about the maxima of the periodic potential in the spinor phase, \(\phis\approx (2n+1)\pi\), \(n\in\mathbb{Z}\), and in powers of 
\begin{align}
  1+\cos([2n+1]\pi+\delta\phis)
  =2\sin^{2}(\delta\phis/2)
\end{align}
up to \(\mathcal{O}(\sin^{4}(\delta\phis/2))\). 
Neglecting again any terms of the order \(\dot{\varphi}_i(1+\cos\phis)\) and \((\nabla\varphi_i)^2(1+\cos\phis)\), $i=\mathrm{s,L}$, as well as higher than quadratic terms in the derivatives, the theories for \(\phis\) and \(\varphi_\mathrm{L}\) decouple and we yet again obtain a free theory for \(\varphi_\mathrm{L}\) with
{
\begin{align}
    \bar{\mathcal{L}}^\mathrm{eff}_{\varphi_\text{L}} 
    = 
   \frac{1-\qbar}{16}\left[\left(\partial_{\bar{t}}{\varphi}_\mathrm{L}\right)^2
   -8\left({\nabla_{\bar{x}}\varphi_\mathrm{L}}\right)^2\right]
    \label{eq:efftheorykxiphiL}
\end{align}
}
and a DSG theory for \(\phis\),
\begin{align}
    \Re\bar{\mathcal{L}}^\mathrm{eff}_{\phis} 
    &= 
    \frac{1-\qbar^{2}}{16}\left[\left(\partial_{\bar{t}}{\varphi}_\mathrm{s}\right)^2
    -\left({\nabla_{\bar{x}}\phis}\right)^2\right]
    + \bar{A}_R \cos\phis
    - \bar{B}_R \sin^2\phis
    \,,    
    \label{eq:efftheorykxiphiS}
\end{align}
with coefficients
\begin{align}
    \bar{A}_R 
    &= \frac{1-\qbar^{2}}{4}\left[1-2\qbar^2+4\qbar^2(1-\qbar^2)+8\qbar^2(1-\qbar^2)^2\right] \,,
    \\
    \bar{B}_R
    &= \qbar^2\frac{1-\qbar^{2}}{4}\left[1+4(1-\qbar^2)+4(1-\qbar^2)^2\right] \,.
\end{align}
This again represents a double sine-Gordon Lagrangian, albeit with different `couplings'.
In the following we suppress overbars and assume all quantities to be dimensionless. We stress, however, that this is an approximation used to gain intuitive insight into the effects of the momentum dependence of the DSG couplings and is not intended  to constitute a rigorous derivation.
\section{Semiclassical simulations and experimental measurements}
\renewcommand{\theequation}{C\arabic{equation}}
In this \sectionterm, we discuss the signatures of the early-time mean-field dynamics of the spinor phase following a quench into the easy plane, as it appears both, in the simulations and the experiment.
We furthermore we briefly summarize the Truncated Wigner (TW) simulation method as well as our experimental procedures. 
We finally define the initial states for the TW simulations and provide further details on the TW results for DSG model in one and two spatial dimensions.

\subsection{Early-time spinor phase dynamics after a quench}
\label{app:EarlyTime}
%
\begin{figure*}[t]
    \centering
    \includegraphics[width=\linewidth]{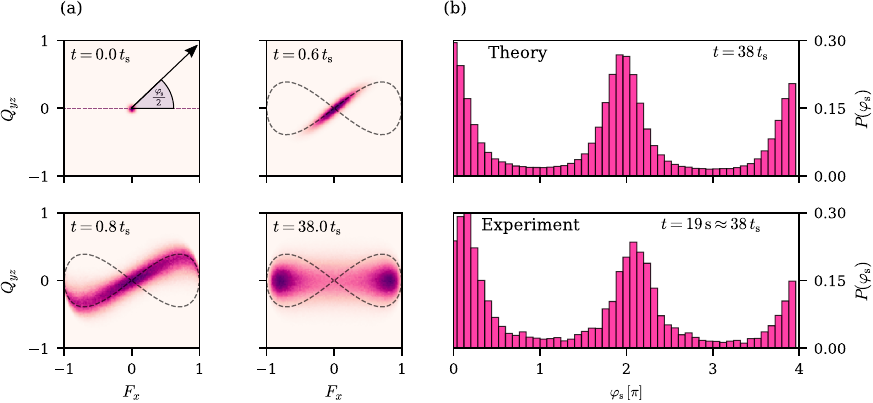}
    \caption{Spinor phase dynamics after a quench from the polar phase into the easy-plane phase. (a) Short-time evolution of the spinor phase probability distribution in the $F_x$-$Q_{yz}$ plane. The upper left panel gives the visual interpretation of the spinor phase. The dashed black lines in the other three panels shows the separatrix on the spin-nematic sphere \cite{Kunkel2019Thesis}. The distribution across a separatrix due to Bogoliubov instabilities ultimately leads to a settling of the field configuration in the values corresponding to the full spin orientation (lower right panel of (a)). Time is given in units of spin healing time
    $t_\mathrm{s} = 2\pi/(\tilde{\rho}|c_1|) \approx 3/(Qc_\mathrm{s})$,
    where $Q$ is the DSG initial-state momentum-box cutoff and $c_\mathrm{s}$ the DSG speed of sound, as used in the main text.
    (b) The theoretical probability distribution function extracted via the angle in the $F_x$-$Q_{yz}$ plane compared to the experimental one. Ths figure shows a larger occupation between the periodic potential minima, due to the method of extraction.}
    \label{fig:spinorphasedynamics}
\end{figure*}
\begin{figure*}[t]
    \centering
    \includegraphics[width=0.85\linewidth]{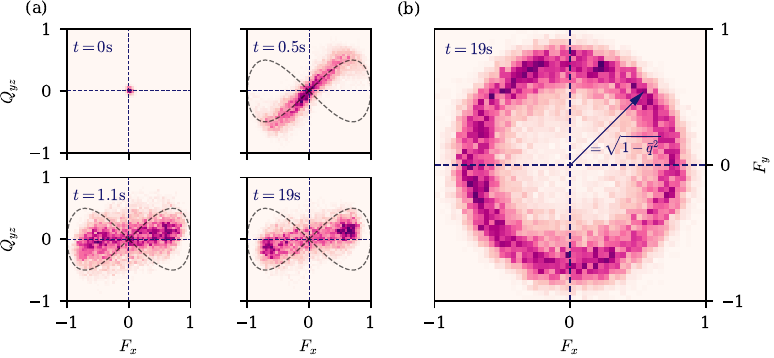}
    \caption{Experimental data after a quench from the polar into the easy-plane phase.
    (a) Time evolution of the probability distribution function in the $F_x$-$Q_{yz}$ plane. The short-time dynamics are characterized by a redistribution along the separatrix, followed by a settling down near the mean-field expectation value, as seen in the lower right panel.
    Notice a systematic distortion for long evolution times compared to \Fig{spinorphasedynamics} which is attributed to a readout calibration error, see \App{ExptMethods}.
    (b) Probability distribution function in the transverse spin plane. The ring-shaped distribution of field values shows that the system is in the easy-plane phase. The spin length $\abs{F_\perp}$ (the radius of the ring) allows for the estimation of the quadratic Zeeman shift according to $\abs{F_\perp}=[1-\bar{q}^2]^{1/2}$, with $\bar{q} = q/q_\mathrm{c} = q/(2\tilde{\rho}|c_1|)$.}
    \label{fig:ExpData}
\end{figure*}

The $\mathrm{U}(3)$ manifold is spanned by a total of $8$ generators, leading to the formation of several $\mathrm{SU}(2)$ subspaces. 
Particular subspaces, under the assumption of $\expval{F_z}=0$, are $\{F_x, Q_{yz},Q_0\}$ and $\{F_y, Q_{xz},Q_0\}$, with the nematic operator $Q_0 = -\frac{1}{3}\mathbb{1} - Q_{zz}$, in terms of the quadrupole operators $Q_{ij} = f_if_j + f_j f_i - 4\delta_{ij}/3$.
For brevity, we constrain the discussion here to the former subsphere, where the spinor phase represents the orientation on the $F_x$-$Q_{yz}$ plane as seen in the upper left panel of \Fig{spinorphasedynamics}a.
The extraction of the spinor phase can be done by numerically directly accessing the complex phases of the fundamental fields. Yet, one may also employ the spin-nematic sphere and read out the orientation of the field in the $F_x$-$Q_{yz}$ plane. The latter is the procedure which is implemented in the experiment.
It is important to note that the full spinor phase dynamics is given only by considering both spin-nematic subspheres simultaneously, thus eliminating the effect of the Larmor phase. By performing the readout of the coordinates in only one sphere, we obtain a nonvanishing probability of field configurations around $\phis\approx \pi$. 
\Fig{spinorphasedynamics}b showcases that this procedure reproduces the pedestals obtained from the experimental data in theory, cf.~\Fig{numhist} in the main text, corroborating a distribution of the experimental data according to the double sine-Gordon model. 
Note that the experimental data shows a systematic shift to higher field values. 
This was taken into account in the lower panel of \Fig{numhist}, where the histogram data shown in the upper panel of \Fig{spinorphasedynamics}b is inserted as a dashed line. 
For more details on the experimental shift in the data, see the following section on experimental methods and \Fig{ExpData}.
The uneven distribution of $\phis$ due to the finite size of the system and its nonvanishing energy causes the extracted effective potential to gain an additional mean-field shift that has to be taken into account, when regarding the potential in \Eq{DSGLagrangian}. 
Such a mean-field shift raises the potential slightly and was added to the full DSG potential to match the data.

\subsection{Truncated-Wigner simulations}
\label{app:TWA}
\subsubsection{Spin-1 Bose gas}
The dynamics of the spin-1 Bose gas is simulated using the Truncated-Wigner (TW) method. The numerical integration of the system gives the time evolution of the full spinor state $\vb*{\Psi} = (\psi_1, \psi_0, \psi_{-1})^T$ comprised of the complex scalar Bose fields describing the three magnetic components of the spin-1 manifold. 
We prepare the system in its corresponding zero-temperature mean-field ground state, i.e., either polar, \Eq{PolarMFGS}, or easy-plane, \eq{EPMFGS}, where for the latter we choose the densities to correspond to the correct $q$ value.
Upon such initialization, we add quantum noise sampled from the Wigner distribution of the coherent state to the Bogoliubov modes of the condensate \cite{Schmied:2018osf.PhysRevA.99.033611}.
We then propagate the initial field configuration by means of the classical field equations of motion derived by the Hamiltonian \eq{Spin1Hamiltonian},
\begin{align}
    \mathrm{i} \partial_t \vb*{\Psi}(x,t) 
    = \biggl[&-\frac{\partial_x^2}{2M} + qf_z^2+ {c_0}\totd(x,t) 
    + {c_1} \vb*{F}(x,t)\cdot \vb*{f}\biggr]\vb*{\Psi}(x,t)
    \,.
    \label{eq:Spin1GPE}
\end{align}
The physical parameters of the simulations are aimed at resembling a cloud of $^{87}\mathrm{Rb}$ atoms in a one-dimensional geometry as performed in the experiments \cite{Prufer:2018hto,Lannig2023a.2306.16497}, the main differences being a purely one-dimensional system and an increased homogeneous density as compared with the one realized in the strongly confined elongated trap in the experiments.
We give spatial length in terms of the spin healing length 
$\xi_\mathrm{s}=(2M\nc)^{-1/2}$
and time in units of the characteristic spin-changing collision time 
$t_\mathrm{s} = 2\pi/(\nc)=2\pi \xi_\mathrm{s}/c_\mathrm{S}$, 
with spin wave velocity 
$c_\mathrm{S}=(\tilde{\rho}|c_1|/2M)^{1/2}$.
Furthermore, the field operators are normalized with respect to the total density, 
$\Tilde{\Psi}_m=\Psi_m/\sqrt{\totd}$,
which also results in a normalization of the spin vector,
$\widetilde{\vb*{F}}=\vb*{F}/\totd$.
In the further discussion here and in the main text, the tilde is omitted and all values are to be understood as dimensionless, unless explicitly stated otherwise.
\subsubsection{Double sine-Gordon model}
Considering the previously derived DSG model with real-valued Lagrangian density
\begin{align}
    \mathcal{L} = \frac{1}{2}\dot{\varphi}^2 - \frac{c_\mathrm{s}^2}{2}(\nabla\varphi)^2 
    + \lambda \cos\varphi + \lambda_\mathrm{s}\sin^2\varphi
    \,,
\label{eq:DSGforNumerics}
\end{align}
with the free speed of sound $c_\mathrm{s}$ and DSG couplings $\lambda$ and $\lambda_\mathrm{s}$, we prepare the field $\varphi$ and its conjugate momentum $\dot{\varphi}$ in a far-from-equilibrium state corresponding to a box distribution in momentum space, with noise added to each mode, 
\begin{align}
    \varphi(x,0) &= \varphi_0 + \int\limits_{\infty}^\infty \frac{\dd{k}}{2\pi}\sqrt{\frac{f_k+1/2}{\omega_k}}c_k\e^{\i kx} 
    \,,\\
    \dot{\varphi}(x,0) &= \dot{\varphi}_0 + \int\limits_{\infty}^\infty \frac{\dd{k}}{2\pi}\sqrt{(f_k+1/2)\omega_k}\tilde{c}_k\e^{\i kx}
    \,,
\end{align}
where $\omega_k = \sqrt{k^2+M^2}$, and the noise coefficients $c_k$, $\tilde{c}_k$ satisfy the relations
\begin{align}
    \expval{c_k c^*_{k'}} = 2\pi \delta(k-k'), \quad \expval{c_k c_{k'}} = \expval{c^*_k c^*_{k'}} = 0\,.
\end{align}
The initial momentum distribution $f_k$ takes the form 
\begin{align}
    f_k = 
\begin{cases}
        &const. \quad |k|<Q \\
        &0  \qquad\quad \text{elsewhere}
    \end{cases}\,.
\end{align}
We then propagate the system according to its classical equations of motion
\begin{align}
    \ddot{\varphi} = c_\mathrm{s}^2 \Delta \varphi -\lambda \sin\varphi + \lambda_\mathrm{s}\sin(2\varphi)  \,.
    \label{eq:DSGEOM}
\end{align}
%
Our one-dimensional numerical grid for subdiffusive scaling comprises of $N=4096$ points with $5\cdot 10^5$ particles with $\lambda = 4\cdot 10^{-4} = 10\,\lambda_\mathrm{s}$ and $c_s^2 = 0.0262$ in numerical units, which differs from the full spin-1 values taken from \Eq{DSGLagrangian}, which give $\lambda_{\mathrm{spin-1}} = 1.9\cdot 10^{-4} \approx 5.8\lambda_{\mathrm{s}, \mathrm{spin-1}}$ and $c_s^2 \approx 0.01$.
The values of the couplings $\lambda$ and $\lambda_\mathrm{s}$ were chosen such as to achieve reliable self-similar scaling in the DSG system.
While the couplings of the DSG model can, in principle, be extracted from the full spin-1 theory, the initial condition for $\phis$ plays a crucial role in determining the system's behavior far from equilibrium. 
However, $\phis$ is not well-defined in the polar phase, and its full density matrix is not known, making its precise initialization nontrivial. 
Hence, to achieve a suitable far-from-equilibrium initial condition, we employ a momentum-box initial condition for $\phis$ as described below. 
This requires adjusting the couplings such as to ensure comparable scaling behavior. 
Despite these differences, the fundamental scaling mechanisms are expected to remain consistent.

For the diffusion-type scaling simulations, we chose $\lambda = 2.5\cdot 10^{-4} = \lambda_\mathrm{s} /2$, such that the potential landscape changes to exhibit a local maximum at $\varphi=0$ and two adjacent degenerate minima. 

\begin{figure*}[t]
    \centering
    \includegraphics[width=1\textwidth]{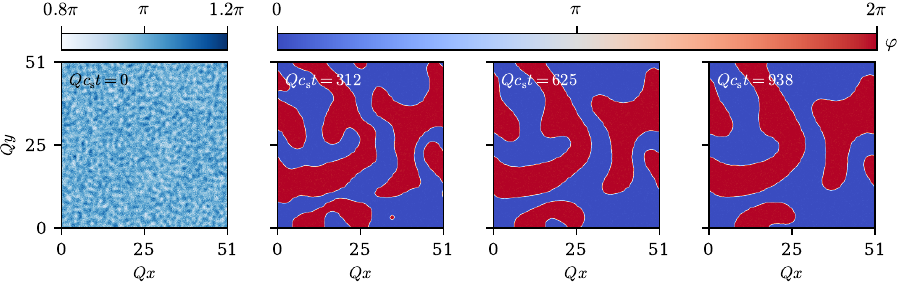}
    \caption{Self-similar scaling of the DSG model in (2+1)D.
    Snapshots of the 2D system at 4 different evolution times $Qc_\text{s}t\in\{0,312,625,938\}$ of the coarsening evolution. 
    In the initial state, the spinor phase is randomly distributed about $\varphi=\pi$, the value at a maximum of the DSG potential, fluctuating according to the box distribution shown in the left panel of \Fig{2DScaling}.
    The system early-on develops closed domains where $\varphi$ fluctuates around either of the two values $0$ and $2\pi$.
    With time proceeding these domains grow in size and eventually merge.
    }
    \label{fig:2DHistogram}
\end{figure*}
The two-dimensional numerical grid, in our simulations, comprises $N=8192\cross 8192$ points with $4\cdot10^6$ particles and couplings $\lambda = 1.6 = 100\,\lambda_\mathrm{s}$ in numerical units. 
The propagation of \Eq{DSGEOM} is done via a second-order leap-frog algorithm computed in parallel on graphics processing units (GPUs), where the observables are averaged over about $10^3$ realizations for one dimension and $10^2$ for two dimensions.
\subsection{Experimental methods}
\label{app:ExptMethods}
We briefly discuss the experimental system and methods that were employed for the acquisition of the experimental data shown in \Fig{numhist}. 
We prepare a Bose-Einstein condensate of $10^5$ $^{87}$Rb atoms in a quasi-one-dimensional box-like trapping potential, for more details see, e.g., \cite{Prufer:2022therm}. 
The experiments are performed in a homogeneous magnetic offset field of $\approx0.9\,$G, which gives rise to a second-order Zeeman shift 
$q_\text{i}\approx 2\pi\times58\,\mathrm{Hz} \gg 2 \tilde{\rho} |c_1|$,
cf.~\Eq{spin-1_lagrange_main}.

We prepare all atoms in the state 
$F=1$, $m_\text{F} = 0$
and initiate spin dynamics by quenching the quadratic Zeeman shift to 
$q_\text{f} \approx \totd |c_1|$
via off-resonant microwave dressing. The observables are extracted from the measured atomic densities by employing a POVM-readout, see \cite{Kunkel2019a.PhysRevLett.123.063603}.
We extract the one-dimensional spatial profiles of $F_x$ and $Q_{yz}$ simultaneously in every experimental realization. Many repetitions give rise to the phase-space distributions depicted in \Fig{ExpData}a. 
We bin the data according to the optical resolution of $\approx 1\,\mu$m and treat each bin as a separate point in the phase space spanned by $F_x$ and $Q_{yz}$.

The system is initialized in a symmetric coherent state and, for short evolution times up to $0.5$s, the measured distributions in $F_x$ and $Q_{yz}$ follow the so-called separatrix of the corresponding mean-field phase space trajectories~\cite{Hamley2012:NaturePhysics}, cf.~Figs.~\fig{spinorphasedynamics}a and \fig{ExpData}a.
For long evolution times, the system settles into a distribution with nonzero mean transversal spin length $F_\perp$, which can also be seen in \Fig{ExpData}b in the phase-space spanned by $F_x$ and $F_y$.
The dynamics of the measured phase-space distributions is in good qualitative agreement with the numerical simulations, as can be seen by comparing \Fig{spinorphasedynamics}  and \Fig{ExpData}.

We estimate the value of $q_f$ from the data shown in \Fig{ExpData}b by assuming that the configuration of the system has relaxed close to the mean-field ground state for late times.
The positions of the minima of the mean-field potential in the easy-plane phase depend on $q$ via 
$\abs{F_\perp}_{\rm{min}}=[1-\qbar^2]^{1/2}$.
As the distribution in $F_\perp$ has a maximum at 
$\abs{F_\perp} \approx 0.75$ at time $t=19$s,
we estimate 
$q_f \approx \totd |c_1|$.

Note that, in contrast to the numerical data shown in figure \Fig{spinorphasedynamics}, the measured phase-space distribution in $F_x$ and $Q_{yz}$ is systematically tilted from the $F_x$-axis for late evolution times.
We attribute this tilt to a systematic calibration error in the readout scheme. As a result, the readout axes are not perfectly orthogonal, which induces a distortion of the experimental distributions.
For the spinor-phase histogram shown in \Fig{numhist} this leads to a shift of $\approx 0.083(3)\,\pi$ and was taken into account by shifting the numerical curve accordingly (dashed line in \Fig{numhist}).

\end{appendix}

\bibliographystyle{apsrev4-1}

\begin{thebibliography}{81}%
\makeatletter
\providecommand \@ifxundefined [1]{%
 \@ifx{#1\undefined}
}%
\providecommand \@ifnum [1]{%
 \ifnum #1\expandafter \@firstoftwo
 \else \expandafter \@secondoftwo
 \fi
}%
\providecommand \@ifx [1]{%
 \ifx #1\expandafter \@firstoftwo
 \else \expandafter \@secondoftwo
 \fi
}%
\providecommand \natexlab [1]{#1}%
\providecommand \enquote  [1]{``#1''}%
\providecommand \bibnamefont  [1]{#1}%
\providecommand \bibfnamefont [1]{#1}%
\providecommand \citenamefont [1]{#1}%
\providecommand \href@noop [0]{\@secondoftwo}%
\providecommand \href [0]{\begingroup \@sanitize@url \@href}%
\providecommand \@href[1]{\@@startlink{#1}\@@href}%
\providecommand \@@href[1]{\endgroup#1\@@endlink}%
\providecommand \@sanitize@url [0]{\catcode `\\12\catcode `\$12\catcode
  `\&12\catcode `\#12\catcode `\^12\catcode `\_12\catcode `\%12\relax}%
\providecommand \@@startlink[1]{}%
\providecommand \@@endlink[0]{}%
\providecommand \url  [0]{\begingroup\@sanitize@url \@url }%
\providecommand \@url [1]{\endgroup\@href {#1}{\urlprefix }}%
\providecommand \urlprefix  [0]{URL }%
\providecommand \Eprint [0]{\href }%
\providecommand \doibase [0]{http://dx.doi.org/}%
\providecommand \selectlanguage [0]{\@gobble}%
\providecommand \bibinfo  [0]{\@secondoftwo}%
\providecommand \bibfield  [0]{\@secondoftwo}%
\providecommand \translation [1]{[#1]}%
\providecommand \BibitemOpen [0]{}%
\providecommand \bibitemStop [0]{}%
\providecommand \bibitemNoStop [0]{.\EOS\space}%
\providecommand \EOS [0]{\spacefactor3000\relax}%
\providecommand \BibitemShut  [1]{\csname bibitem#1\endcsname}%
\let\auto@bib@innerbib\@empty
\bibitem [{\citenamefont {Vinen}(2006)}]{Vinen2006a}%
  \BibitemOpen
  \bibfield  {author} {\bibinfo {author} {\bibfnamefont {W.}~\bibnamefont
  {Vinen}},\ }\href {\doibase 10.1007/s10909-006-9240-6} {\bibfield  {journal}
  {\bibinfo  {journal} {J. Low Temp. Phys.}\ }\textbf {\bibinfo {volume}
  {145}},\ \bibinfo {pages} {7} (\bibinfo {year} {2006})}\BibitemShut {NoStop}%
\bibitem [{\citenamefont {Tsubota}(2008)}]{Tsubota2008a}%
  \BibitemOpen
  \bibfield  {author} {\bibinfo {author} {\bibfnamefont {M.}~\bibnamefont
  {Tsubota}},\ }\href {\doibase 10.1143/JPSJ.77.111006} {\bibfield  {journal}
  {\bibinfo  {journal} {J. Phys. Soc. Jpn.}\ }\textbf {\bibinfo {volume}
  {77}},\ \bibinfo {pages} {111006} (\bibinfo {year} {2008})},\ \Eprint
  {http://arxiv.org/abs/0806.2737} {arXiv:0806.2737 [cond-mat.other]}
  \BibitemShut {NoStop}%
\bibitem [{\citenamefont {Zakharov}\ \emph {et~al.}(1992)\citenamefont
  {Zakharov}, \citenamefont {{L'vov}},\ and\ \citenamefont
  {Falkovich}}]{Zakharov1992a}%
  \BibitemOpen
  \bibfield  {author} {\bibinfo {author} {\bibfnamefont {V.~E.}\ \bibnamefont
  {Zakharov}}, \bibinfo {author} {\bibfnamefont {V.~S.}\ \bibnamefont
  {{L'vov}}}, \ and\ \bibinfo {author} {\bibfnamefont {G.}~\bibnamefont
  {Falkovich}},\ }\href {\doibase 10.1007/978-3-642-50052-7} {\emph {\bibinfo
  {title} {Kolmogorov Spectra of Turbulence I: Wave Turbulence}}}\ (\bibinfo
  {publisher} {Springer, Berlin},\ \bibinfo {year} {1992})\BibitemShut
  {NoStop}%
\bibitem [{\citenamefont {Nazarenko}(2011)}]{Nazarenko2011a}%
  \BibitemOpen
  \bibfield  {author} {\bibinfo {author} {\bibfnamefont {S.}~\bibnamefont
  {Nazarenko}},\ }\href {\doibase 10.1007/978-3-642-15942-8} {\emph {\bibinfo
  {title} {Wave turbulence}}},\ \bibinfo {series} {Lecture Notes in Physics}\
  No.\ \bibinfo {number} {825}\ (\bibinfo  {publisher} {Springer},\ \bibinfo
  {address} {Heidelberg},\ \bibinfo {year} {2011})\ pp.\ \bibinfo {pages} {XVI,
  279 S.}\BibitemShut {Stop}%
\bibitem [{\citenamefont {Berges}\ \emph {et~al.}(2008)\citenamefont {Berges},
  \citenamefont {Rothkopf},\ and\ \citenamefont {Schmidt}}]{Berges:2008wm}%
  \BibitemOpen
  \bibfield  {author} {\bibinfo {author} {\bibfnamefont {J.}~\bibnamefont
  {Berges}}, \bibinfo {author} {\bibfnamefont {A.}~\bibnamefont {Rothkopf}}, \
  and\ \bibinfo {author} {\bibfnamefont {J.}~\bibnamefont {Schmidt}},\ }\href
  {\doibase 10.1103/PhysRevLett.101.041603} {\bibfield  {journal} {\bibinfo
  {journal} {Phys. Rev. Lett.}\ }\textbf {\bibinfo {volume} {101}},\ \bibinfo
  {pages} {041603} (\bibinfo {year} {2008})},\ \Eprint
  {http://arxiv.org/abs/0803.0131} {arXiv:0803.0131 [hep-ph]} \BibitemShut
  {NoStop}%
\bibitem [{\citenamefont {Schole}\ \emph {et~al.}(2012)\citenamefont {Schole},
  \citenamefont {Nowak},\ and\ \citenamefont {Gasenzer}}]{Schole:2012kt}%
  \BibitemOpen
  \bibfield  {author} {\bibinfo {author} {\bibfnamefont {J.}~\bibnamefont
  {Schole}}, \bibinfo {author} {\bibfnamefont {B.}~\bibnamefont {Nowak}}, \
  and\ \bibinfo {author} {\bibfnamefont {T.}~\bibnamefont {Gasenzer}},\ }\href
  {\doibase 10.1103/PhysRevA.86.013624} {\bibfield  {journal} {\bibinfo
  {journal} {Phys. Rev. A}\ }\textbf {\bibinfo {volume} {86}},\ \bibinfo
  {pages} {013624} (\bibinfo {year} {2012})},\ \Eprint
  {http://arxiv.org/abs/1204.2487} {arXiv:1204.2487 [cond-mat.quant-gas]}
  \BibitemShut {NoStop}%
\bibitem [{\citenamefont {Pi{\~n}eiro~Orioli}\ \emph
  {et~al.}(2015)\citenamefont {Pi{\~n}eiro~Orioli}, \citenamefont
  {Boguslavski},\ and\ \citenamefont {Berges}}]{PineiroOrioli:2015dxa}%
  \BibitemOpen
  \bibfield  {author} {\bibinfo {author} {\bibfnamefont {A.}~\bibnamefont
  {Pi{\~n}eiro~Orioli}}, \bibinfo {author} {\bibfnamefont {K.}~\bibnamefont
  {Boguslavski}}, \ and\ \bibinfo {author} {\bibfnamefont {J.}~\bibnamefont
  {Berges}},\ }\href {\doibase 10.1103/PhysRevD.92.025041} {\bibfield
  {journal} {\bibinfo  {journal} {Phys. Rev. D}\ }\textbf {\bibinfo {volume}
  {92}},\ \bibinfo {pages} {025041} (\bibinfo {year} {2015})},\ \Eprint
  {http://arxiv.org/abs/1503.02498} {arXiv:1503.02498 [hep-ph]} \BibitemShut
  {NoStop}%
\bibitem [{\citenamefont {Chantesana}\ \emph {et~al.}(2019)\citenamefont
  {Chantesana}, \citenamefont {Pi{\~n}eiro~Orioli},\ and\ \citenamefont
  {Gasenzer}}]{Chantesana:2018qsb.PhysRevA.99.043620}%
  \BibitemOpen
  \bibfield  {author} {\bibinfo {author} {\bibfnamefont {I.}~\bibnamefont
  {Chantesana}}, \bibinfo {author} {\bibfnamefont {A.}~\bibnamefont
  {Pi{\~n}eiro~Orioli}}, \ and\ \bibinfo {author} {\bibfnamefont
  {T.}~\bibnamefont {Gasenzer}},\ }\href {\doibase 10.1103/PhysRevA.99.043620}
  {\bibfield  {journal} {\bibinfo  {journal} {Phys. Rev. A}\ }\textbf {\bibinfo
  {volume} {99}},\ \bibinfo {pages} {043620} (\bibinfo {year} {2019})},\
  \Eprint {http://arxiv.org/abs/1801.09490} {arXiv:1801.09490
  [cond-mat.quant-gas]} \BibitemShut {NoStop}%
\bibitem [{\citenamefont {Mikheev}\ \emph {et~al.}(2019)\citenamefont
  {Mikheev}, \citenamefont {Schmied},\ and\ \citenamefont
  {Gasenzer}}]{Mikheev:2018adp}%
  \BibitemOpen
  \bibfield  {author} {\bibinfo {author} {\bibfnamefont {A.~N.}\ \bibnamefont
  {Mikheev}}, \bibinfo {author} {\bibfnamefont {C.-M.}\ \bibnamefont
  {Schmied}}, \ and\ \bibinfo {author} {\bibfnamefont {T.}~\bibnamefont
  {Gasenzer}},\ }\href {\doibase 10.1103/PhysRevA.99.063622} {\bibfield
  {journal} {\bibinfo  {journal} {Phys. Rev. A}\ }\textbf {\bibinfo {volume}
  {99}},\ \bibinfo {pages} {063622} (\bibinfo {year} {2019})},\ \Eprint
  {http://arxiv.org/abs/1807.10228} {arXiv:1807.10228 [cond-mat.quant-gas]}
  \BibitemShut {NoStop}%
\bibitem [{\citenamefont {Hohenberg}\ and\ \citenamefont
  {Halperin}(1977)}]{Hohenberg1977a}%
  \BibitemOpen
  \bibfield  {author} {\bibinfo {author} {\bibfnamefont {P.~C.}\ \bibnamefont
  {Hohenberg}}\ and\ \bibinfo {author} {\bibfnamefont {B.~I.}\ \bibnamefont
  {Halperin}},\ }\href {\doibase 10.1103/RevModPhys.49.435} {\bibfield
  {journal} {\bibinfo  {journal} {Rev. Mod. Phys.}\ }\textbf {\bibinfo {volume}
  {49}},\ \bibinfo {pages} {435} (\bibinfo {year} {1977})}\BibitemShut
  {NoStop}%
\bibitem [{\citenamefont {Janssen}(1979)}]{Janssen1979a}%
  \BibitemOpen
  \bibfield  {author} {\bibinfo {author} {\bibfnamefont {H.}~\bibnamefont
  {Janssen}},\ }in\ \href {\doibase 10.1007/3-540-09523-3_2} {\emph {\bibinfo
  {booktitle} {Dynamical critical phenomena and related topics, Lecture Notes
  in Physics, vol. 104}}}\ (\bibinfo  {publisher} {Springer, Heidelberg},\
  \bibinfo {year} {1979})\ p.~\bibinfo {pages} {26}\BibitemShut {NoStop}%
\bibitem [{\citenamefont {Diehl}(1986)}]{Diehl1986a}%
  \BibitemOpen
  \bibfield  {author} {\bibinfo {author} {\bibfnamefont {H.~W.}\ \bibnamefont
  {Diehl}},\ }in\ \href@noop {} {\emph {\bibinfo {booktitle} {Phase Transitions
  and Critical Phenomena}}}\ (\bibinfo  {publisher} {Academic Press, London},\
  \bibinfo {year} {1986})\BibitemShut {NoStop}%
\bibitem [{\citenamefont {Janssen}(1992)}]{Janssen1992a}%
  \BibitemOpen
  \bibfield  {author} {\bibinfo {author} {\bibfnamefont {H.}~\bibnamefont
  {Janssen}},\ }in\ \href {\doibase 10.1142/1633} {\emph {\bibinfo {booktitle}
  {From phase transitions to chaos}}}\ (\bibinfo  {publisher} {World
  Scientific, Singapore},\ \bibinfo {year} {1992})\ p.~\bibinfo {pages}
  {68}\BibitemShut {NoStop}%
\bibitem [{\citenamefont {Bray}(1994)}]{Bray1994a.AdvPhys.43.357}%
  \BibitemOpen
  \bibfield  {author} {\bibinfo {author} {\bibfnamefont {A.~J.}\ \bibnamefont
  {Bray}},\ }\href {\doibase 10.1080/00018739400101505} {\bibfield  {journal}
  {\bibinfo  {journal} {Adv. Phys.}\ }\textbf {\bibinfo {volume} {43}},\
  \bibinfo {pages} {357} (\bibinfo {year} {1994})}\BibitemShut {NoStop}%
\bibitem [{\citenamefont {Puri}\ and\ \citenamefont
  {Wadhawan}(2009)}]{Puri2019a.KineticsOfPT}%
  \BibitemOpen
  \bibinfo {editor} {\bibfnamefont {S.}~\bibnamefont {Puri}}\ and\ \bibinfo
  {editor} {\bibfnamefont {V.}~\bibnamefont {Wadhawan}},\ eds.,\ \href
  {\doibase 10.1201/9781420008364} {\emph {\bibinfo {title} {Kinetics of Phase
  Transitions}}}\ (\bibinfo  {publisher} {Taylor \& Francis Group, Boca
  Raton},\ \bibinfo {year} {2009})\BibitemShut {NoStop}%
\bibitem [{\citenamefont
  {{Cugliandolo}}(2015)}]{Cugliandolo2014arXiv1412.0855C}%
  \BibitemOpen
  \bibfield  {author} {\bibinfo {author} {\bibfnamefont {L.~F.}\ \bibnamefont
  {{Cugliandolo}}},\ }\href {\doibase 10.1016/j.crhy.2015.02.005} {\bibfield
  {journal} {\bibinfo  {journal} {C. R. Phys.}\ }\textbf {\bibinfo {volume}
  {16}},\ \bibinfo {pages} {257} (\bibinfo {year} {2015})},\ \Eprint
  {http://arxiv.org/abs/1412.0855} {arXiv:1412.0855 [cond-mat.stat-mech]}
  \BibitemShut {NoStop}%
\bibitem [{\citenamefont {Henn}\ \emph {et~al.}(2009)\citenamefont {Henn},
  \citenamefont {Seman}, \citenamefont {Roati}, \citenamefont {Magalh\~aes},\
  and\ \citenamefont {Bagnato}}]{Henn2009a.PhysRevLett.103.045301}%
  \BibitemOpen
  \bibfield  {author} {\bibinfo {author} {\bibfnamefont {E.~A.~L.}\
  \bibnamefont {Henn}}, \bibinfo {author} {\bibfnamefont {J.~A.}\ \bibnamefont
  {Seman}}, \bibinfo {author} {\bibfnamefont {G.}~\bibnamefont {Roati}},
  \bibinfo {author} {\bibfnamefont {K.~M.~F.}\ \bibnamefont {Magalh\~aes}}, \
  and\ \bibinfo {author} {\bibfnamefont {V.~S.}\ \bibnamefont {Bagnato}},\
  }\href {\doibase 10.1103/PhysRevLett.103.045301} {\bibfield  {journal}
  {\bibinfo  {journal} {Phys. Rev. Lett.}\ }\textbf {\bibinfo {volume} {103}},\
  \bibinfo {pages} {045301} (\bibinfo {year} {2009})}\BibitemShut {NoStop}%
\bibitem [{\citenamefont {Gring}\ \emph {et~al.}(2012)\citenamefont {Gring},
  \citenamefont {Kuhnert}, \citenamefont {Langen}, \citenamefont {Kitagawa},
  \citenamefont {Rauer}, \citenamefont {Schreitl}, \citenamefont {Mazets},
  \citenamefont {Smith}, \citenamefont {Demler},\ and\ \citenamefont
  {Schmiedmayer}}]{Gring2011a}%
  \BibitemOpen
  \bibfield  {author} {\bibinfo {author} {\bibfnamefont {M.}~\bibnamefont
  {Gring}}, \bibinfo {author} {\bibfnamefont {M.}~\bibnamefont {Kuhnert}},
  \bibinfo {author} {\bibfnamefont {T.}~\bibnamefont {Langen}}, \bibinfo
  {author} {\bibfnamefont {T.}~\bibnamefont {Kitagawa}}, \bibinfo {author}
  {\bibfnamefont {B.}~\bibnamefont {Rauer}}, \bibinfo {author} {\bibfnamefont
  {M.}~\bibnamefont {Schreitl}}, \bibinfo {author} {\bibfnamefont
  {I.}~\bibnamefont {Mazets}}, \bibinfo {author} {\bibfnamefont {D.~A.}\
  \bibnamefont {Smith}}, \bibinfo {author} {\bibfnamefont {E.}~\bibnamefont
  {Demler}}, \ and\ \bibinfo {author} {\bibfnamefont {J.}~\bibnamefont
  {Schmiedmayer}},\ }\href {\doibase 10.1126/science.1224953} {\bibfield
  {journal} {\bibinfo  {journal} {Science}\ }\textbf {\bibinfo {volume}
  {337}},\ \bibinfo {pages} {1318} (\bibinfo {year} {2012})}\BibitemShut
  {NoStop}%
\bibitem [{\citenamefont {Smith}\ \emph {et~al.}(2013)\citenamefont {Smith},
  \citenamefont {Gring}, \citenamefont {Langen}, \citenamefont {Kuhnert},
  \citenamefont {Rauer}, \citenamefont {Geiger}, \citenamefont {Kitagawa},
  \citenamefont {Mazets}, \citenamefont {Demler},\ and\ \citenamefont
  {Schmiedmayer}}]{AduSmith2013a}%
  \BibitemOpen
  \bibfield  {author} {\bibinfo {author} {\bibfnamefont {D.~A.}\ \bibnamefont
  {Smith}}, \bibinfo {author} {\bibfnamefont {M.}~\bibnamefont {Gring}},
  \bibinfo {author} {\bibfnamefont {T.}~\bibnamefont {Langen}}, \bibinfo
  {author} {\bibfnamefont {M.}~\bibnamefont {Kuhnert}}, \bibinfo {author}
  {\bibfnamefont {B.}~\bibnamefont {Rauer}}, \bibinfo {author} {\bibfnamefont
  {R.}~\bibnamefont {Geiger}}, \bibinfo {author} {\bibfnamefont
  {T.}~\bibnamefont {Kitagawa}}, \bibinfo {author} {\bibfnamefont
  {I.}~\bibnamefont {Mazets}}, \bibinfo {author} {\bibfnamefont
  {E.}~\bibnamefont {Demler}}, \ and\ \bibinfo {author} {\bibfnamefont
  {J.}~\bibnamefont {Schmiedmayer}},\ }\href
  {http://stacks.iop.org/1367-2630/15/i=7/a=075011} {\bibfield  {journal}
  {\bibinfo  {journal} {New J. Phys.}\ }\textbf {\bibinfo {volume} {15}},\
  \bibinfo {pages} {075011} (\bibinfo {year} {2013})}\BibitemShut {NoStop}%
\bibitem [{\citenamefont {Langen}\ \emph {et~al.}(2015)\citenamefont {Langen},
  \citenamefont {Erne}, \citenamefont {Geiger}, \citenamefont {Rauer},
  \citenamefont {Schweigler}, \citenamefont {Kuhnert}, \citenamefont
  {Rohringer}, \citenamefont {Mazets}, \citenamefont {Gasenzer},\ and\
  \citenamefont {Schmiedmayer}}]{Langen2015b.Science348.207}%
  \BibitemOpen
  \bibfield  {author} {\bibinfo {author} {\bibfnamefont {T.}~\bibnamefont
  {Langen}}, \bibinfo {author} {\bibfnamefont {S.}~\bibnamefont {Erne}},
  \bibinfo {author} {\bibfnamefont {R.}~\bibnamefont {Geiger}}, \bibinfo
  {author} {\bibfnamefont {B.}~\bibnamefont {Rauer}}, \bibinfo {author}
  {\bibfnamefont {T.}~\bibnamefont {Schweigler}}, \bibinfo {author}
  {\bibfnamefont {M.}~\bibnamefont {Kuhnert}}, \bibinfo {author} {\bibfnamefont
  {W.}~\bibnamefont {Rohringer}}, \bibinfo {author} {\bibfnamefont {I.~E.}\
  \bibnamefont {Mazets}}, \bibinfo {author} {\bibfnamefont {T.}~\bibnamefont
  {Gasenzer}}, \ and\ \bibinfo {author} {\bibfnamefont {J.}~\bibnamefont
  {Schmiedmayer}},\ }\href {\doibase 10.1126/science.1257026} {\bibfield
  {journal} {\bibinfo  {journal} {Science}\ }\textbf {\bibinfo {volume}
  {348}},\ \bibinfo {pages} {207} (\bibinfo {year} {2015})}\BibitemShut
  {NoStop}%
\bibitem [{\citenamefont {{Navon}}\ \emph {et~al.}(2015)\citenamefont
  {{Navon}}, \citenamefont {{Gaunt}}, \citenamefont {{Smith}},\ and\
  \citenamefont {{Hadzibabic}}}]{Navon2015a.Science.347.167N}%
  \BibitemOpen
  \bibfield  {author} {\bibinfo {author} {\bibfnamefont {N.}~\bibnamefont
  {{Navon}}}, \bibinfo {author} {\bibfnamefont {A.~L.}\ \bibnamefont
  {{Gaunt}}}, \bibinfo {author} {\bibfnamefont {R.~P.}\ \bibnamefont
  {{Smith}}}, \ and\ \bibinfo {author} {\bibfnamefont {Z.}~\bibnamefont
  {{Hadzibabic}}},\ }\href {\doibase 10.1126/science.1258676} {\bibfield
  {journal} {\bibinfo  {journal} {Science}\ }\textbf {\bibinfo {volume}
  {347}},\ \bibinfo {pages} {167} (\bibinfo {year} {2015})},\ \Eprint
  {http://arxiv.org/abs/1410.8487} {arXiv:1410.8487 [cond-mat.quant-gas]}
  \BibitemShut {NoStop}%
\bibitem [{\citenamefont {{Navon}}\ \emph {et~al.}(2016)\citenamefont
  {{Navon}}, \citenamefont {{Gaunt}}, \citenamefont {{Smith}},\ and\
  \citenamefont {{Hadzibabic}}}]{Navon2016a.Nature.539.72}%
  \BibitemOpen
  \bibfield  {author} {\bibinfo {author} {\bibfnamefont {N.}~\bibnamefont
  {{Navon}}}, \bibinfo {author} {\bibfnamefont {A.~L.}\ \bibnamefont
  {{Gaunt}}}, \bibinfo {author} {\bibfnamefont {R.~P.}\ \bibnamefont
  {{Smith}}}, \ and\ \bibinfo {author} {\bibfnamefont {Z.}~\bibnamefont
  {{Hadzibabic}}},\ }\href {\doibase 10.1038/nature20114} {\bibfield  {journal}
  {\bibinfo  {journal} {Nature}\ }\textbf {\bibinfo {volume} {539}},\ \bibinfo
  {pages} {72} (\bibinfo {year} {2016})},\ \Eprint
  {http://arxiv.org/abs/1609.01271} {arXiv:1609.01271 [cond-mat.quant-gas]}
  \BibitemShut {NoStop}%
\bibitem [{\citenamefont {Rauer}\ \emph {et~al.}(2018)\citenamefont {Rauer},
  \citenamefont {Erne}, \citenamefont {Schweigler}, \citenamefont {Cataldini},
  \citenamefont {Tajik},\ and\ \citenamefont
  {Schmiedmayer}}]{Rauer2017a.arXiv170508231R.Science360.307}%
  \BibitemOpen
  \bibfield  {author} {\bibinfo {author} {\bibfnamefont {B.}~\bibnamefont
  {Rauer}}, \bibinfo {author} {\bibfnamefont {S.}~\bibnamefont {Erne}},
  \bibinfo {author} {\bibfnamefont {T.}~\bibnamefont {Schweigler}}, \bibinfo
  {author} {\bibfnamefont {F.}~\bibnamefont {Cataldini}}, \bibinfo {author}
  {\bibfnamefont {M.}~\bibnamefont {Tajik}}, \ and\ \bibinfo {author}
  {\bibfnamefont {J.}~\bibnamefont {Schmiedmayer}},\ }\href 
  {\doibase 10.1126/science.aan7938} {\bibfield  {journal} {\bibinfo  {journal}
  {Science}\ }\textbf {\bibinfo {volume} {360}},\ \bibinfo {pages} {307}
  (\bibinfo {year} {2018})}\BibitemShut {NoStop}%
\bibitem [{\citenamefont {Gauthier}\ \emph {et~al.}(2019)\citenamefont
  {Gauthier}, \citenamefont {Reeves}, \citenamefont {Yu}, \citenamefont
  {Bradley}, \citenamefont {Baker}, \citenamefont {Bell}, \citenamefont
  {Rubinsztein-Dunlop}, \citenamefont {Davis},\ and\ \citenamefont
  {Neely}}]{Gauthier2019a.Science.364.1264}%
  \BibitemOpen
  \bibfield  {author} {\bibinfo {author} {\bibfnamefont {G.}~\bibnamefont
  {Gauthier}}, \bibinfo {author} {\bibfnamefont {M.~T.}\ \bibnamefont
  {Reeves}}, \bibinfo {author} {\bibfnamefont {X.}~\bibnamefont {Yu}}, \bibinfo
  {author} {\bibfnamefont {A.~S.}\ \bibnamefont {Bradley}}, \bibinfo {author}
  {\bibfnamefont {M.~A.}\ \bibnamefont {Baker}}, \bibinfo {author}
  {\bibfnamefont {T.~A.}\ \bibnamefont {Bell}}, \bibinfo {author}
  {\bibfnamefont {H.}~\bibnamefont {Rubinsztein-Dunlop}}, \bibinfo {author}
  {\bibfnamefont {M.~J.}\ \bibnamefont {Davis}}, \ and\ \bibinfo {author}
  {\bibfnamefont {T.~W.}\ \bibnamefont {Neely}},\ }\href 
  {\doibase 10.1126/science.aat5718} {\bibfield  {journal} {\bibinfo  {journal}
  {Science}\ }\textbf {\bibinfo {volume} {364}},\ \bibinfo {pages} {1264}
  (\bibinfo {year} {2019})},\ \Eprint {http://arxiv.org/abs/1801.06951}
  {arXiv:1801.06951 [cond-mat. quant-gas]} \BibitemShut {NoStop}%
\bibitem [{\citenamefont {Johnstone}\ \emph {et~al.}(2019)\citenamefont
  {Johnstone}, \citenamefont {Groszek}, \citenamefont {Starkey}, \citenamefont
  {Billington}, \citenamefont {Simula},\ and\ \citenamefont
  {Helmerson}}]{Johnstone2019a.Science.364.1267}%
  \BibitemOpen
  \bibfield  {author} {\bibinfo {author} {\bibfnamefont {S.~P.}\ \bibnamefont
  {Johnstone}}, \bibinfo {author} {\bibfnamefont {A.~J.}\ \bibnamefont
  {Groszek}}, \bibinfo {author} {\bibfnamefont {P.~T.}\ \bibnamefont
  {Starkey}}, \bibinfo {author} {\bibfnamefont {C.~J.}\ \bibnamefont
  {Billington}}, \bibinfo {author} {\bibfnamefont {T.~P.}\ \bibnamefont
  {Simula}}, \ and\ \bibinfo {author} {\bibfnamefont {K.}~\bibnamefont
  {Helmerson}},\ }\href {\doibase 10.1126/science.aat5793} {\bibfield
  {journal} {\bibinfo  {journal} {Science}\ }\textbf {\bibinfo {volume}
  {364}},\ \bibinfo {pages} {1267} (\bibinfo {year} {2019})},\ \Eprint
  {http://arxiv.org/abs/1801.06952v2} {arXiv:1801.06952v2 [cond-mat.quant-gas]}
  \BibitemShut {NoStop}%
\bibitem [{\citenamefont {{Eigen}}\ \emph {et~al.}(2018)\citenamefont
  {{Eigen}}, \citenamefont {{Glidden}}, \citenamefont {{Lopes}}, \citenamefont
  {{Cornell}}, \citenamefont {{Smith}},\ and\ \citenamefont
  {{Hadzibabic}}}]{Eigen2018a.arXiv180509802E}%
  \BibitemOpen
  \bibfield  {author} {\bibinfo {author} {\bibfnamefont {C.}~\bibnamefont
  {{Eigen}}}, \bibinfo {author} {\bibfnamefont {J.~A.~P.}\ \bibnamefont
  {{Glidden}}}, \bibinfo {author} {\bibfnamefont {R.}~\bibnamefont {{Lopes}}},
  \bibinfo {author} {\bibfnamefont {E.~A.}\ \bibnamefont {{Cornell}}}, \bibinfo
  {author} {\bibfnamefont {R.~P.}\ \bibnamefont {{Smith}}}, \ and\ \bibinfo
  {author} {\bibfnamefont {Z.}~\bibnamefont {{Hadzibabic}}},\ }\href 
  {\doibase 10.1038/s41586-018-0674-1} {\bibfield  {journal} {\bibinfo  {journal}
  {Nature}\ }\textbf {\bibinfo {volume} {563}},\ \bibinfo {pages} {221}
  (\bibinfo {year} {2018})},\ \Eprint {http://arxiv.org/abs/1805.09802}
  {arXiv:1805.09802 [cond-mat.quant-gas]} \BibitemShut {NoStop}%
\bibitem [{\citenamefont {Pr{\"u}fer}\ \emph {et~al.}(2018)\citenamefont
  {Pr{\"u}fer}, \citenamefont {Kunkel}, \citenamefont {Strobel}, \citenamefont
  {Lannig}, \citenamefont {Linnemann}, \citenamefont {Schmied}, \citenamefont
  {Berges}, \citenamefont {Gasenzer},\ and\ \citenamefont
  {Oberthaler}}]{Prufer:2018hto}%
  \BibitemOpen
  \bibfield  {author} {\bibinfo {author} {\bibfnamefont {M.}~\bibnamefont
  {Pr{\"u}fer}}, \bibinfo {author} {\bibfnamefont {P.}~\bibnamefont {Kunkel}},
  \bibinfo {author} {\bibfnamefont {H.}~\bibnamefont {Strobel}}, \bibinfo
  {author} {\bibfnamefont {S.}~\bibnamefont {Lannig}}, \bibinfo {author}
  {\bibfnamefont {D.}~\bibnamefont {Linnemann}}, \bibinfo {author}
  {\bibfnamefont {C.-M.}\ \bibnamefont {Schmied}}, \bibinfo {author}
  {\bibfnamefont {J.}~\bibnamefont {Berges}}, \bibinfo {author} {\bibfnamefont
  {T.}~\bibnamefont {Gasenzer}}, \ and\ \bibinfo {author} {\bibfnamefont
  {M.~K.}\ \bibnamefont {Oberthaler}},\ }\href 
  {\doibase 10.1038/s41586-018-0659-0} {\bibfield  {journal} {\bibinfo  {journal}
  {Nature}\ }\textbf {\bibinfo {volume} {563}},\ \bibinfo {pages} {217}
  (\bibinfo {year} {2018})},\ \Eprint {http://arxiv.org/abs/1805.11881}
  {arXiv:1805.11881 [cond-mat.quant-gas]} \BibitemShut {NoStop}%
\bibitem [{\citenamefont {Erne}\ \emph {et~al.}(2018)\citenamefont {Erne},
  \citenamefont {B{\"u}cker}, \citenamefont {Gasenzer}, \citenamefont
  {Berges},\ and\ \citenamefont {Schmiedmayer}}]{Erne:2018gmz}%
  \BibitemOpen
  \bibfield  {author} {\bibinfo {author} {\bibfnamefont {S.}~\bibnamefont
  {Erne}}, \bibinfo {author} {\bibfnamefont {R.}~\bibnamefont {B{\"u}cker}},
  \bibinfo {author} {\bibfnamefont {T.}~\bibnamefont {Gasenzer}}, \bibinfo
  {author} {\bibfnamefont {J.}~\bibnamefont {Berges}}, \ and\ \bibinfo {author}
  {\bibfnamefont {J.}~\bibnamefont {Schmiedmayer}},\ }\href 
  {\doibase 10.1038/s41586-018-0667-0} {\bibfield  {journal} {\bibinfo  {journal}
  {Nature}\ }\textbf {\bibinfo {volume} {563}},\ \bibinfo {pages} {225}
  (\bibinfo {year} {2018})},\ \Eprint {http://arxiv.org/abs/1805.12310}
  {arXiv:1805.12310 [cond-mat. quant-gas]} \BibitemShut {NoStop}%
\bibitem [{\citenamefont {Navon}\ \emph {et~al.}(2019)\citenamefont {Navon},
  \citenamefont {Eigen}, \citenamefont {Zhang}, \citenamefont {Lopes},
  \citenamefont {Gaunt}, \citenamefont {Fujimoto}, \citenamefont {Tsubota},
  \citenamefont {Smith},\ and\ \citenamefont
  {Hadzibabic}}]{Navon2018a.Science.366.382}%
  \BibitemOpen
  \bibfield  {author} {\bibinfo {author} {\bibfnamefont {N.}~\bibnamefont
  {Navon}}, \bibinfo {author} {\bibfnamefont {C.}~\bibnamefont {Eigen}},
  \bibinfo {author} {\bibfnamefont {J.}~\bibnamefont {Zhang}}, \bibinfo
  {author} {\bibfnamefont {R.}~\bibnamefont {Lopes}}, \bibinfo {author}
  {\bibfnamefont {A.~L.}\ \bibnamefont {Gaunt}}, \bibinfo {author}
  {\bibfnamefont {K.}~\bibnamefont {Fujimoto}}, \bibinfo {author}
  {\bibfnamefont {M.}~\bibnamefont {Tsubota}}, \bibinfo {author} {\bibfnamefont
  {R.~P.}\ \bibnamefont {Smith}}, \ and\ \bibinfo {author} {\bibfnamefont
  {Z.}~\bibnamefont {Hadzibabic}},\ }\href {\doibase 10.1126/science.aau6103}
  {\bibfield  {journal} {\bibinfo  {journal} {Science}\ }\textbf {\bibinfo
  {volume} {366}},\ \bibinfo {pages} {382} (\bibinfo {year}
  {2019})}\BibitemShut {NoStop}%
\bibitem [{\citenamefont {Glidden}\ \emph {et~al.}(2021)\citenamefont
  {Glidden}, \citenamefont {Eigen}, \citenamefont {Dogra}, \citenamefont
  {Hilker}, \citenamefont {Smith},\ and\ \citenamefont
  {Hadzibabic}}]{Glidden:2020qmu}%
  \BibitemOpen
  \bibfield  {author} {\bibinfo {author} {\bibfnamefont {J.~A.~P.}\
  \bibnamefont {Glidden}}, \bibinfo {author} {\bibfnamefont {C.}~\bibnamefont
  {Eigen}}, \bibinfo {author} {\bibfnamefont {L.~H.}\ \bibnamefont {Dogra}},
  \bibinfo {author} {\bibfnamefont {T.~A.}\ \bibnamefont {Hilker}}, \bibinfo
  {author} {\bibfnamefont {R.~P.}\ \bibnamefont {Smith}}, \ and\ \bibinfo
  {author} {\bibfnamefont {Z.}~\bibnamefont {Hadzibabic}},\ }\href 
  {\doibase 10.1038/s41567-020-01114-x} {\bibfield  {journal} {\bibinfo  {journal}
  {Nature Phys.}\ }\textbf {\bibinfo {volume} {17}},\ \bibinfo {pages} {457}
  (\bibinfo {year} {2021})},\ \Eprint {http://arxiv.org/abs/2006.01118}
  {arXiv: 2006.01118 [cond-mat.quant-gas]} \BibitemShut {NoStop}%
\bibitem [{\citenamefont {Garc\'{\i}a-Orozco}\ \emph
  {et~al.}(2022)\citenamefont {Garc\'{\i}a-Orozco}, \citenamefont {Madeira},
  \citenamefont {Moreno-Armijos}, \citenamefont {Fritsch}, \citenamefont
  {Tavares}, \citenamefont {Castilho}, \citenamefont {Cidrim}, \citenamefont
  {Roati},\ and\ \citenamefont
  {Bagnato}}]{GarciaOrozco2021a.PhysRevA.106.023314}%
  \BibitemOpen
  \bibfield  {author} {\bibinfo {author} {\bibfnamefont {A.~D.}\ \bibnamefont
  {Garc\'{\i}a-Orozco}}, \bibinfo {author} {\bibfnamefont {L.}~\bibnamefont
  {Madeira}}, \bibinfo {author} {\bibfnamefont {M.~A.}\ \bibnamefont
  {Moreno-Armijos}}, \bibinfo {author} {\bibfnamefont {A.~R.}\ \bibnamefont
  {Fritsch}}, \bibinfo {author} {\bibfnamefont {P.~E.~S.}\ \bibnamefont
  {Tavares}}, \bibinfo {author} {\bibfnamefont {P.~C.~M.}\ \bibnamefont
  {Castilho}}, \bibinfo {author} {\bibfnamefont {A.}~\bibnamefont {Cidrim}},
  \bibinfo {author} {\bibfnamefont {G.}~\bibnamefont {Roati}}, \ and\ \bibinfo
  {author} {\bibfnamefont {V.~S.}\ \bibnamefont {Bagnato}},\ }\href 
  {\doibase 10.1103/PhysRevA.106.023314} {\bibfield  {journal} {\bibinfo  {journal}
  {Phys. Rev. A}\ }\textbf {\bibinfo {volume} {106}},\ \bibinfo {pages}
  {023314} (\bibinfo {year} {2022})},\ \Eprint
  {http://arxiv.org/abs/2107.07421} {arXiv:2107.07421 [cond-mat.quant-gas]}
  \BibitemShut {NoStop}%
\bibitem [{\citenamefont {Lannig}\ \emph {et~al.}(2023)\citenamefont {Lannig},
  \citenamefont {Pr\"ufer}, \citenamefont {Deller}, \citenamefont {Siovitz},
  \citenamefont {Dreher}, \citenamefont {Gasenzer}, \citenamefont {Strobel},\
  and\ \citenamefont {Oberthaler}}]{Lannig2023a.2306.16497}%
  \BibitemOpen
  \bibfield  {author} {\bibinfo {author} {\bibfnamefont {S.}~\bibnamefont
  {Lannig}}, \bibinfo {author} {\bibfnamefont {M.}~\bibnamefont {Pr\"ufer}},
  \bibinfo {author} {\bibfnamefont {Y.}~\bibnamefont {Deller}}, \bibinfo
  {author} {\bibfnamefont {I.}~\bibnamefont {Siovitz}}, \bibinfo {author}
  {\bibfnamefont {J.}~\bibnamefont {Dreher}}, \bibinfo {author} {\bibfnamefont
  {T.}~\bibnamefont {Gasenzer}}, \bibinfo {author} {\bibfnamefont
  {H.}~\bibnamefont {Strobel}}, \ and\ \bibinfo {author} {\bibfnamefont
  {M.~K.}\ \bibnamefont {Oberthaler}},\ }\href@noop {} {\  (\bibinfo {year}
  {2023})},\ \Eprint {http://arxiv.org/abs/2306.16497} {arXiv:2306.16497
  [cond-mat.quant-gas]} \BibitemShut {NoStop}%
\bibitem [{\citenamefont {Martirosyan}\ \emph
  {et~al.}(2024{\natexlab{a}})\citenamefont {Martirosyan}, \citenamefont {Ho},
  \citenamefont {Etrych}, \citenamefont {Zhang}, \citenamefont {Cao},
  \citenamefont {Hadzibabic},\ and\ \citenamefont
  {Eigen}}]{Martirosyan:2023mml}%
  \BibitemOpen
  \bibfield  {author} {\bibinfo {author} {\bibfnamefont {G.}~\bibnamefont
  {Martirosyan}}, \bibinfo {author} {\bibfnamefont {C.~J.}\ \bibnamefont {Ho}},
  \bibinfo {author} {\bibfnamefont {J.}~\bibnamefont {Etrych}}, \bibinfo
  {author} {\bibfnamefont {Y.}~\bibnamefont {Zhang}}, \bibinfo {author}
  {\bibfnamefont {A.}~\bibnamefont {Cao}}, \bibinfo {author} {\bibfnamefont
  {Z.}~\bibnamefont {Hadzibabic}}, \ and\ \bibinfo {author} {\bibfnamefont
  {C.}~\bibnamefont {Eigen}},\ }\href {\doibase 10.1103/PhysRevLett.132.113401}
  {\bibfield  {journal} {\bibinfo  {journal} {Phys. Rev. Lett.}\ }\textbf
  {\bibinfo {volume} {132}},\ \bibinfo {pages} {113401} (\bibinfo {year}
  {2024}{\natexlab{a}})},\ \Eprint {http://arxiv.org/abs/2304.06697}
  {arXiv:2304.06697 [cond-mat.quant-gas]} \BibitemShut {NoStop}%
\bibitem [{\citenamefont {Gazo}\ \emph {et~al.}(2023)\citenamefont {Gazo},
  \citenamefont {Karailiev}, \citenamefont {Satoor}, \citenamefont {Eigen},
  \citenamefont {Ga\l{}ka},\ and\ \citenamefont {Hadzibabic}}]{Gazo:2023exc}%
  \BibitemOpen
  \bibfield  {author} {\bibinfo {author} {\bibfnamefont {M.}~\bibnamefont
  {Gazo}}, \bibinfo {author} {\bibfnamefont {A.}~\bibnamefont {Karailiev}},
  \bibinfo {author} {\bibfnamefont {T.}~\bibnamefont {Satoor}}, \bibinfo
  {author} {\bibfnamefont {C.}~\bibnamefont {Eigen}}, \bibinfo {author}
  {\bibfnamefont {M.}~\bibnamefont {Ga\l{}ka}}, \ and\ \bibinfo {author}
  {\bibfnamefont {Z.} \bibnamefont {Hadzibabic}},\ }\href@noop {} {\  (\bibinfo
  {year} {2023})},\ \Eprint {http://arxiv.org/abs/2312.09248} {arXiv:2312.09248
  [cond-mat.quant-gas]} \BibitemShut {NoStop}%
\bibitem [{\citenamefont {Moreno-Armijos}\ \emph {et~al.}(2025)\citenamefont
  {Moreno-Armijos}, \citenamefont {Fritsch}, \citenamefont
  {Garc\'{\i}a-Orozco}, \citenamefont {Sab}, \citenamefont {Telles},
  \citenamefont {Zhu}, \citenamefont {Madeira}, \citenamefont {Nazarenko},
  \citenamefont {Yukalov},\ and\ \citenamefont
  {Bagnato}}]{MorenoArmijos2024a.PhysRevLett.134.023401}%
  \BibitemOpen
  \bibfield  {author} {\bibinfo {author} {\bibfnamefont {M.~A.}\ \bibnamefont
  {Moreno-Armijos}}, \bibinfo {author} {\bibfnamefont {A.~R.}\ \bibnamefont
  {Fritsch}}, \bibinfo {author} {\bibfnamefont {A.~D.}\ \bibnamefont
  {Garc\'{\i}a-Orozco}}, \bibinfo {author} {\bibfnamefont {S.}~\bibnamefont
  {Sab}}, \bibinfo {author} {\bibfnamefont {G.}~\bibnamefont {Telles}},
  \bibinfo {author} {\bibfnamefont {Y.}~\bibnamefont {Zhu}}, \bibinfo {author}
  {\bibfnamefont {L.}~\bibnamefont {Madeira}}, \bibinfo {author} {\bibfnamefont
  {S.}~\bibnamefont {Nazarenko}}, \bibinfo {author} {\bibfnamefont {V.~I.}\
  \bibnamefont {Yukalov}}, \ and\ \bibinfo {author} {\bibfnamefont {V.~S.}\
  \bibnamefont {Bagnato}},\ }\href {\doibase 10.1103/PhysRevLett.134.023401}
  {\bibfield  {journal} {\bibinfo  {journal} {Phys. Rev. Lett.}\ }\textbf
  {\bibinfo {volume} {134}},\ \bibinfo {pages} {023401} (\bibinfo {year}
  {2025})},\ \Eprint {http://arxiv.org/abs/2407.11237} {arXiv:2407.11237
  [cond-mat.quant-gas]} \BibitemShut {NoStop}%
\bibitem [{\citenamefont {Martirosyan}\ \emph
  {et~al.}(2024{\natexlab{b}})\citenamefont {Martirosyan}, \citenamefont
  {Gazo}, \citenamefont {Etrych}, \citenamefont {Fischer}, \citenamefont
  {Morris}, \citenamefont {Ho}, \citenamefont {Eigen},\ and\ \citenamefont
  {Hadzibabic}}]{Martirosyan:2024rxm}%
  \BibitemOpen
  \bibfield  {author} {\bibinfo {author} {\bibfnamefont {G.}~\bibnamefont
  {Martirosyan}}, \bibinfo {author} {\bibfnamefont {M.}~\bibnamefont {Gazo}},
  \bibinfo {author} {\bibfnamefont {J.}~\bibnamefont {Etrych}}, \bibinfo
  {author} {\bibfnamefont {S.~M.}\ \bibnamefont {Fischer}}, \bibinfo {author}
  {\bibfnamefont {S.~J.}\ \bibnamefont {Morris}}, \bibinfo {author}
  {\bibfnamefont {C.~J.}\ \bibnamefont {Ho}}, \bibinfo {author} {\bibfnamefont
  {C.}~\bibnamefont {Eigen}}, \ and\ \bibinfo {author} {\bibfnamefont
  {Z.}~\bibnamefont {Hadzibabic}},\ }\href@noop {} {\  (\bibinfo {year}
  {2024}{\natexlab{b}})},\ \Eprint {http://arxiv.org/abs/2410.08204}
  {arXiv: 2410. 08204 [cond-mat.quant-gas]} \BibitemShut {NoStop}%
\bibitem [{\citenamefont {Berges}\ \emph {et~al.}(2004)\citenamefont {Berges},
  \citenamefont {Borsanyi},\ and\ \citenamefont {Wetterich}}]{Berges:2004ce}%
  \BibitemOpen
  \bibfield  {author} {\bibinfo {author} {\bibfnamefont {J.}~\bibnamefont
  {Berges}}, \bibinfo {author} {\bibfnamefont {S.}~\bibnamefont {Borsanyi}}, \
  and\ \bibinfo {author} {\bibfnamefont {C.}~\bibnamefont {Wetterich}},\ }\href
  {\doibase 10.1103/PhysRevLett.93.142002} {\bibfield  {journal} {\bibinfo
  {journal} {Phys. Rev. Lett.}\ }\textbf {\bibinfo {volume} {93}},\ \bibinfo
  {pages} {142002} (\bibinfo {year} {2004})},\ \Eprint
  {http://arxiv.org/abs/hep-ph/0403234} {arXiv:hep-ph/0403234 [hep-ph]}
  \BibitemShut {NoStop}%
\bibitem [{\citenamefont {Kodama}\ \emph {et~al.}(2005)\citenamefont {Kodama},
  \citenamefont {Elze}, \citenamefont {Aguiar},\ and\ \citenamefont
  {Koide}}]{Kodama:2004dk}%
  \BibitemOpen
  \bibfield  {author} {\bibinfo {author} {\bibfnamefont {T.}~\bibnamefont
  {Kodama}}, \bibinfo {author} {\bibfnamefont {H.~T.}\ \bibnamefont {Elze}},
  \bibinfo {author} {\bibfnamefont {C.~E.}\ \bibnamefont {Aguiar}}, \ and\
  \bibinfo {author} {\bibfnamefont {T.}~\bibnamefont {Koide}},\ }\href
  {\doibase 10.1209/epl/i2004-10506-9} {\bibfield  {journal} {\bibinfo
  {journal} {Europhys. Lett.}\ }\textbf {\bibinfo {volume} {70}},\ \bibinfo
  {pages} {439} (\bibinfo {year} {2005})},\ \Eprint
  {http://arxiv.org/abs/cond-mat/0406732} {arXiv:cond-mat/0406732 [cond-mat]}
  \BibitemShut {NoStop}%
\bibitem [{\citenamefont {Barnett}\ \emph {et~al.}(2011)\citenamefont
  {Barnett}, \citenamefont {Polkovnikov},\ and\ \citenamefont
  {Vengalattore}}]{Barnett2011a}%
  \BibitemOpen
  \bibfield  {author} {\bibinfo {author} {\bibfnamefont {R.}~\bibnamefont
  {Barnett}}, \bibinfo {author} {\bibfnamefont {A.}~\bibnamefont
  {Polkovnikov}}, \ and\ \bibinfo {author} {\bibfnamefont {M.}~\bibnamefont
  {Vengalattore}},\ }\href {\doibase 10.1103/PhysRevA.84.023606} {\bibfield
  {journal} {\bibinfo  {journal} {Phys. Rev. A}\ }\textbf {\bibinfo {volume}
  {84}},\ \bibinfo {pages} {023606} (\bibinfo {year} {2011})}\BibitemShut
  {NoStop}%
\bibitem [{\citenamefont {Marcuzzi}\ \emph {et~al.}(2013)\citenamefont
  {Marcuzzi}, \citenamefont {Marino}, \citenamefont {Gambassi},\ and\
  \citenamefont {Silva}}]{Marcuzzi2013a.PhysRevLett.111.197203}%
  \BibitemOpen
  \bibfield  {author} {\bibinfo {author} {\bibfnamefont {M.}~\bibnamefont
  {Marcuzzi}}, \bibinfo {author} {\bibfnamefont {J.}~\bibnamefont {Marino}},
  \bibinfo {author} {\bibfnamefont {A.}~\bibnamefont {Gambassi}}, \ and\
  \bibinfo {author} {\bibfnamefont {A.}~\bibnamefont {Silva}},\ }\href
  {\doibase 10.1103/PhysRevLett.111.197203} {\bibfield  {journal} {\bibinfo
  {journal} {Phys. Rev. Lett.}\ }\textbf {\bibinfo {volume} {111}},\ \bibinfo
  {pages} {197203} (\bibinfo {year} {2013})}\BibitemShut {NoStop}%
\bibitem [{\citenamefont {{Langen}}\ \emph {et~al.}(2013)\citenamefont
  {{Langen}}, \citenamefont {{Gring}}, \citenamefont {{Kuhnert}}, \citenamefont
  {{Rauer}}, \citenamefont {{Geiger}}, \citenamefont {{Smith}}, \citenamefont
  {{Mazets}},\ and\ \citenamefont {{Schmiedmayer}}}]{Langen2013a.EPJST.217}%
  \BibitemOpen
  \bibfield  {author} {\bibinfo {author} {\bibfnamefont {T.}~\bibnamefont
  {{Langen}}}, \bibinfo {author} {\bibfnamefont {M.}~\bibnamefont {{Gring}}},
  \bibinfo {author} {\bibfnamefont {M.}~\bibnamefont {{Kuhnert}}}, \bibinfo
  {author} {\bibfnamefont {B.}~\bibnamefont {{Rauer}}}, \bibinfo {author}
  {\bibfnamefont {R.}~\bibnamefont {{Geiger}}}, \bibinfo {author}
  {\bibfnamefont {D.~A.}\ \bibnamefont {{Smith}}}, \bibinfo {author}
  {\bibfnamefont {I.~E.}\ \bibnamefont {{Mazets}}}, \ and\ \bibinfo {author}
  {\bibfnamefont {J.}~\bibnamefont {{Schmiedmayer}}},\ }\href 
  {\doibase 10.1140/epjst/e2013-01752-0} {\bibfield  {journal} {\bibinfo  {journal} {Eur.
  Phys. J. ST}\ }\textbf {\bibinfo {volume} {217}},\ \bibinfo {pages} {43}
  (\bibinfo {year} {2013})},\ \Eprint {http://arxiv.org/abs/1211.0016}
  {arXiv:1211.0016 [cond-mat.quant-gas]} \BibitemShut {NoStop}%
\bibitem [{\citenamefont {Nessi}\ \emph {et~al.}(2014)\citenamefont {Nessi},
  \citenamefont {Iucci},\ and\ \citenamefont
  {Cazalilla}}]{Nessi2014a.PhysRevLett.113.210402}%
  \BibitemOpen
  \bibfield  {author} {\bibinfo {author} {\bibfnamefont {N.}~\bibnamefont
  {Nessi}}, \bibinfo {author} {\bibfnamefont {A.}~\bibnamefont {Iucci}}, \ and\
  \bibinfo {author} {\bibfnamefont {M.~A.}\ \bibnamefont {Cazalilla}},\ }\href
  {\doibase 10.1103/PhysRevLett.113.210402} {\bibfield  {journal} {\bibinfo
  {journal} {Phys. Rev. Lett.}\ }\textbf {\bibinfo {volume} {113}},\ \bibinfo
  {pages} {210402} (\bibinfo {year} {2014})}\BibitemShut {NoStop}%
\bibitem [{\citenamefont {Gagel}\ \emph {et~al.}(2014)\citenamefont {Gagel},
  \citenamefont {Orth},\ and\ \citenamefont
  {Schmalian}}]{Gagel2014a.PhysRevLett.113.220401}%
  \BibitemOpen
  \bibfield  {author} {\bibinfo {author} {\bibfnamefont {P.}~\bibnamefont
  {Gagel}}, \bibinfo {author} {\bibfnamefont {P.~P.}\ \bibnamefont {Orth}}, \
  and\ \bibinfo {author} {\bibfnamefont {J.}~\bibnamefont {Schmalian}},\ }\href
  {\doibase 10.1103/PhysRevLett.113.220401} {\bibfield  {journal} {\bibinfo
  {journal} {Phys. Rev. Lett.}\ }\textbf {\bibinfo {volume} {113}},\ \bibinfo
  {pages} {220401} (\bibinfo {year} {2014})}\BibitemShut {NoStop}%
\bibitem [{\citenamefont {Bertini}\ \emph {et~al.}(2015)\citenamefont
  {Bertini}, \citenamefont {Essler}, \citenamefont {Groha},\ and\ \citenamefont
  {Robinson}}]{Bertini2015a.PhysRevLett.115.180601}%
  \BibitemOpen
  \bibfield  {author} {\bibinfo {author} {\bibfnamefont {B.}~\bibnamefont
  {Bertini}}, \bibinfo {author} {\bibfnamefont {F.~H.~L.}\ \bibnamefont
  {Essler}}, \bibinfo {author} {\bibfnamefont {S.}~\bibnamefont {Groha}}, \
  and\ \bibinfo {author} {\bibfnamefont {N.~J.}\ \bibnamefont {Robinson}},\
  }\href {\doibase 10.1103/PhysRevLett.115.180601} {\bibfield  {journal}
  {\bibinfo  {journal} {Phys. Rev. Lett.}\ }\textbf {\bibinfo {volume} {115}},\
  \bibinfo {pages} {180601} (\bibinfo {year} {2015})}\BibitemShut {NoStop}%
\bibitem [{\citenamefont {Babadi}\ \emph {et~al.}(2015)\citenamefont {Babadi},
  \citenamefont {Demler},\ and\ \citenamefont
  {Knap}}]{Babadi2015a.PhysRevX.5.041005}%
  \BibitemOpen
  \bibfield  {author} {\bibinfo {author} {\bibfnamefont {M.}~\bibnamefont
  {Babadi}}, \bibinfo {author} {\bibfnamefont {E.}~\bibnamefont {Demler}}, \
  and\ \bibinfo {author} {\bibfnamefont {M.}~\bibnamefont {Knap}},\ }\href
  {\doibase 10.1103/PhysRevX.5.041005} {\bibfield  {journal} {\bibinfo
  {journal} {Phys. Rev. X}\ }\textbf {\bibinfo {volume} {5}},\ \bibinfo {pages}
  {041005} (\bibinfo {year} {2015})}\BibitemShut {NoStop}%
\bibitem [{\citenamefont {Buchhold}\ \emph {et~al.}(2016)\citenamefont
  {Buchhold}, \citenamefont {Heyl},\ and\ \citenamefont
  {Diehl}}]{Buchhold2015a.PhysRevA.94.013601}%
  \BibitemOpen
  \bibfield  {author} {\bibinfo {author} {\bibfnamefont {M.}~\bibnamefont
  {Buchhold}}, \bibinfo {author} {\bibfnamefont {M.}~\bibnamefont {Heyl}}, \
  and\ \bibinfo {author} {\bibfnamefont {S.}~\bibnamefont {Diehl}},\ }\href
  {\doibase 10.1103/PhysRevA.94.013601} {\bibfield  {journal} {\bibinfo
  {journal} {Phys. Rev. A}\ }\textbf {\bibinfo {volume} {94}},\ \bibinfo
  {pages} {013601} (\bibinfo {year} {2016})}\BibitemShut {NoStop}%
\bibitem [{\citenamefont {Berges}\ and\ \citenamefont
  {Hoffmeister}(2009)}]{Berges:2008sr}%
  \BibitemOpen
  \bibfield  {author} {\bibinfo {author} {\bibfnamefont {J.}~\bibnamefont
  {Berges}}\ and\ \bibinfo {author} {\bibfnamefont {G.}~\bibnamefont
  {Hoffmeister}},\ }\href {\doibase 10.1016/j.nuclphysb.2008.12.017} {\bibfield
   {journal} {\bibinfo  {journal} {Nucl. Phys.}\ }\textbf {\bibinfo {volume}
  {B813}},\ \bibinfo {pages} {383} (\bibinfo {year} {2009})},\ \Eprint
  {http://arxiv.org/abs/0809.5208} {arXiv:0809.5208 [hep-th]} \BibitemShut
  {NoStop}%
\bibitem [{\citenamefont {{Nowak}}\ \emph {et~al.}(2014)\citenamefont
  {{Nowak}}, \citenamefont {{Schole}},\ and\ \citenamefont
  {{Gasenzer}}}]{Nowak:2012gd}%
  \BibitemOpen
  \bibfield  {author} {\bibinfo {author} {\bibfnamefont {B.}~\bibnamefont
  {{Nowak}}}, \bibinfo {author} {\bibfnamefont {J.}~\bibnamefont {{Schole}}}, \
  and\ \bibinfo {author} {\bibfnamefont {T.}~\bibnamefont {{Gasenzer}}},\
  }\href {\doibase 10.1088/1367-2630/16/9/093052} {\bibfield  {journal}
  {\bibinfo  {journal} {New J. Phys.}\ }\textbf {\bibinfo {volume} {16}},\
  \bibinfo {pages} {093052} (\bibinfo {year} {2014})},\ \Eprint
  {http://arxiv.org/abs/1206.3181v2} {arXiv:1206.3181v2 [cond-mat.quant-gas]}
  \BibitemShut {NoStop}%
\bibitem [{\citenamefont {{Hofmann}}\ \emph {et~al.}(2014)\citenamefont
  {{Hofmann}}, \citenamefont {{Natu}},\ and\ \citenamefont {{Das
  Sarma}}}]{Hofmann2014a}%
  \BibitemOpen
  \bibfield  {author} {\bibinfo {author} {\bibfnamefont {J.}~\bibnamefont
  {{Hofmann}}}, \bibinfo {author} {\bibfnamefont {S.~S.}\ \bibnamefont
  {{Natu}}}, \ and\ \bibinfo {author} {\bibfnamefont {S.}~\bibnamefont {{Das
  Sarma}}},\ }\href {\doibase 10.1103/PhysRevLett.113.095702} {\bibfield
  {journal} {\bibinfo  {journal} {Phys. Rev. Lett.}\ }\textbf {\bibinfo
  {volume} {113}},\ \bibinfo {eid} {095702} (\bibinfo {year} {2014})},\ \Eprint
  {http://arxiv.org/abs/1403.1284} {arXiv:1403.1284 [cond-mat.quant-gas]}
  \BibitemShut {NoStop}%
\bibitem [{\citenamefont {Maraga}\ \emph {et~al.}(2015)\citenamefont {Maraga},
  \citenamefont {Chiocchetta}, \citenamefont {Mitra},\ and\ \citenamefont
  {Gambassi}}]{Maraga2015a.PhysRevE.92.042151}%
  \BibitemOpen
  \bibfield  {author} {\bibinfo {author} {\bibfnamefont {A.}~\bibnamefont
  {Maraga}}, \bibinfo {author} {\bibfnamefont {A.}~\bibnamefont {Chiocchetta}},
  \bibinfo {author} {\bibfnamefont {A.}~\bibnamefont {Mitra}}, \ and\ \bibinfo
  {author} {\bibfnamefont {A.}~\bibnamefont {Gambassi}},\ }\href 
  {\doibase 10.1103/PhysRevE.92.042151} {\bibfield  {journal} {\bibinfo  {journal} {Phys.
  Rev. E}\ }\textbf {\bibinfo {volume} {92}},\ \bibinfo {pages} {042151}
  (\bibinfo {year} {2015})}\BibitemShut {NoStop}%
\bibitem [{\citenamefont {Williamson}\ and\ \citenamefont
  {Blakie}(2016{\natexlab{a}})}]{Williamson2016a.PhysRevLett.116.025301}%
  \BibitemOpen
  \bibfield  {author} {\bibinfo {author} {\bibfnamefont {L.~A.}\ \bibnamefont
  {Williamson}}\ and\ \bibinfo {author} {\bibfnamefont {P.~B.}\ \bibnamefont
  {Blakie}},\ }\href {\doibase 10.1103/PhysRevLett.116.025301} {\bibfield
  {journal} {\bibinfo  {journal} {Phys. Rev. Lett.}\ }\textbf {\bibinfo
  {volume} {116}},\ \bibinfo {pages} {025301} (\bibinfo {year}
  {2016}{\natexlab{a}})}\BibitemShut {NoStop}%
\bibitem [{\citenamefont {Williamson}\ and\ \citenamefont
  {Blakie}(2016{\natexlab{b}})}]{Williamson2016a.PhysRevA.94.023608}%
  \BibitemOpen
  \bibfield  {author} {\bibinfo {author} {\bibfnamefont {L.~A.}\ \bibnamefont
  {Williamson}}\ and\ \bibinfo {author} {\bibfnamefont {P.~B.}\ \bibnamefont
  {Blakie}},\ }\href {\doibase 10.1103/PhysRevA.94.023608} {\bibfield
  {journal} {\bibinfo  {journal} {Phys. Rev. A}\ }\textbf {\bibinfo {volume}
  {94}},\ \bibinfo {pages} {023608} (\bibinfo {year}
  {2016}{\natexlab{b}})}\BibitemShut {NoStop}%
\bibitem [{\citenamefont {Bourges}\ and\ \citenamefont
  {Blakie}(2017)}]{Bourges2016a.arXiv161108922B.PhysRevA.95.023616}%
  \BibitemOpen
  \bibfield  {author} {\bibinfo {author} {\bibfnamefont {A.}~\bibnamefont
  {Bourges}}\ and\ \bibinfo {author} {\bibfnamefont {P.~B.}\ \bibnamefont
  {Blakie}},\ }\href {\doibase 10.1103/PhysRevA.95.023616} {\bibfield
  {journal} {\bibinfo  {journal} {Phys. Rev. A}\ }\textbf {\bibinfo {volume}
  {95}},\ \bibinfo {pages} {023616} (\bibinfo {year} {2017})}\BibitemShut
  {NoStop}%
\bibitem [{\citenamefont {Chiocchetta}\ \emph {et~al.}(2016)\citenamefont
  {Chiocchetta}, \citenamefont {Gambassi}, \citenamefont {Diehl},\ and\
  \citenamefont {Marino}}]{Chiocchetta:2016waa.PhysRevB.94.174301}%
  \BibitemOpen
  \bibfield  {author} {\bibinfo {author} {\bibfnamefont {A.}~\bibnamefont
  {Chiocchetta}}, \bibinfo {author} {\bibfnamefont {A.}~\bibnamefont
  {Gambassi}}, \bibinfo {author} {\bibfnamefont {S.}~\bibnamefont {Diehl}}, \
  and\ \bibinfo {author} {\bibfnamefont {J.}~\bibnamefont {Marino}},\ }\href
  {\doibase 10.1103/PhysRevB.94.174301} {\bibfield  {journal} {\bibinfo
  {journal} {Phys. Rev. B}\ }\textbf {\bibinfo {volume} {94}},\ \bibinfo
  {pages} {174301} (\bibinfo {year} {2016})}\BibitemShut {NoStop}%
\bibitem [{\citenamefont {Karl}\ and\ \citenamefont
  {Gasenzer}(2017)}]{Karl2017b.NJP19.093014}%
  \BibitemOpen
  \bibfield  {author} {\bibinfo {author} {\bibfnamefont {M.}~\bibnamefont
  {Karl}}\ and\ \bibinfo {author} {\bibfnamefont {T.}~\bibnamefont
  {Gasenzer}},\ }\href {\doibase 10.1088/1367-2630/aa7eeb} {\bibfield
  {journal} {\bibinfo  {journal} {New J. Phys.}\ }\textbf {\bibinfo {volume}
  {19}},\ \bibinfo {pages} {093014} (\bibinfo {year} {2017})},\ \Eprint
  {http://arxiv.org/abs/1611.01163} {arXiv:1611.01163 [cond-mat.quant-gas]}
  \BibitemShut {NoStop}%
\bibitem [{\citenamefont {Schachner}\ \emph {et~al.}(2017)\citenamefont
  {Schachner}, \citenamefont {Pi{\~n}eiro~Orioli},\ and\ \citenamefont
  {Berges}}]{Schachner:2016frd}%
  \BibitemOpen
  \bibfield  {author} {\bibinfo {author} {\bibfnamefont {A.}~\bibnamefont
  {Schachner}}, \bibinfo {author} {\bibfnamefont {A.}~\bibnamefont
  {Pi{\~n}eiro~Orioli}}, \ and\ \bibinfo {author} {\bibfnamefont
  {J.}~\bibnamefont {Berges}},\ }\href {\doibase 10.1103/PhysRevA.95.053605}
  {\bibfield  {journal} {\bibinfo  {journal} {Phys. Rev. A}\ }\textbf {\bibinfo
  {volume} {95}},\ \bibinfo {pages} {053605} (\bibinfo {year} {2017})},\
  \Eprint {http://arxiv.org/abs/1612.03038} {arXiv:1612.03038
  [cond-mat.quant-gas]} \BibitemShut {NoStop}%
\bibitem [{\citenamefont {Walz}\ \emph {et~al.}(2018)\citenamefont {Walz},
  \citenamefont {Boguslavski},\ and\ \citenamefont
  {Berges}}]{Walz:2017ffj.PhysRevD.97.116011}%
  \BibitemOpen
  \bibfield  {author} {\bibinfo {author} {\bibfnamefont {R.}~\bibnamefont
  {Walz}}, \bibinfo {author} {\bibfnamefont {K.}~\bibnamefont {Boguslavski}}, \
  and\ \bibinfo {author} {\bibfnamefont {J.}~\bibnamefont {Berges}},\ }\href
  {\doibase 10.1103/PhysRevD.97.116011} {\bibfield  {journal} {\bibinfo
  {journal} {Phys. Rev. D}\ }\textbf {\bibinfo {volume} {97}},\ \bibinfo
  {pages} {116011} (\bibinfo {year} {2018})}\BibitemShut {NoStop}%
\bibitem [{\citenamefont {Schmied}\ \emph
  {et~al.}(2019{\natexlab{a}})\citenamefont {Schmied}, \citenamefont
  {Mikheev},\ and\ \citenamefont
  {Gasenzer}}]{Schmied:2018upn.PhysRevLett.122.170404}%
  \BibitemOpen
  \bibfield  {author} {\bibinfo {author} {\bibfnamefont {C.-M.}\ \bibnamefont
  {Schmied}}, \bibinfo {author} {\bibfnamefont {A.~N.}\ \bibnamefont
  {Mikheev}}, \ and\ \bibinfo {author} {\bibfnamefont {T.}~\bibnamefont
  {Gasenzer}},\ }\href {\doibase 10.1103/PhysRevLett.122.170404} {\bibfield
  {journal} {\bibinfo  {journal} {Phys. Rev. Lett.} \ }\textbf {\bibinfo
  {volume} {122}},\ \bibinfo {pages} {170404} (\bibinfo {year}
  {2019}{\natexlab{a}})},\ \Eprint {http://arxiv.org/abs/1807.07514}
  {arXiv:1807.07514 [cond-mat.quant-gas]} \BibitemShut {NoStop}%
\bibitem [{\citenamefont {Mazeliauskas}\ and\ \citenamefont
  {Berges}(2019)}]{Mazeliauskas:2018yef}%
  \BibitemOpen
  \bibfield  {author} {\bibinfo {author} {\bibfnamefont {A.}~\bibnamefont
  {Mazeliauskas}}\ and\ \bibinfo {author} {\bibfnamefont {J.}~\bibnamefont
  {Berges}},\ }\href {\doibase 10.1103/PhysRevLett.122.122301} {\bibfield
  {journal} {\bibinfo  {journal} {Phys. Rev. Lett.}\ }\textbf {\bibinfo
  {volume} {122}},\ \bibinfo {pages} {122301} (\bibinfo {year} {2019})},\
  \Eprint {http://arxiv.org/abs/1810.10554} {arXiv:1810.10554 [hep-ph]}
  \BibitemShut {NoStop}%
\bibitem [{\citenamefont {Schmied}\ \emph
  {et~al.}(2019{\natexlab{c}})\citenamefont {Schmied}, \citenamefont
  {Pr\"ufer}, \citenamefont {Oberthaler},\ and\ \citenamefont
  {Gasenzer}}]{Schmied:2018osf.PhysRevA.99.033611}%
  \BibitemOpen
  \bibfield  {author} {\bibinfo {author} {\bibfnamefont {C.-M.}\ \bibnamefont
  {Schmied}}, \bibinfo {author} {\bibfnamefont {M.}~\bibnamefont {Pr\"ufer}},
  \bibinfo {author} {\bibfnamefont {M.~K.}\ \bibnamefont {Oberthaler}}, \ and\
  \bibinfo {author} {\bibfnamefont {T.}~\bibnamefont {Gasenzer}},\ }\href
  {\doibase 10.1103/PhysRevA.99.033611} {\bibfield  {journal} {\bibinfo
  {journal} {Phys. Rev. A}\ }\textbf {\bibinfo {volume} {99}},\ \bibinfo
  {pages} {033611} (\bibinfo {year} {2019}{\natexlab{c}})}\BibitemShut
  {NoStop}%
\bibitem [{\citenamefont {{Williamson}}\ and\ \citenamefont
  {{Blakie}}(2019)}]{Williamson2019a.ScPP7.29}%
  \BibitemOpen
  \bibfield  {author} {\bibinfo {author} {\bibfnamefont {L.~A.}\ \bibnamefont
  {{Williamson}}}\ and\ \bibinfo {author} {\bibfnamefont {P.~B.}\ \bibnamefont
  {{Blakie}}},\ }\href {\doibase 10.21468/SciPostPhys.7.3.029} {\bibfield
  {journal} {\bibinfo  {journal} {SciPost Physics}\ }\textbf {\bibinfo {volume}
  {7}},\ \bibinfo {eid} {029} (\bibinfo {year} {2019})},\ \Eprint
  {http://arxiv.org/abs/1902.10792} {arXiv:1902.10792 [cond-mat.quant-gas]}
  \BibitemShut {NoStop}%
\bibitem [{\citenamefont {Schmied}\ \emph
  {et~al.}(2019{\natexlab{d}})\citenamefont {Schmied}, \citenamefont
  {Gasenzer},\ and\ \citenamefont {Blakie}}]{Schmied:2019abm}%
  \BibitemOpen
  \bibfield  {author} {\bibinfo {author} {\bibfnamefont {C.~M.}\ \bibnamefont
  {Schmied}}, \bibinfo {author} {\bibfnamefont {T.}~\bibnamefont {Gasenzer}}, \
  and\ \bibinfo {author} {\bibfnamefont {P.~B.}\ \bibnamefont {Blakie}},\
  }\href {\doibase 10.1103/PhysRevA.100.033603} {\bibfield  {journal} {\bibinfo
   {journal} {Phys. Rev. A}\ }\textbf {\bibinfo {volume} {100}},\ \bibinfo
  {pages} {033603} (\bibinfo {year} {2019}{\natexlab{d}})},\ \Eprint
  {http://arxiv.org/abs/1904.13222} {arXiv:1904.13222 [cond-mat.quant-gas]}
  \BibitemShut {NoStop}%
\bibitem [{\citenamefont {Spitz}\ \emph {et~al.}(2021)\citenamefont {Spitz},
  \citenamefont {Berges}, \citenamefont {Oberthaler},\ and\ \citenamefont
  {Wienhard}}]{Spitz2021a.SciPostPhys11.3.060}%
  \BibitemOpen
  \bibfield  {author} {\bibinfo {author} {\bibfnamefont {D.}~\bibnamefont
  {Spitz}}, \bibinfo {author} {\bibfnamefont {J.}~\bibnamefont {Berges}},
  \bibinfo {author} {\bibfnamefont {M.}~\bibnamefont {Oberthaler}}, \ and\
  \bibinfo {author} {\bibfnamefont {A.}~\bibnamefont {Wienhard}},\ }\href
  {\doibase 10.21468/scipostphys.11.3.060} {\bibfield  {journal} {\bibinfo
  {journal} {SciPost Physics}\ }\textbf {\bibinfo {volume} {11}},\ \bibinfo
  {pages} {060} (\bibinfo {year} {2021})}\BibitemShut {NoStop}%
\bibitem [{\citenamefont {Gao}\ \emph {et~al.}(2020)\citenamefont {Gao},
  \citenamefont {Sun}, \citenamefont {Zhang},\ and\ \citenamefont
  {Zhai}}]{Gao2020a.PhysRevLett.124.040403}%
  \BibitemOpen
  \bibfield  {author} {\bibinfo {author} {\bibfnamefont {C.}~\bibnamefont
  {Gao}}, \bibinfo {author} {\bibfnamefont {M.}~\bibnamefont {Sun}}, \bibinfo
  {author} {\bibfnamefont {P.}~\bibnamefont {Zhang}}, \ and\ \bibinfo {author}
  {\bibfnamefont {H.}~\bibnamefont {Zhai}},\ }\href 
  {\doibase 10.1103/PhysRevLett.124.040403} {\bibfield  {journal} {\bibinfo  {journal}
  {Phys. Rev. Lett.}\ }\textbf {\bibinfo {volume} {124}},\ \bibinfo {pages}
  {040403} (\bibinfo {year} {2020})}\BibitemShut {NoStop}%
\bibitem [{\citenamefont {{Wheeler}}\ \emph {et~al.}(2021)\citenamefont
  {{Wheeler}}, \citenamefont {{Salman}},\ and\ \citenamefont
  {{Borgh}}}]{Wheeler2021a.EPL135.30004}%
  \BibitemOpen
  \bibfield  {author} {\bibinfo {author} {\bibfnamefont {M.~T.}\ \bibnamefont
  {{Wheeler}}}, \bibinfo {author} {\bibfnamefont {H.}~\bibnamefont {{Salman}}},
  \ and\ \bibinfo {author} {\bibfnamefont {M.~O.}\ \bibnamefont {{Borgh}}},\
  }\href {\doibase 10.1209/0295-5075/ac2c53} {\bibfield  {journal} {\bibinfo
  {journal} {EPL}\ }\textbf {\bibinfo {volume} {135}},\ \bibinfo {eid} {30004}
  (\bibinfo {year} {2021})},\ \Eprint {http://arxiv.org/abs/2110.02671}
  {arXiv:2110.02671 [cond-mat.quant-gas]} \BibitemShut {NoStop}%
\bibitem [{\citenamefont {Gresista}\ \emph {et~al.}(2022)\citenamefont
  {Gresista}, \citenamefont {Zache},\ and\ \citenamefont
  {Berges}}]{Gresista:2021qqa}%
  \BibitemOpen
  \bibfield  {author} {\bibinfo {author} {\bibfnamefont {L.}~\bibnamefont
  {Gresista}}, \bibinfo {author} {\bibfnamefont {T.~V.}\ \bibnamefont {Zache}},
  \ and\ \bibinfo {author} {\bibfnamefont {J.}~\bibnamefont {Berges}},\ }\href
  {\doibase 10.1103/PhysRevA.105.013320} {\bibfield  {journal} {\bibinfo
  {journal} {Phys. Rev. A}\ }\textbf {\bibinfo {volume} {105}},\ \bibinfo
  {pages} {013320} (\bibinfo {year} {2022})},\ \Eprint
  {http://arxiv.org/abs/2107.11749} {arXiv:2107.11749 [cond-mat.quant-gas]}
  \BibitemShut {NoStop}%
\bibitem [{\citenamefont {{Rodriguez-Nieva}}\ \emph {et~al.}(2022)\citenamefont
  {{Rodriguez-Nieva}}, \citenamefont {{Pi{\~n}eiro Orioli}},\ and\
  \citenamefont {{Marino}}}]{RodriguezNieva2021a.arXiv210600023R}%
  \BibitemOpen
  \bibfield  {author} {\bibinfo {author} {\bibfnamefont {J.~F.}\ \bibnamefont
  {{Rodriguez-Nieva}}}, \bibinfo {author} {\bibfnamefont {A.}~\bibnamefont
  {{Pi{\~n}eiro Orioli}}}, \ and\ \bibinfo {author} {\bibfnamefont
  {J.}~\bibnamefont {{Marino}}},\ }\href {\doibase 10.1073/pnas.2122599119}
  {\bibfield  {journal} {\bibinfo  {journal} {PNAS}\ }\textbf {\bibinfo
  {volume} {119}},\ \bibinfo {pages} {e2122599119} (\bibinfo {year} {2022})},\
  \Eprint {http://arxiv.org/abs/2106.00023} {arXiv:2106.00023
  [cond-mat.stat-mech]} \BibitemShut {NoStop}%
\bibitem [{\citenamefont {Preis}\ \emph {et~al.}(2023)\citenamefont {Preis},
  \citenamefont {Heller},\ and\ \citenamefont
  {Berges}}]{Preis2023a.PhysRevLett.130.031602}%
  \BibitemOpen
  \bibfield  {author} {\bibinfo {author} {\bibfnamefont {T.}~\bibnamefont
  {Preis}}, \bibinfo {author} {\bibfnamefont {M.~P.}\ \bibnamefont {Heller}}, \
  and\ \bibinfo {author} {\bibfnamefont {J.}~\bibnamefont {Berges}},\ }\href
  {\doibase 10.1103/PhysRevLett.130.031602} {\bibfield  {journal} {\bibinfo
  {journal} {Phys. Rev. Lett.}\ }\textbf {\bibinfo {volume} {130}}, \ \bibinfo
  {pages} {031602} (\bibinfo {year} {2023})}\BibitemShut {NoStop}%
\bibitem [{\citenamefont {Liu}\ \emph {et~al.}(2023)\citenamefont {Liu},
  \citenamefont {Proukakis},\ and\ \citenamefont {Rigopoulos}}]{Liu:2022rss}%
  \BibitemOpen
  \bibfield  {author} {\bibinfo {author} {\bibfnamefont {I.-K.}\ \bibnamefont
  {Liu}}, \bibinfo {author} {\bibfnamefont {N.~P.}\ \bibnamefont {Proukakis}},
  \ and\ \bibinfo {author} {\bibfnamefont {G.}~\bibnamefont {Rigopoulos}},\
  }\href {\doibase 10.1093/mnras/stad591} {\bibfield  {journal} {\bibinfo
  {journal} {Mon. Not. Roy. Astron. Soc.}\ }\textbf {\bibinfo {volume} {521}},\
  \bibinfo {pages} {3625} (\bibinfo {year} {2023})},\ \Eprint
  {http://arxiv.org/abs/2211.02565} {arXiv:2211.02565 [astro-ph.CO]}
  \BibitemShut {NoStop}%
\bibitem [{\citenamefont {Heinen}\ \emph {et~al.}(2022)\citenamefont {Heinen},
  \citenamefont {Mikheev}, \citenamefont {Schmied},\ and\ \citenamefont
  {Gasenzer}}]{Heinen:2022rew}%
  \BibitemOpen
  \bibfield  {author} {\bibinfo {author} {\bibfnamefont {P.}~\bibnamefont
  {Heinen}}, \bibinfo {author} {\bibfnamefont {A.~N.}\ \bibnamefont {Mikheev}},
  \bibinfo {author} {\bibfnamefont {C.-M.}\ \bibnamefont {Schmied}}, \ and\
  \bibinfo {author} {\bibfnamefont {T.}~\bibnamefont {Gasenzer}},\ }\href
  {https://arxiv.org/abs/2212.01162} {\  (\bibinfo {year} {2022})},\ \Eprint
  {http://arxiv.org/abs/2212.01162} {arXiv:2212.01162 [cond-mat.quant-gas]}
  \BibitemShut {NoStop}%
\bibitem [{\citenamefont {Heinen}\ \emph {et~al.}(2023)\citenamefont {Heinen},
  \citenamefont {Mikheev},\ and\ \citenamefont
  {Gasenzer}}]{Heinen2023a.PhysRevA.107.043303}%
  \BibitemOpen
  \bibfield  {author} {\bibinfo {author} {\bibfnamefont {P.}~\bibnamefont
  {Heinen}}, \bibinfo {author} {\bibfnamefont {A.~N.}\ \bibnamefont {Mikheev}},
  \ and\ \bibinfo {author} {\bibfnamefont {T.}~\bibnamefont {Gasenzer}},\
  }\href {\doibase 10.1103/PhysRevA.107.043303} {\bibfield  {journal} {\bibinfo
   {journal} {Phys. Rev. A}\ }\textbf {\bibinfo {volume} {107}},\ \bibinfo
  {pages} {043303} (\bibinfo {year} {2023})},\ \Eprint
  {http://arxiv.org/abs/2212.01163} {arXiv:2212.01163 [cond-mat.quant-gas]}
  \BibitemShut {NoStop}%
\bibitem [{\citenamefont {Siovitz}\ \emph {et~al.}(2023)\citenamefont
  {Siovitz}, \citenamefont {Lannig}, \citenamefont {Deller}, \citenamefont
  {Strobel}, \citenamefont {Oberthaler},\ and\ \citenamefont
  {Gasenzer}}]{Siovitz:2023ius.PhysRevLett.131.183402}%
  \BibitemOpen
  \bibfield  {author} {\bibinfo {author} {\bibfnamefont {I.}~\bibnamefont
  {Siovitz}}, \bibinfo {author} {\bibfnamefont {S.}~\bibnamefont {Lannig}},
  \bibinfo {author} {\bibfnamefont {Y.}~\bibnamefont {Deller}}, \bibinfo
  {author} {\bibfnamefont {H.}~\bibnamefont {Strobel}}, \bibinfo {author}
  {\bibfnamefont {M.~K.}\ \bibnamefont {Oberthaler}}, \ and\ \bibinfo {author}
  {\bibfnamefont {T.}~\bibnamefont {Gasenzer}},\ }\href 
  {\doibase 10.1103/PhysRevLett.131.183402} {\bibfield  {journal} {\bibinfo  {journal}
  {Phys. Rev. Lett.}\ }\textbf {\bibinfo {volume} {131}},\ \bibinfo {pages}
  {183402} (\bibinfo {year} {2023})}\BibitemShut {NoStop}%
\bibitem [{\citenamefont {Gliott}\ \emph {et~al.}(2024)\citenamefont {Gliott},
  \citenamefont {Ran\ifmmode~\mbox{\c{c}}\else \c{c}\fi{}on},\ and\
  \citenamefont {Cherroret}}]{Gliott2024a.PhysRevLett.133.233403}%
  \BibitemOpen
  \bibfield  {author} {\bibinfo {author} {\bibfnamefont {E.}~\bibnamefont
  {Gliott}}, \bibinfo {author} {\bibfnamefont {A.}~\bibnamefont
  {Ran\ifmmode~\mbox{\c{c}}\else \c{c}\fi{}on}}, \ and\ \bibinfo {author}
  {\bibfnamefont {N.}~\bibnamefont {Cherroret}},\ }\href 
  {\doibase 10.1103/PhysRevLett.133.233403} {\bibfield  {journal} {\bibinfo  {journal}
  {Phys. Rev. Lett.}\ }\textbf {\bibinfo {volume} {133}},\ \bibinfo {pages}
  {233403} (\bibinfo {year} {2024})}\BibitemShut {NoStop}%
\bibitem [{\citenamefont {Noel}\ and\ \citenamefont
  {Spitz}(2024)}]{Noel2024:PhysRevD.109.056011}%
  \BibitemOpen
  \bibfield  {author} {\bibinfo {author} {\bibfnamefont {V.}~\bibnamefont
  {Noel}}\ and\ \bibinfo {author} {\bibfnamefont {D.}~\bibnamefont {Spitz}},\
  }\href {\doibase 10.1103/PhysRevD.109.056011} {\bibfield  {journal} {\bibinfo
   {journal} {Phys. Rev. D}\ }\textbf {\bibinfo {volume} {109}},\ \bibinfo
  {pages} {056011} (\bibinfo {year} {2024})}\BibitemShut {NoStop}%
\bibitem [{\citenamefont {Noel}\ \emph {et~al.}(2025)\citenamefont {Noel},
  \citenamefont {Gasenzer},\ and\ \citenamefont {Boguslavski}}]{Noel:2025mtb}%
  \BibitemOpen
  \bibfield  {author} {\bibinfo {author} {\bibfnamefont {V.}~\bibnamefont
  {Noel}}, \bibinfo {author} {\bibfnamefont {T.}~\bibnamefont {Gasenzer}}, \
  and\ \bibinfo {author} {\bibfnamefont {K.}~\bibnamefont {Boguslavski}},\
  }\href@noop {} {\  (\bibinfo {year} {2025})},\ \Eprint
  {http://arxiv.org/abs/2503.01771} {arXiv: 2503.01771 [cond-mat.quant-gas]}
  \BibitemShut {NoStop}%
\bibitem [{\citenamefont {Berges}(2016)}]{Berges:2015kfa}%
  \BibitemOpen
  \bibfield  {author} {\bibinfo {author} {\bibfnamefont {J.}~\bibnamefont
  {Berges}},\ }in\ \href {\doibase 10.1093/acprof:oso/9780198768166.001.0001}
  {\emph {\bibinfo {booktitle} {Proc. Int. School on Strongly Interacting
  Quantum Systems Out of Equilibrium, Les Houches}}},\ \bibinfo {editor}
  {edited by\ \bibinfo {editor} {\bibfnamefont {T.}~\bibnamefont {{Giamarchi et
  al.}}}}\ (\bibinfo  {publisher} {OUP, Oxford},\ \bibinfo {year} {2016})\ pp.\
  \bibinfo {pages} {69--206},\ \Eprint {http://arxiv.org/abs/1503.02907}
  {arXiv:1503.02907 [hep-ph]} \BibitemShut {NoStop}%
\bibitem [{\citenamefont {Langen}\ \emph {et~al.}(2016)\citenamefont {Langen},
  \citenamefont {Gasenzer},\ and\ \citenamefont
  {Schmiedmayer}}]{Langen:2016vdb}%
  \BibitemOpen
  \bibfield  {author} {\bibinfo {author} {\bibfnamefont {T.}~\bibnamefont
  {Langen}}, \bibinfo {author} {\bibfnamefont {T.}~\bibnamefont {Gasenzer}}, \
  and\ \bibinfo {author} {\bibfnamefont {J.}~\bibnamefont {Schmiedmayer}},\
  }\href {\doibase 10.1088/1742-5468/2016/06/064009} {\bibfield  {journal}
  {\bibinfo  {journal} {J. Stat. Mech.}\ }\textbf {\bibinfo {volume} {1606}},\
  \bibinfo {pages} {064009} (\bibinfo {year} {2016})},\ \Eprint
  {http://arxiv.org/abs/1603.09385} {arXiv:1603.09385 [cond-mat.quant-gas]}
  \BibitemShut {NoStop}%
\bibitem [{\citenamefont {Schmied}\ \emph
  {et~al.}(2019{\natexlab{b}})\citenamefont {Schmied}, \citenamefont
  {Mikheev},\ and\ \citenamefont {Gasenzer}}]{Schmied:2018mte}%
  \BibitemOpen
  \bibfield  {author} {\bibinfo {author} {\bibfnamefont {C.-M.}\ \bibnamefont
  {Schmied}}, \bibinfo {author} {\bibfnamefont {A.~N.}\ \bibnamefont
  {Mikheev}}, \ and\ \bibinfo {author} {\bibfnamefont {T.}~\bibnamefont
  {Gasenzer}},\ }\href {\doibase 10.1142/S0217751X19410069} {\bibfield
  {journal} {\bibinfo  {journal} {Int. J. Mod. Phys. A}\ }\textbf {\bibinfo
  {volume} {34}},\ \bibinfo {pages} {1941006} (\bibinfo {year}
  {2019}{\natexlab{b}})},\ \Eprint {http://arxiv.org/abs/1810.08143}
  {arXiv:1810.08143 [cond-mat.quant-gas]} \BibitemShut {NoStop}%
\bibitem [{\citenamefont {Mikheev}\ \emph {et~al.}(2023)\citenamefont
  {Mikheev}, \citenamefont {Siovitz},\ and\ \citenamefont
  {Gasenzer}}]{Mikheev:2023juq}%
  \BibitemOpen
  \bibfield  {author} {\bibinfo {author} {\bibfnamefont {A.~N.}\ \bibnamefont
  {Mikheev}}, \bibinfo {author} {\bibfnamefont {I.}~\bibnamefont {Siovitz}}, \
  and\ \bibinfo {author} {\bibfnamefont {T.}~\bibnamefont {Gasenzer}},\ }\href
  {\doibase 10.1140/epjs/s11734-023-00974-7} {\bibfield  {journal} {\bibinfo
  {journal} {Eur. Phys. J. Spec. Top.}\ }\textbf {\bibinfo {volume} {232}},\
  \bibinfo {pages} {3393} (\bibinfo {year} {2023})},\ \Eprint
  {http://arxiv.org/abs/2304.12464} {arXiv:2304.12464 [cond-mat.quant-gas]}
  \BibitemShut {NoStop}%
\bibitem [{\citenamefont {K{\"o}per}(2023)}]{Koeper2023a}%
  \BibitemOpen
  \bibfield  {author} {\bibinfo {author} {\bibfnamefont {H.}~\bibnamefont
  {K{\"o}per}},\ }\emph {\bibinfo {title} {Approximations to the two-particle
  irreducible quantum effective action}},\ \href@noop {} {\bibinfo {type}
  {MSc thesis}},\ \bibinfo  {school} {Universit{\"a}t Heidelberg} (\bibinfo
  {year} {2023})\BibitemShut {NoStop}%
\bibitem [{\citenamefont {Kawaguchi}\ and\ \citenamefont
  {Ueda}(2012)}]{Kawaguchi2012a.PhyRep.520.253}%
  \BibitemOpen
  \bibfield  {author} {\bibinfo {author} {\bibfnamefont {Y.}~\bibnamefont
  {Kawaguchi}}\ and\ \bibinfo {author} {\bibfnamefont {M.}~\bibnamefont
  {Ueda}},\ }\href {\doibase 10.1016/j.physrep.2012.07.005} {\bibfield
  {journal} {\bibinfo  {journal} {Phys. Rep.}\ }\textbf {\bibinfo {volume}
  {520}},\ \bibinfo {pages} {253 } (\bibinfo {year} {2012})}\BibitemShut
  {NoStop}%
\bibitem [{\citenamefont {Kunkel}(2019)}]{Kunkel2019Thesis}%
  \BibitemOpen
  \bibfield  {author} {\bibinfo {author} {\bibfnamefont {P.}~\bibnamefont
  {Kunkel}},\ }\emph {\bibinfo {title} {Splitting a Bose-Einstein condensate
  enables EPR steering and simultaneous readout of noncommuting observables}},\
  \href{\doibase 10.11588/heidok.00027462} {Ph{D} thesis},\ \bibinfo  {school} 
  {Universität Heidelberg}
  (\bibinfo {year} {2019})\BibitemShut {NoStop}%
\bibitem [{\citenamefont {Gritsev}\ \emph {et~al.}(2007)\citenamefont
  {Gritsev}, \citenamefont {Polkovnikov},\ and\ \citenamefont
  {Demler}}]{Gritsev2007a.PhysRevB.75.174511}%
  \BibitemOpen
  \bibfield  {author} {\bibinfo {author} {\bibfnamefont {V.}~\bibnamefont
  {Gritsev}}, \bibinfo {author} {\bibfnamefont {A.}~\bibnamefont
  {Polkovnikov}}, \ and\ \bibinfo {author} {\bibfnamefont {E.}~\bibnamefont
  {Demler}},\ }\href {\doibase 10.1103/PhysRevB.75.174511} {\bibfield
  {journal} {\bibinfo  {journal} {Phys. Rev. B}\ }\textbf {\bibinfo {volume}
  {75}},\ \bibinfo {pages} {174511} (\bibinfo {year} {2007})}\BibitemShut
  {NoStop}%
\bibitem [{\citenamefont {Coleman}(1975)}]{Coleman1975a.PRD11.2088}%
  \BibitemOpen
  \bibfield  {author} {\bibinfo {author} {\bibfnamefont {S.}~\bibnamefont
  {Coleman}},\ }\href {\doibase 10.1103/PhysRevD.11.2088} {\bibfield  {journal}
  {\bibinfo  {journal} {Phys. Rev. D}\ }\textbf {\bibinfo {volume} {11}},\
  \bibinfo {pages} {2088} (\bibinfo {year} {1975})}\BibitemShut {NoStop}%
\bibitem [{\citenamefont {Kosterlitz}\ and\ \citenamefont
  {Thouless}(1973)}]{Kosterlitz1973a}%
  \BibitemOpen
  \bibfield  {author} {\bibinfo {author} {\bibfnamefont {J.}~\bibnamefont
  {Kosterlitz}}\ and\ \bibinfo {author} {\bibfnamefont {D.}~\bibnamefont
  {Thouless}},\ }\href {\doibase 10.1088/0022-3719/6/7/010} {\bibfield
  {journal} {\bibinfo  {journal} {J. Phys. C: Sol. St. Phys.}\ }\textbf
  {\bibinfo {volume} {6}},\ \bibinfo {pages} {1181} (\bibinfo {year}
  {1973})}\BibitemShut {NoStop}%
\bibitem [{\citenamefont {Amit}\ \emph {et~al.}(1980)\citenamefont {Amit},
  \citenamefont {Goldschmidt},\ and\ \citenamefont
  {Grinstein}}]{Amit1980a.JPA13.585}%
  \BibitemOpen
  \bibfield  {author} {\bibinfo {author} {\bibfnamefont {D.}~\bibnamefont
  {Amit}}, \bibinfo {author} {\bibfnamefont {Y.}~\bibnamefont {Goldschmidt}}, \
  and\ \bibinfo {author} {\bibfnamefont {S.}~\bibnamefont {Grinstein}},\ }\href
  {\doibase 10.1088/0305-4470/13/2/024} {\bibfield  {journal} {\bibinfo
  {journal} {J. Phys. A: Math. Gen.}\ }\textbf {\bibinfo {volume}
  {13}},\ \bibinfo {pages} {585} (\bibinfo {year} {1980})}\BibitemShut
  {NoStop}%
\bibitem [{\citenamefont {Giamarchi}(2003)}]{Giamarchi2003a}%
  \BibitemOpen
  \bibfield  {author} {\bibinfo {author} {\bibfnamefont {T.}~\bibnamefont
  {Giamarchi}},\ }\href {\doibase 10.1093/acprof:oso/9780198525004.001.0001}
  {\emph {\bibinfo {title} {{Quantum Physics in One Dimension}}}}\ (\bibinfo
  {publisher} {Oxford University Press},\ \bibinfo {year} {2003})\BibitemShut
  {NoStop}%
\bibitem [{\citenamefont {Cazalilla}\ \emph {et~al.}(2011)\citenamefont
  {Cazalilla}, \citenamefont {Citro}, \citenamefont {Giamarchi}, \citenamefont
  {Orignac},\ and\ \citenamefont {Rigol}}]{Cazalilla2011a}%
  \BibitemOpen
  \bibfield  {author} {\bibinfo {author} {\bibfnamefont {M.~A.}\ \bibnamefont
  {Cazalilla}}, \bibinfo {author} {\bibfnamefont {R.}~\bibnamefont {Citro}},
  \bibinfo {author} {\bibfnamefont {T.}~\bibnamefont {Giamarchi}}, \bibinfo
  {author} {\bibfnamefont {E.}~\bibnamefont {Orignac}}, \ and\ \bibinfo
  {author} {\bibfnamefont {M.}~\bibnamefont {Rigol}},\ }\href 
  {\doibase 10.1103/RevModPhys.83.1405} {\bibfield  {journal} {\bibinfo  {journal} {Rev.
  Mod. Phys.}\ }\textbf {\bibinfo {volume} {83}},\ \bibinfo {pages} {1405}
  (\bibinfo {year} {2011})},\ \Eprint {http://arxiv.org/abs/1101.5337}
  {arXiv:1101.5337 [cond-mat.str-el]} \BibitemShut {NoStop}%
\bibitem [{\citenamefont {Mikheev}(2023)}]{Mikheev2023a}%
  \BibitemOpen
  \bibfield  {author} {\bibinfo {author} {\bibfnamefont {A.~N.}\ \bibnamefont
  {Mikheev}},\ }\emph {\bibinfo {title} {Far-from-equilibrium universal scaling
  dynamics in ultracold atomic systems and heavy-ion collisions}},\ \href
  {\doibase 10.11588/heidok.00032924} {\bibinfo {type} {Ph{D} thesis}},\
  \bibinfo  {school} {Ruprecht-Karls Universit{\"a}t Heidelberg} (\bibinfo
  {year} {2023})\BibitemShut {NoStop}%
\bibitem [{\citenamefont {Cahn}\ and\ \citenamefont
  {Hilliard}(1958)}]{Cahn1958a.JChemPhys.28.258}%
  \BibitemOpen
  \bibfield  {author} {\bibinfo {author} {\bibfnamefont {J.~W.}\ \bibnamefont
  {Cahn}}\ and\ \bibinfo {author} {\bibfnamefont {J.~E.}\ \bibnamefont
  {Hilliard}},\ }\href {\doibase 10.1063/1.1744102} {\bibfield  {journal}
  {\bibinfo  {journal} {The Journal of Chemical Physics}\ }\textbf {\bibinfo
  {volume} {28}},\ \bibinfo {pages} {258} (\bibinfo {year} {1958})}\BibitemShut
  {NoStop}%
\bibitem [{\citenamefont {Yu}\ and\ \citenamefont
  {Blakie}(2021)}]{blakie_2021;PhysRevResearch.3.023043}%
  \BibitemOpen
  \bibfield  {author} {\bibinfo {author} {\bibfnamefont {X.}~\bibnamefont
  {Yu}}\ and\ \bibinfo {author} {\bibfnamefont {P.B.}~\bibnamefont
  {Blakie}},\ }\href {\doibase 10.1103/PhysRevResearch.3.023043} {\bibfield
  {journal} {\bibinfo  {journal} {Phys. Rev. Res.}\ }\textbf {\bibinfo {volume}
  {3}},\ \bibinfo {pages} {023043} (\bibinfo {year} {2021})},\ \Eprint
  {https://arxiv.org/abs/2008.08175} {arXiv:2008.08175v2 [cond-mat.quant-gas]} \BibitemShut
  {NoStop}%
\bibitem [{\citenamefont {Dmitriev}\ , \emph
  {et~al.}(2008)\citenamefont {Dmitriev}, \citenamefont
  {Panayotis},\ and\ \citenamefont {Kivshar}}]{dmitriev_2008;PhysRevE.78.046604}%
  \BibitemOpen
  \bibfield  {author} {\bibinfo {author} {\bibfnamefont {S.~V.}\ \bibnamefont
  {Dmitriev}}, \bibinfo {author} {\bibfnamefont {P.~G.}~\bibnamefont {Panayotis}}, \
  and\ \bibinfo {author} {\bibfnamefont {Y.~S.}\ \bibnamefont {Kivshar}},\
  }\href {\doibase 10.1103/PhysRevE.78.046604} {\bibfield  {journal} {\bibinfo
   {journal} {Phys. Rev. E}\ }\textbf {\bibinfo {volume} {78}},\ \bibinfo
  {pages} {046604} (\bibinfo {year} {2008}{\natexlab{d}})}\
  \BibitemShut {NoStop}%
\bibitem [{\citenamefont {Campbell}\ , \emph
  {et~al.}(1986)\citenamefont {Campbell}, \citenamefont
  {Peyrard},\ and\ \citenamefont {Sodano}}]{campbell_1986}%
  \BibitemOpen
  \bibfield  {author} {\bibinfo {author} {\bibfnamefont {D.~K.}\ \bibnamefont
  {Campbell}}, \bibinfo {author} {\bibfnamefont {M.}~\bibnamefont {Peyrard}}, \
  and\ \bibinfo {author} {\bibfnamefont {P.}\ \bibnamefont {Sodano}},\
  }\href {\doibase https://doi.org/10.1016/0167-2789(86)90019-9} {\bibfield  {journal} {\bibinfo
   {journal} {Physica D: Nonlin.~Phen.}\ }\textbf {\bibinfo {volume} {19}},\ \bibinfo
  {pages} {165} (\bibinfo {year} {1986}{\natexlab{d}})}\
  \BibitemShut {NoStop}%
\bibitem [{\citenamefont {Prüfer}\ \emph {et~al.}(2022)\citenamefont
  {Prüfer}, \citenamefont {Spitz}, \citenamefont {Lannig}, \citenamefont
  {Strobel}, \citenamefont {Berges},\ and\ \citenamefont
  {Oberthaler}}]{Prufer:2022therm}%
  \BibitemOpen
  \bibfield  {author} {\bibinfo {author} {\bibfnamefont {M.}~\bibnamefont
  {Prüfer}}, \bibinfo {author} {\bibfnamefont {D.}~\bibnamefont {Spitz}},
  \bibinfo {author} {\bibfnamefont {S.}~\bibnamefont {Lannig}}, \bibinfo
  {author} {\bibfnamefont {H.}~\bibnamefont {Strobel}}, \bibinfo {author}
  {\bibfnamefont {J.}~\bibnamefont {Berges}}, \ and\ \bibinfo {author}
  {\bibfnamefont {M.~K.}\ \bibnamefont {Oberthaler}},\ }\href {\doibase
  10.1038/s41567-022-01779-6} {\bibfield  {journal} {\bibinfo  {journal}
  {Nature Physics}\ }\textbf {\bibinfo {volume} {18}},\ \bibinfo {pages}
  {1459–1463} (\bibinfo {year} {2022})}\BibitemShut {NoStop}%
\bibitem [{\citenamefont {Kunkel}\ \emph {et~al.}(2019)\citenamefont {Kunkel},
  \citenamefont {Pr\"ufer}, \citenamefont {Lannig}, \citenamefont
  {Rosa-Medina}, \citenamefont {Bonnin}, \citenamefont {G\"arttner},
  \citenamefont {Strobel},\ and\ \citenamefont
  {Oberthaler}}]{Kunkel2019a.PhysRevLett.123.063603}%
  \BibitemOpen
  \bibfield  {author} {\bibinfo {author} {\bibfnamefont {P.}~\bibnamefont
  {Kunkel}}, \bibinfo {author} {\bibfnamefont {M.}~\bibnamefont {Pr\"ufer}},
  \bibinfo {author} {\bibfnamefont {S.}~\bibnamefont {Lannig}}, \bibinfo
  {author} {\bibfnamefont {R.}~\bibnamefont {Rosa-Medina}}, \bibinfo {author}
  {\bibfnamefont {A.}~\bibnamefont {Bonnin}}, \bibinfo {author} {\bibfnamefont
  {M.}~\bibnamefont {G\"arttner}}, \bibinfo {author} {\bibfnamefont
  {H.}~\bibnamefont {Strobel}}, \ and\ \bibinfo {author} {\bibfnamefont
  {M.~K.}\ \bibnamefont {Oberthaler}},\ }\href {\doibase
  10.1103/PhysRevLett.123.063603} {\bibfield  {journal} {\bibinfo  {journal}
  {Phys. Rev. Lett.}\ }\textbf {\bibinfo {volume} {123}},\ \bibinfo {pages}
  {063603} (\bibinfo {year} {2019})}\BibitemShut {NoStop}%
\bibitem [{\citenamefont {Hamley}\ \emph {et~al.}(2012)\citenamefont {Hamley},
  \citenamefont {Gerving}, \citenamefont {Hoang}, \citenamefont {Bookjans},\
  and\ \citenamefont {Chapman}}]{Hamley2012:NaturePhysics}%
  \BibitemOpen
  \bibfield  {author} {\bibinfo {author} {\bibfnamefont {C.~D.}\ \bibnamefont
  {Hamley}}, \bibinfo {author} {\bibfnamefont {C.}~\bibnamefont {Gerving}},
  \bibinfo {author} {\bibfnamefont {T.~M.}\ \bibnamefont {Hoang}}, \bibinfo
  {author} {\bibfnamefont {E.~M.}\ \bibnamefont {Bookjans}}, \ and\ \bibinfo
  {author} {\bibfnamefont {M.~S.}\ \bibnamefont {Chapman}},\ }\href {\doibase
  10.1038/nphys2245} {\bibfield  {journal} {\bibinfo  {journal} {Nature
  Physics}\ }\textbf {\bibinfo {volume} {8}},\ \bibinfo {pages} {305} (\bibinfo
  {year} {2012})}\BibitemShut {NoStop}%
\end{thebibliography}
%
\end{document}